%% file: revtomo.tex
\documentclass{report}
\usepackage{epsf,amssymb,psfig}
\begin{document}
\author{G. Mauro D'Ariano, Matteo G. A. Paris, and Massimiliano F. Sacchi}
\title{Quantum~Tomography}
\maketitle
\pagenumbering{arabic}
\tableofcontents
\newpage
\include{cap1}
\thispagestyle{empty}
\cleardoublepage
\include{cap2}

\thispagestyle{empty}
\cleardoublepage
\include{cap3}
\thispagestyle{empty}
\cleardoublepage
\include{cap4}

\thispagestyle{empty}
\cleardoublepage
\include{cap5}
\thispagestyle{empty}
\cleardoublepage
\include{cap6}
\thispagestyle{empty}
\cleardoublepage
\include{cap7}
\thispagestyle{empty}
\cleardoublepage
\include{cap8}

\thispagestyle{empty}
\cleardoublepage
\include{cap9}

\thispagestyle{empty}
\cleardoublepage

\include{biblio}
\end{document}

%% file: cap1.tex
\vfill
\section*{Acknowledgments}
The writing of the present Review has been co-sponsored by: the Italian {\em
Ministero dell'Istru\-zione, dell'Universita' e della Ricerca} (MIUR)
under the Cofinanziamento 2002 {\em Entanglement assisted high
precision measurements}, the {\em Istituto Nazionale di
Fisica della Materia} under the project PRA-2002-CLON, and 
by the European Community programs ATESIT (Contract No. IST-2000-29681) 
and EQUIP (Contract No. IST-1999-11053).
G. M. D. acknowledges partial support by the Department of Defense 
Multidisciplinary University Research Initiative (MURI) program
administered by the Army Research Office under Grant
No. DAAD19-00-1-0177. M. G. A. P. is research fellow at {\em Collegio 
Alessandro Volta}. 
\newpage 
\chapter{Introduction}
The state of a physical system is the mathematical description that 
provides a complete information on the system. Its knowledge is
equivalent to know the result of any possible measurement on the
system. In Classical Mechanics it is always possible, at least in
principle, to devise a procedure made of multiple measurements which
fully recovers the state of the system. In Quantum Mechanics, on the
contrary, this is not possible, due to the fundamental limitations
related to the Heisenberg uncertainty principle \cite{heis,vonNeumann} and the
no-cloning theorem \cite{noclon}. In fact, on one hand one
cannot perform an arbitrary sequence of measurements on a single
system without inducing on it a back-action of some sort. On the other
hand, the no-cloning theorem forbids to create a perfect copy of the
system without already knowing its state in advance. Thus, there is no
way out, not even in principle, to infer the quantum state of a single
system without having some prior knowledge on it \cite{imposs}. 

\par It is possible to estimate the unknown quantum state of a system
when many identical copies are available in the same state, so that a
different measurement can be performed on each copy. 
A procedure of such kind is called {\em quantum tomography}.  
The problem of finding a procedure to determine the state of a system
from multiple copies  was first addressed in 1957 by Fano \cite{fano},
who called {\em quorum} a set of observables sufficient for a complete
determination of the density matrix.  However, since for a
particle it is difficult to devise concretely measurable observables
other than position, momentum and energy, the fundamental problem of
measuring the quantum state has remained at the level of mere
speculation up to almost ten years ago, when the issue finally entered
the realm of experimental physics with the pioneering experiments by Raymer's 
group \cite{raymer} in the domain of quantum optics. In quantum optics,
in fact, using a balanced homodyne detector one has the unique
opportunity of measuring all possible linear combinations of position
and momentum of a harmonic oscillator, which here represents a single
mode of the electromagnetic field. 

\par The first technique to reconstruct the density matrix from
homodyne measurements --- so called {\em homodyne tomography} ---
originated from the observation by Vogel and Risken \cite{vogel} that
the collection of probability distributions achieved by homodyne
detection is just the Radon transform of the Wigner function
$W$. Therefore, as in classical imaging, by Radon transform inversion
one can obtain $W$, and then from $W$ the matrix elements of the
density operator. This first method, however, was affected by
uncontrollable approximations, since arbitrary smoothing parameters
are needed for the inverse Radon transform. In Ref. \cite{dmp} the
first exact technique was given for measuring experimentally the
matrix elements of the density operator in the photon-number
representation, by simply averaging functions of homodyne data. After
that, the method was further simplified \cite{dlp}, and the
feasibility for non-unit quantum efficiency of detectors---above some
bounds---was established.

The exact homodyne method has been implemented experimentally to
measure the photon statistics of a semiconductor laser
\cite{raymer95}, and the density matrix of a squeezed vacuum
\cite{sch}.  The success of optical homodyne tomography has then
stimulated the development of state-reconstruction procedures for
atomic beams \cite{atobeams}, the experimental determination of the
vibrational state of a molecule \cite{dunn}, of an ensemble of helium
atoms \cite{exper-freeatoms}, and of a single ion in a Paul trap
\cite{leib}.

\par Through quantum tomography the state is perfectly recovered in
the limit of infinite number of measurements, while in the practical
finite-measurements case, one can always estimate the statistical
error that affects the reconstruction.  For infinite
dimensions the propagation of statistical errors of the density matrix
elements make them useless for estimating the ensemble average of
unbounded operators, and a method for estimating the ensemble average
of arbitrary observable of the field without using the density matrix
elements has been derived \cite{tokio}. Further insight on the
general method of state reconstruction has lead to generalize homodyne
tomography to any number of modes \cite{homtom}, and then to extend 
the tomographic method from the harmonic oscillator to an arbitrary
quantum system using group theory \cite{chicago,paini-rete,deconv,math}.
A general data analysis method has been designed in order to unbias
the estimation procedure from any known instrumental noise
\cite{deconv}. Moreover, algorithms have been engineered to improve
the statistical errors on a given sample of experimental data---the
so-called adaptive tomography \cite{adapt}---and then max-likelihood
strategies \cite{maxlik} have been used that improved dramatically
statistical errors, however, at the expense of some bias in the
infinite dimensional case, and of exponential complexity versus $N$
for the joint tomography of $N$ quantum systems. The latest technical  
developments \cite{orth} derive the general tomographic method from
spanning sets of operators, the previous group theoretical approaches
\cite{chicago,paini-rete,deconv,math} being just a particular case of this
general method, where the group representation is just a device to
find suitable operator ``orthogonality'' and ``completeness''
relations in the linear algebra of operators. Finally, very recently,
a method for tomographic estimation of the unknown quantum operation
of a quantum device has been derived \cite{cptomo}, which uses a
single fixed input entangled state, which plays the role of
all possible input states in quantum parallel on the tested device,
making finally the method a true ``quantum radiography'' of the
functioning of a device.

\par In this Review we will give a self-contained and unified
derivation of the methods of quantum tomography, with examples of
applications to different kinds of quantum systems, and with
particular focus on quantum optics, where also some results from
experiments are reexamined. The Review is organized as follows.  

\par In Chapter 2 we introduce the
generalized Wigner functions \cite{wig,cgl2} and we provide the basic
elements of detection theory in quantum optics, by giving the
description of photodetection, homodyne detection, and heterodyne
detection. As we will see, heterodyne detection also provides a
method for estimating the ensemble average of polynomials in the field
operators, however, it is unsuitable for the density matrix elements
in the photon-number representation. 
The effect of non unit quantum efficiency is taken into
account for all such detection schemes.
\par
In Chapter 3 we give a brief history of quantum tomography, starting with the
first proposal of Vogel and Risken \cite{vogel} as the extension to the domain
of quantum optics of the conventional tomographic imaging. As already
mentioned, this method indirectly recovers the state of
the system through the reconstruction of the Wigner function, and is
affected by uncontrollable bias.  The exact homodyne tomography method
of Ref. \cite{dmp} (successively simplified in Ref. \cite{dlp}) is
here presented on the basis of the general tomographic method of
spanning sets of operators of Ref. \cite{orth}. As another application
of the general method, the tomography of spin systems \cite{spinmac}
is provided from the group theoretical method of
Refs. \cite{chicago,paini-rete, deconv}. In this chapter we include also further
developments to improve the method, such as the deconvolution
techniques of \cite{deconv} to correct the effects of experimental
noise by data processing, and the adaptive tomography \cite{adapt} to
reduce the statistical fluctuations of tomographic estimators.
\par
Chapter 4 is devoted to the evaluation from Ref. \cite{tokio} of the
expectation value of arbitrary operators of a single-mode radiation
field via homodyne tomography. Here we also report from
Ref. \cite{added} the estimation of the added noise with respect to
the perfect measurement of field observables, for some relevant
observables, along with a comparison with the noise that would have
been obtained using heterodyne detection.
\par
The generalization of Ref. \cite{homtom} of homodyne tomography to
many modes of radiation is reviewed in Chapter 5, where it is shown how
tomography of a multimode field can be performed by using only a
single local oscillator with a tunable field mode. Some results of
Monte Carlo simulations from Ref. \cite{homtom} are also shown for the
state that describes light from parametric downconversion.
\par
Chapter 6 reviews some applications of quantum homodyne tomography to perform
fundamental test of quantum mechanics.  The first is the proposal of
Ref. \cite{noncl}
to measure the nonclassicality of radiation field. The second is the
scheme of Ref. \cite{sr} to test the state reduction rule using light
from parametric downconversion. Finally, we review some experimental
results about tomography of coherent signals with applications to the
estimation of losses introduced by simple optical components \cite{TomoNa}.
\par
Chapter 7 reviews the tomographic method of Ref. \cite{cptomo} to
reconstruct the quantum operation of a device, such as an amplifier or
a measuring device, using a single fixed input entangled state, which
plays the role of all possible input states in a quantum parallel fashion.
\par
Chapter 8 is devoted to the reconstruction technique of
Ref. \cite{maxlik} based on the maximum likelihood principle. As
mentioned, for infinite dimensions this method is necessarily biased,
however, it is more suited to the estimation of a finite number of
parameters, as proposed in Ref. \cite{parlik}, or to the state
determination in the presence of very low number of experimental data
\cite{maxlik}. Unfortunately, the algorithm of this method has
exponential complexity versus the number of quantum systems for a 
joint tomography of many systems.
\par
Finally, in Chapter 9 we briefly review Ref. \cite{fict}, showing how
quantum tomography could be profitably used as a tool for
reconstruction and compression in classical imaging.   

%% file: cap2.tex
\chapter{Wigner functions and elements of detection theory}
In this chapter we review some simple formulas from Ref. \cite{mio13}
that connect the generalized Wigner functions for $s$-ordering with
the density matrix, and {\em vice-versa}. These formulas prove very
useful for quantum mechanical applications as, for example, for connecting master
equations with Fokker-Planck equations, or for evaluating the quantum
state from Monte Carlo simulations of Fokker-Planck equations, and
finally for studying positivity of the generalized Wigner functions in
the complex plane.  Moreover, as we will show in Chapter 3, the first
proposal of quantum state reconstruction \cite{vogel} used the Wigner
function as an intermediate step.
\par In the second part of the chapter, we evaluate the probability
distribution of the photocurrent of photodetectors, balanced homodyne
detectors, and heterodyne detectors. We show that under suitable
limits the respective photocurrents provide the measurement of the
photon number distribution, of the quadrature, and of the complex
amplitude of a single mode of the electromagnetic field. When the
effect of non-unit quantum efficiency is taken into account an
additional noise affects the measurement, giving a Bernoulli
convolution for photo-detection, and a Gaussian convolution for
homodyne and heterodyne detection.  Extensive use of the results in
this chapter will be made in the next chapters devoted to quantum
homodyne tomography.  \par
\section{Wigner functions \label{wigsec}}
\par Since Wigner's pioneering work \cite{wig}, generalized
phase-space tech\-niques have prov\-ed very useful in various branches
of physics \cite{phs}.  As a method to express the density
operator in terms of c-number functions, the Wigner functions often
lead to considerable simplification of the quantum equations of
motion, as for example, for transforming master equations in operator
form into more amenable Fokker-Planck differential equations (see, for
example, Ref. \cite{gard}).  Using the Wigner function one can express
quantum-mechanical expectation values in form of averages over the
complex plane (the classical phase-space), the Wigner function playing
the role of a c-number quasi-probability distribution, which generally
can also have negative values.  More precisely, the original Wigner
function allows to easily evaluate expectations of symmetrically
ordered products of the field operators, corresponding to the Weyl's
quantization procedure \cite{weyl}. However, with a slight change of
the original definition, one defines generalized $s$-ordered Wigner
function $W_s(\alpha,\alpha ^*)$, as follows \cite{cgl2}
\begin{eqnarray}
W_s(\alpha,\alpha ^*)=\int _{\mathbb C}\frac{d^2\lambda}{\pi^2}
e^{\alpha\lambda^*-\alpha ^*\lambda+{ \frac s 2}|\lambda|^2}
\mbox{Tr}[D(\lambda)\rho ]\;,\label{Ws}
\end{eqnarray}
where $\alpha ^*$ denotes the complex conjugate of $\alpha $, 
the integral is performed on the complex plane with
measure $d^2\lambda=d{\hbox {Re}}\lambda\,d{\hbox {Im}}\lambda$, 
$\rho $ represents the density operator, and 
\begin{eqnarray}
D(\alpha )\equiv \exp(\alpha a^\dag  -\alpha ^*a)
\;\label{disp}
\end{eqnarray}
denotes the 
displacement operator, where $a$ and $a^{\dag}$ ($[a,a^{\dag}]=1$) are the
annihilation and creation operators of the field mode of interest. 
The Wigner functions in Eq. (\ref{Ws}) allow to evaluate 
$s$-ordered expectation values of the field operators 
through the following relation 
\begin{eqnarray}
\mbox{Tr}[\mbox{{\bf :}} (a^{\dag})^n a^m \mbox{{\bf :}}_s\; \rho ]
=\int _{\mathbb C} d^2\alpha\,W_s(\alpha,
\alpha ^* )\,\alpha ^{*n}\alpha^m\;.\label{ex}
\end{eqnarray}
The particular cases $s=-1,0,1$ correspond to {\em anti-normal}, {\em
symmetrical}, and {\em normal} ordering, respectively. In these cases
the generalized Wigner function $W_s(\alpha, \alpha ^*)$ are usually
denoted by the following symbols and names
\begin{eqnarray}
\begin{array}{ll}
{\frac 1 \pi}Q(\alpha,\alpha ^*) \qquad &\mbox{ for }s=-1
\mbox{ ``$Q$ function''} \\
W(\alpha,\alpha ^*)\qquad & \mbox{ for }s=0\mbox{ (usual Wigner function)}
\\  P(\alpha,\alpha ^*)\qquad &\mbox{ for }s=1\mbox{ ``$P$ function''}
\end{array}
\;\label{3}
\end{eqnarray}
For the normal ($s=1$) and anti-normal ($s=-1$) orderings, the
following  simple relations with the density matrix are well known 
\begin{eqnarray}
&&Q(\alpha,\alpha ^*)\equiv\langle \alpha|\rho |\alpha\rangle \;,\label{QQ}\\
&&\rho =\int _{\mathbb C}d^2\alpha\,P(\alpha,\alpha ^*)\,|\alpha\rangle \langle \alpha|\;,
\label{P}\end{eqnarray}
where $|\alpha \rangle $ denotes the customary coherent state $|\alpha
\rangle =D(\alpha )|0\rangle $, $|0\rangle $ being the vacuum state of the field.  Among
the three particular representations (\ref{3}), the $Q$ function is
positively definite and infinitely differentiable (it actually
represents the probability distribution for ideal joint measurements
of position and momentum of the harmonic oscillator: see
Sec. \ref{hetsec}). On the other hand, the $P$ function is known to be
possibly highly singular, and the only pure states for which it is
positive are the coherent states \cite{cah}. Finally, the usual Wigner
function has the remarkable property of providing the probability
distribution of the quadratures of the field in form of marginal
distribution, namely
\begin{eqnarray}
\int _{-\infty}^{\infty} 
d\, {\hbox {Im}}\alpha\,W(\alpha e^{i\varphi },\alpha ^*e^{-i\varphi })
={}_{\varphi }\langle  {{\hbox {Re}}\alpha|\rho |{\hbox
    {Re}}\alpha} \rangle _{\varphi }   \;,\label{margw}
\end{eqnarray}
where $|x \rangle _{\varphi }$ denotes the (unnormalizable) eigenstate of the field 
quadrature 
\begin{eqnarray}
X_{\varphi }=\frac {{a^\dag } e^{i\varphi }+ a e^{-i \varphi }}{2}\;\label{xfi}
\end{eqnarray}
with real eigenvalue $x$.  Notice that any couple of quadratures
$X_{\varphi }$, $X_{\varphi +\pi/2}$ is canonically conjugate, namely
$[ X_{\varphi },X_{\varphi +\pi/2}]=i/2$, and it is equivalent to
position and momentum of a harmonic oscillator. Usually, negative
values of the Wigner function are viewed as signature of a
non-classical state,  the most eloquent example being the
Schr\"odinger-cat state \cite{cat}, whose Wigner function is
characterized by rapid oscillations around the origin of the complex
plane.  From Eq. (\ref{Ws}) one can notice that all $s$-ordered
Wigner functions are related to each other through Gaussian convolution
\begin{eqnarray}
W_s(\alpha,\alpha ^*)&=&\int _{\mathbb C} d^2\beta\,
W_{s'}(\beta,\beta^*)
\frac{2}{\pi(s'-s)}\exp\left(-\frac{2}{s'-s}|\alpha-\beta|^2\right)
\label{convw}
\\
&=&\exp\left(\frac{s'-s}{2}
\frac{\partial^2}{\partial\alpha\partial\alpha ^*}\right)
W_{s'}(\alpha,\alpha ^*)\;,\quad (s'>s)\;.
\end{eqnarray}
Equation (\ref{convw}) shows the positivity of the generalized
Wigner function for $s<-1$, as a consequence of the positivity of
the $Q$ function.  From a qualitative point of view, the maximum value
of $s$ keeping the generalized Wigner functions as positive can be
considered as an indication of the classical nature of the physical
state \cite{lee1}.

An equivalent expression for $W_s(\alpha ,\alpha ^* )$ can be derived
as follows \cite{mio13}. Eq. (\ref{Ws}) can be rewritten as 
\begin{eqnarray}
W_s(\alpha ,\alpha ^* ) =
\mbox{Tr}[\rho D(\alpha ) \hat W_s D^\dag (\alpha )]\;,\label{Ws2}
\end{eqnarray}
where 
\begin{eqnarray}
\hat W_s= \int _{\mathbb C} \frac{d^2\lambda }{\pi^2}
e^{ \frac s 2|\lambda|^2}
\,D(\lambda)\;.\label{ws3}
\end{eqnarray}
Through the customary Baker-Campbell-Hausdorff (BCH) formula 
\begin{eqnarray}
\exp{A}\exp{  B}=\exp\left(  A+  B+{\frac 1 2}[  A,  B]\right)
\;,\label{BCH_easy}
\end{eqnarray}
which holds when $[A,[A,B]]=[B,[A,B]]=0$, one writes the displacement
in normal order, and integrating on $\arg(\lambda )$ and $|\lambda |$
one obtains
\begin{eqnarray}
\hat W_s= \frac {2}{\pi (1-s)} \sum _{n=0}^\infty \frac{1}{n!} \left( 
\frac {2}{s-1}\right)^n\,a^{\dag n}a^n 
=\frac {2}{\pi (1-s)} \left( \frac {s+1}{s-1}\right)^{a^\dag a}
\;,
\end{eqnarray}
 where we used the normal-ordered forms
\begin{eqnarray}
\mbox{{\bf :}}({a^\dag } a)^n \mbox{{\bf :}}
=({a^\dag } )^n a^n ={a^\dag } a({a^\dag } a-1)\ldots({a^\dag } a-n+1)\;,\label{recurr}
\end{eqnarray}
and the identity
\begin{eqnarray}
\mbox{{\bf :}}e^{-x{a^\dag } a}\mbox{{\bf :}}=
\sum_{l=0}^{\infty}{\frac{(-x)^l}{l!}}({a^\dag })^l a^l
=(1-x)^{{a^\dag } a}.\label{Louisell}
\end{eqnarray}
\par 
The density matrix can be recovered from 
the generalized Wigner functions using the following expression
\begin{eqnarray}
\rho =\frac{2}{1+s}\int  _{\mathbb C}  d^2\alpha\, W_s(\alpha,\alpha ^*)
e^{-\frac{2}{1+s}|\alpha|^2}\,e^{\frac{2\alpha}{1+s} a^{\dag}}\,
\left(\frac{s-1}{s+1}\right)^{a^{\dag}a}\,
e^{\frac{2\alpha ^*}{1+s}a}\;.\label{mybest2}
\end{eqnarray}
For the proof of Eq. (\ref{mybest2}) the reader is referred to
Ref. \cite{mio13}. In particular, for $s=0$ one has
the inverse of the Glauber formula
\begin{eqnarray}
\rho =2 \int  _{\mathbb C} d^2 \alpha\, W(\alpha,\alpha ^*) D(2\alpha)
(-)^{a^{\dag}a}\;,
\end{eqnarray}
whereas for $s=1$ one recovers Eq. (\ref{P}) that defines the
$P$ function. 

\section{Photodetection}\label{phdet}
Light is revealed by exploiting its interaction with
atoms/molecules or electrons in a solid, and, essentially, each photon
ionizes a single atom or promotes an electron to a conduction band,
and the resulting charge is then amplified to produce a measurable
pulse. In practice, however, available photodetectors are not
ideally counting all photons, and their performances is limited by a
non-unit quantum efficiency $\zeta$, namely only a fraction $\zeta$ of
the incoming photons lead to an electric signal, and ultimately to a
{\em count}: some photons are either reflected from the surface of the
detector, or are absorbed without being transformed into electric pulses.

Let us consider a light beam entering a photodetector of quantum
efficiency $\zeta$, {\em i.e.} a detector that transforms just a
fraction $\zeta$ of the incoming light pulse into electric signal. If
the detector is small with respect to the coherence length of
radiation and its window is open for a time interval $T$, then the 
Poissonian process of
counting gives a probability $p(m;T)$ of revealing $m$ photons that
writes \cite{kelley}
\begin{eqnarray}
p(m;T) = \hbox{Tr}\left[\rho \mbox{{\bf :}}{\frac {[\zeta I (T)T]^m }{m!}}
\exp [-\zeta I (T)T]\mbox{{\bf :}} \right]        
\label{pc-gen}\;, 
\end{eqnarray}
where $\rho $ is the quantum state of light, $\mbox{{\bf :}}\ \mbox{{\bf :}}$ 
denotes the normal ordering of field operators, and 
$I(T)$ is the beam intensity
\begin{eqnarray}
 I (T)={\frac {2\epsilon_0 c}{T}}\int_0^T {\mbox{\bf E}}^{(-)} (\mbox{\bf r},t)\cdot
{\mbox{\bf E}}^{(+)} (\mbox{\bf r},t) dt
\label{intensity}\;,
\end{eqnarray}
given in terms of the 
positive (negative) frequency part of the electric field operator
${\mbox{\bf E}}^{(+)} (\mbox{\bf r},t)$ (${\mbox{\bf E}}^{(-)} (\mbox{\bf r},t)$). 
The quantity $p(t)=\zeta \hbox{Tr}\left[\rho I (T) \right ]$ 
equals the probability of a single count during the time interval $(t,t+dt)$.
Let us now focus our attention to the case of the radiation field 
excited in a 
stationary state of a single mode at frequency $\omega $. 
Eq. (\ref{pc-gen}) can be rewritten as 
\begin{eqnarray}
p_\eta (m) = \hbox{Tr}\left[\rho \,
\mbox{{\bf :}}{\frac {(\eta a^{\dag} a)^m}{m!}}\exp
(-\eta a^{\dag} a)\mbox{{\bf :}} \right] 
\label{pc-singlemode}\;,
\end{eqnarray}
where the parameter $\eta=\zeta c\hbar\omega/V$ denotes the 
overall {\em quantum efficiency} of the photodetector. Using
Eqs. (\ref{recurr}) and (\ref{Louisell}) one obtains 
\begin{eqnarray}
p_\eta (m) =\sum_{n=m}^{\infty} \rho_{nn}
\left(\begin{array}{c} n\\m\end{array}\right) \eta^m (1-\eta )^{n-m}
\label{conv_n}\;,
\end{eqnarray}
where 
\begin{eqnarray}
\rho_{nn}\equiv \langle n|\rho | n \rangle 
=p_{\eta =1}(n)
\;.\label{Pn}
\end{eqnarray}
Hence, for unit quantum efficiency a photodetector measures the photon
number distribution of the state, whereas for non unit quantum
efficiency the output distribution of counts is given by a Bernoulli
convolution of the ideal distribution.  \par The effects of non unit
quantum efficiency on the statistics of a photodetector, {\em i.e.}
Eq. (\ref{conv_n}) for the output distribution, can be also described
by means of a simple model in which the realistic photodetector is
replaced with an ideal photodetector preceded by a beam splitter of
transmissivity $\tau\equiv\eta$.
 The reflected mode is absorbed, whereas the
transmitted mode is photo-detected with unit quantum efficiency.  In
order to obtain the probability of measuring $m$ clicks, notice that,
apart from trivial phase changes, a beam splitter of transmissivity
$\tau$ affects the unitary transformation of fields
\begin{eqnarray}
\left(\begin{array}{c} c\\d\end{array}\right) \equiv
U_\tau^\dag\left(\begin{array}{c} a\\b\end{array}\right)U_\tau= 
\left(\begin{array}{cc} \sqrt \tau& -\sqrt{1-\tau}\\
\sqrt{1-\tau}& \sqrt\tau \end{array}\right)
\left(\begin{array}{c}
a\\b\end{array}\right)\;,\label{bse}
\end{eqnarray} 
where all field modes are considered at the same frequency. 
Hence, the output mode $c$ hitting the detector is given by the 
linear combination
\begin{eqnarray}
c=\sqrt \tau  a-\sqrt{1-\tau} b\;,\label{a''}
\end{eqnarray}
and the probability of counts reads
\begin{eqnarray}
p_\tau (m) &=& \hbox{Tr} \left[ U_\tau   
\rho\otimes|0\rangle \langle  0|  
U^{\dag}_\tau   |m\rangle \langle m| \otimes 1 \right] \nonumber\\
&=&\sum_{n=m}^{\infty}\rho_{nn}
\left(\begin{array}{c} n\\m\end{array}\right) (1-\tau)^{n-m} \tau^m
\label{conv_n1}\;.
\end{eqnarray}
Eq. (\ref{conv_n1}) reproduces the probability distribution of
Eq. (\ref{conv_n}) with $\tau=\eta$. We conclude that a photo-detector
of quantum efficiency $\eta$ is equivalent to a perfect photo-detector
preceded by a beam splitter of transmissivity $\eta$ 
which accounts for the overall losses of the detection process.
\section{Balanced homodyne detection \label{homsec}}
 The balanced homodyne detector provides the measurement of the
quadrature of the field $X_\varphi  $ in Eq. (\ref{xfi}). 
It was proposed by Yuen and Chan \cite{yuenchan}, 
and subsequently demonstrated by Abbas, Chan and Yee \cite{abbas}.
\begin{figure}[htb]
\begin{center}
\epsfxsize=.4\hsize\leavevmode\epsffile{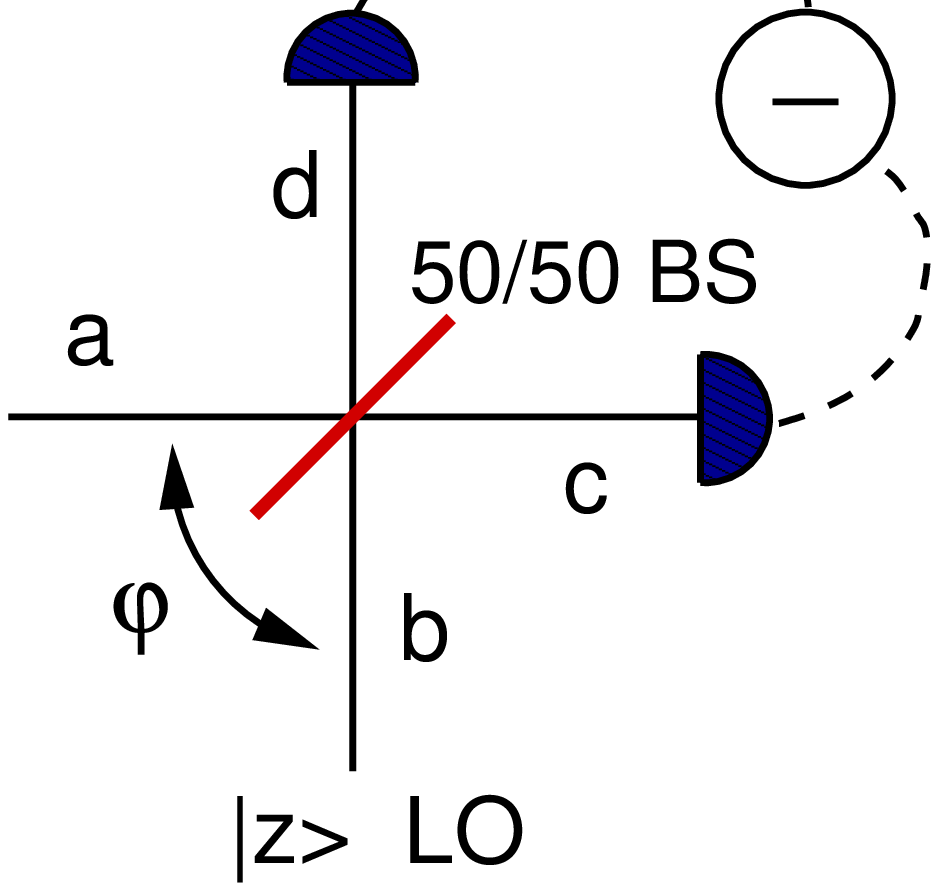}
\end{center}
\caption{Scheme of the balanced homodyne
detector.\label{f:homodyne}}
\end{figure}
\par The scheme of a balanced homodyne detector is depicted in
Fig. \ref{f:homodyne}. The signal mode $a$ interferes 
with a strong laser beam mode $b$ in a balanced 50/50 beam splitter. 
The mode $b$ is the so-called the {\em local oscillator} (LO) 
mode of the detector. It operates at the same frequency of $a$, 
and is excited by the laser in a strong coherent state $|z\rangle $.
Since in all experiments that use homodyne detectors the signal and the LO 
beams are generated by a common source, we assume that they have a fixed 
phase relation. In this case the LO phase provides a reference for the 
quadrature measurement, namely  we identify the 
phase of the LO with the phase difference between the two modes. As we will 
see,  by tuning $\varphi =\arg z$ we can measure the quadrature $X_\varphi $ at 
different phases. 

After the beam splitter the two modes are detected by two identical
photodetectors (usually linear avalanche photodiodes), and finally the
difference of photocurrents at zero frequency is electronically
processed and rescaled by $2|z|$. According to Eqs. (\ref{bse}), the
modes at the output of the $50/50$ beam splitter ($\tau = 1/2$) write\
\begin{eqnarray}
c=\frac{a-b}{\sqrt{2}}\;,  \qquad d=\frac{a+b}{\sqrt{2}} \;, 
\end{eqnarray}
hence the difference of photocurrents is given by the following
operator
\begin{eqnarray}
I={\frac{d^{\dag}d- c^{\dag}c }{2|z|}}=\frac{{a^\dag } b+b^{\dag}a}{2|z|}\;.
\end{eqnarray}
Let us now proceed to evaluate the probability distribution of the output 
photocurrent $I$ for a generic state $\rho$ of the signal mode $a$. 
In the following treatment we will follow Refs. \cite{DARNC,bilk}.  
\par Let us consider the moments generating function of the 
photocurrent $I$
\begin{eqnarray}
\chi(\lambda)=\hbox{Tr}\left[\rho\otimes |z\rangle \langle z|  
\,e^{i\lambda I} \right]
\;,\label{H1}
\end{eqnarray}
which provides the probability distribution of $I$ as the 
Fourier transform 
\begin{eqnarray}
 P(I)= \int_{-\infty}^{+\infty}\frac{d\lambda}{2\pi} 
e^{-i\lambda I} \chi(\lambda)
\;.\label{H1FT}
\end{eqnarray}
Using the BCH formula \cite{Wodk,DIJMP} for the $SU(2)$ group,
namely
\begin{eqnarray}
\exp\left(\xi a b^\dag -\xi^*a^{\dag}b \right)=
e^{\zeta b^{\dag}a}\left(1+|\zeta|^2\right)^{{\frac 12}\left(b^{\dag}b
-a^{\dag} a\right)}e^{-\zeta^* a^{\dag}b},\;
\ \zeta={\frac{\xi}{|\xi|}}\tan 
|\xi|\;,\label{BCH}
\end{eqnarray}
one can write the exponential in Eq.~(\ref{H1}) in normal-ordered form
with respect to mode $b$ as follows
\begin{eqnarray}
\chi(\lambda)=\left\langle e^{
i\tan\left({\frac {\lambda}{2|z|}}\right)b^{\dag}
a }\left[\cos\left({\frac {\lambda}{2|z|}}\right)\right ]^{{a^\dag } a
-b^{\dag}b}e^{i\tan\left({\frac{\lambda}{2|z|}}\right){a^\dag } b }\right\rangle _{ab}\;.
\end{eqnarray}
Since mode $b$ is in a coherent state $|z \rangle $ 
the partial trace over $b$ can be evaluated  as follows
\begin{eqnarray}
\chi(\lambda)&=&\left\langle e^{i\tan\left({\frac{\lambda}{2|z|}}\right)
z^*a} \left[\cos\left({\frac{\lambda}{2|z|}}\right)
\right]^{{a^\dag } a}e^{i\tan\left({\frac {\lambda}{2|z|}}\right)z{a^\dag }  }
\right\rangle _a \nonumber\\
&\times& \left\langle z\Bigg|\left[\cos\left({\frac {\lambda}{2|z|}}
\right)\right]^{-b^{\dag}b}\Bigg|z\right\rangle \;.\label{xx1}
\end{eqnarray}
Using now Eq. (\ref{BCH_easy}), 
one can rewrite Eq. (\ref{xx1}) in normal order with respect to $a$, 
namely
\begin{eqnarray}
\chi(\lambda)=
\left\langle e^{iz\sin\left({\frac {\lambda}{2|z|}}\right)
{a^\dag } }
\exp\left[-2\sin^2\left({\frac{\lambda}{4|z|}}\right)({a^\dag } a+|z|^2)\right]
e^{iz^*\sin\left({\frac {\lambda}{2|z|}}\right) a}
\right\rangle _a\;,\label{anbra}
\end{eqnarray}
In the strong-LO limit $z\to\infty$, only the lowest order terms in
$\lambda/|z|$ are retained, ${a^\dag } a$ is neglected with respect to
$|z|^2$, and Eq. (\ref{anbra}) simplifies as follows
\begin{eqnarray}
\lim_{z\to\infty}\chi(\lambda)=
\left\langle e^{i{\frac {\lambda} {2}}e^{i\varphi }{a^\dag }}\,
\exp\left[-{\frac{\lambda^2} {8}}\right]\,e^{i{\frac {\lambda} {2}}e^{-i\varphi
}a }\right\rangle _a= \left\langle \exp[i\lambda X_\varphi
]\right\rangle _a\;,\label{wwww}
\end{eqnarray}
where $\varphi =\mbox{arg}z$. The generating function in Eq. (\ref{wwww}) 
is then equivalent to the POVM
\begin{eqnarray}
\Pi (x)= \int_{-\infty}^{+\infty} {\frac {d \lambda }{2\pi}} \,
\exp[i\lambda  (X_\varphi -x)]=
\delta(X_\varphi -x)\equiv |x\rangle _\varphi {}_\varphi \langle x|\;,\label{pmx}
\end{eqnarray}
namely the projector on the eigenstate of the quadrature $X_\varphi  $
with eigenvalue $x$.  In conclusion, the balanced homodyne detector
achieves the ideal measurement of the quadrature $ X_\varphi $ in the
strong LO limit.  In this limit, the probability distribution of the
output photocurrent $I$ approaches exactly the probability
distribution $p(x,\varphi  )={}_\varphi \langle x|\rho |x \rangle  _\varphi  $ 
of the quadrature $X_ \varphi  $, and this for
any state $\rho $  of the signal mode $a$. 
\par It is easy to take into account non-unit quantum
efficiency at detectors. According to Eq. (\ref{a''}) one has the
replacements
\begin{eqnarray}
c&\Longrightarrow&\sqrt{\eta} c -\sqrt{1-\eta}u \;,\qquad u,v\hbox{ vacuum modes}\\
d&\Longrightarrow&\sqrt{\eta} d -\sqrt{1-\eta}v \;,
\end{eqnarray}
and now the output current is rescaled by $2|z|\eta$, namely
\begin{eqnarray}
I_\eta \simeq  \frac{1}{2|z|}\left\{\left[ a+\sqrt{\frac{1-\eta
    }{2\eta }}(u +  v )\right]b^\dag + h.c \right\}
\;,\label{IDeta}
\end{eqnarray}
where only terms containing the strong LO mode $b$ are retained. The
POVM is then obtained by replacing 
\begin{eqnarray}
X_\varphi \rightarrow X_\varphi +\sqrt{\frac{1-\eta }{2\eta
}}(u_\varphi +v_\varphi)\;
\end{eqnarray}
in Eq. (\ref{pmx}), with $w_\varphi =(w^\dag e^{i\varphi} +w
e^{-i\varphi})/2 $, $w=u,v$, and tracing the vacuum modes
$u$ and $v$. One then obtains
\begin{eqnarray}
\Pi _{\eta}(x)&=&\int _{-\infty }^{+\infty}
\frac{d\lambda}{2\pi}\,e^{i\lambda  (X_\varphi -x)}
|\langle 0|e^{i\lambda
\sqrt{\frac{1-\eta}{2\eta}}u_{\varphi }}|0\rangle|^2
=\int _{-\infty }^{+\infty}\frac{d\lambda}{2\pi}\,
e^{i\lambda (X_\varphi -x)}
e^{-\lambda^2\frac{1-\eta}{8\eta}}\nonumber\\
&=&\frac{1}{\sqrt{2\pi\Delta^2_{\eta}}}
\exp\left[ -\frac{(x-  X_\varphi )^2}{2\Delta_{\eta}^2}\right]
\nonumber \\&= &\frac{1}{\sqrt{2\pi\Delta^2_{\eta}}}
\int_{-\infty }^{+\infty}dx'\, e^{ -\frac{1}{2\Delta_{\eta}^2}
(x-  x')^2}\,|x'\rangle _\varphi {}_\varphi \langle x'|
\;,\label{OM}
\end{eqnarray}
where 
\begin{eqnarray}
\Delta_{\eta}^2=\frac{1-\eta}{4\eta}\;.\label{Deltaeta}
\end{eqnarray}
Thus the POVM, and in turn the probability distribution of the output 
photocurrent, are just the Gaussian convolution of the ideal ones 
with rms $\Delta_{\eta}=\sqrt{(1-\eta )/(4\eta )}$.
\section{Heterodyne detection \label{hetsec}}
\begin{figure}[htb]
\begin{center}
\epsfxsize=.95\hsize\leavevmode\epsffile{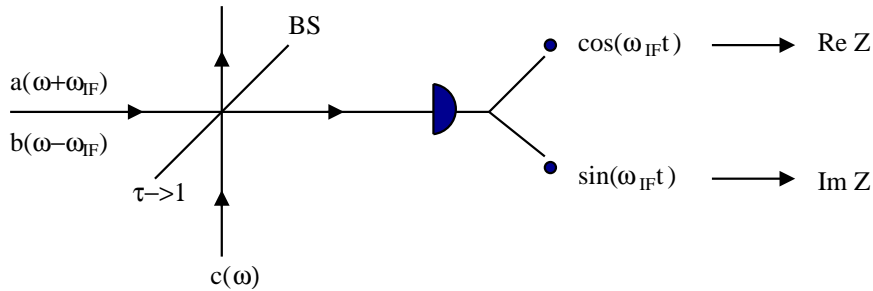}
\end{center}
\caption{Scheme of the heterodyne detector.\label{f:het}}
\end{figure}

Heterodyne detection allows to perform the joint measurement of two
conjugated quadratures of the field \cite{hetyuen1,hetyuen2}. The
scheme of the heterodyne detector is depicted in Fig. \ref{f:het}.
\par A strong local oscillator at frequency $\omega $ in a coherent
state $|\alpha \rangle $ hits a beam splitter with transmissivity
$\tau \to 1$, and with the coherent amplitude $\alpha $ such that $\gamma  
\equiv |\alpha |\sqrt {\tau (1-\tau)}$ is kept constant.  If the
output photocurrent is sampled at the intermediate frequency $\omega
_{IF}$, just the field modes $a$ and $b$ at frequency $\omega \pm
\omega _{IF}$ are selected by the detector.  Modes $a$ and $b$ are
usually referred to as signal band and image band modes,
respectively.  In the strong LO limit, upon tracing the LO mode, 
the output photocurrent $I(\omega _{IF})$ rescaled by $\gamma $ 
is equivalent to the complex operator 
\begin{eqnarray}
Z=\frac{I(\omega _{IF})}{\gamma }=a-b^\dag,\label{zeta}
\end{eqnarray}
 where the arbitrary phases of modes have been suitably chosen.
The heterodyne photocurrent $Z$ is a normal operator, 
equivalent to a couple of commuting
selfadjoint operators 
\begin{eqnarray}
Z=\hbox{Re}Z +i \hbox{Im}Z\;,\qquad [Z,Z^\dag ]= 
[\hbox{Re}Z,\hbox{Im}Z]=0\;.
\; 
\end{eqnarray}
The POVM of the detector is then given by the orthogonal eigenvectors of $Z$.
 It is here convenient to introduce the notation of Ref. \cite{bellobs}
for vectors in the tensor product of Hilbert spaces ${\cal H}\otimes {\cal H}$  
\begin{eqnarray}
|A{\rangle\!\rangle}=\sum_{nm}A_{nm}|n\rangle \otimes|m\rangle \equiv (A\otimes
I)|I{\rangle\!\rangle} \equiv (I\otimes A^\tau )|I{\rangle\!\rangle}
\;,\label{iso} \end{eqnarray} where $A^\tau $ denotes the transposed
operator with respect to some pre-chosen orthonormal basis. Eq. (\ref{iso})
exploits the isomorphism between the Hilbert space of the Hilbert-Schmidt
operators $A,B\in\sf {HS({\cal H})}$ with scalar product $\langle A,B\rangle
=\hbox{Tr}[A^\dag B]$, and the Hilbert space of bipartite vectors
$|A{\rangle\!\rangle},|B{\rangle\!\rangle}\in {\cal H}\otimes{\cal H}$, where
one has ${\langle\!\langle} A|B{\rangle\!\rangle}\equiv\langle  A,B\rangle $.
\par Using the above notation it is easy to write the eigenvectors of $Z$ with
eigenvalue $z$ as $\frac {1}{\sqrt\pi }|D(z) {\rangle\!\rangle}$. In fact one
has \cite{2mode}
\begin{eqnarray}
Z |D(z) {\rangle\!\rangle}&=& (a- b^\dag ) (D_a(z)\otimes I_b)
|I{\rangle\!\rangle} =(D_a(z) \otimes I_b) (a-b^\dag +z)
\sum_{n=0}^\infty |n \rangle \otimes |n \rangle \nonumber \\&= & z (D_a(z)\otimes I_b)
|I{\rangle\!\rangle} =z |D(z){\rangle\!\rangle} \;.
\end{eqnarray}
The orthogonality of such eigenvectors can be verified through the
relation
\begin{eqnarray}
{\langle\!\langle}D(z)|D(z'){\rangle\!\rangle}= \hbox{Tr}[D^\dag (z)D(z')]=\pi \delta ^{(2)}(z-z')\;, 
\end{eqnarray}
where $\delta ^{(2)}(\alpha )$ denotes the Dirac delta function over
the complex plane
\begin{eqnarray}
\delta^{(2)}(\alpha)=\int _{\mathbb C}\frac{
d^2 \gamma}{\pi ^2}\,\exp(\gamma\alpha^*-\gamma ^*\alpha)\;.\label{d2}
\end{eqnarray}
\par In conventional heterodyne detection the image band mode is in the
vacuum state, and one is just interested in measuring the field mode
$a$. In this case we can evaluate the POVM upon tracing on mode $b$. 
One has 
\begin{eqnarray}
\Pi (z , z^* )&=& 
\frac {1}{\pi }\hbox{Tr}_b [|D(z){\rangle\!\rangle}{\langle\!\langle}D(z)|
  I_a \otimes |0 \rangle \langle 0|] 
 \nonumber \\&= &
\frac{1}{\pi} D(z) |0 \rangle \langle 0| D^\dag (z) =
\frac{1}{\pi} |z \rangle \langle z| 
\;,  
\end{eqnarray}
namely one obtain the projectors on coherent states. The
coherent-state POVM provides the optimal joint measurement of
conjugated quadratures of the field \cite{hel}. In fact, heterodyne detection
allows to measure the Q-function in Eq. (\ref{3}). According to
Eq. (\ref{ex}) then it provides the expectation value of anti-normal
ordered field operator. For a state $\rho $
the expectation value of any quadrature $X_\varphi $ is obtained as
\begin{eqnarray}
\langle X_\varphi \rangle =\hbox{Tr}[\rho X_\varphi ]= \int _{\mathbb
C}\frac{d^2 \alpha}{\pi } \hbox{Re}(\alpha e^{-i \varphi}) Q(\alpha
,\alpha ^* ) \;.
\end{eqnarray}
The price to pay for jointly measuring noncommuting observables is an
additional noise. The rms fluctuation is evaluated as follows
\begin{eqnarray}
\int_{\mathbb C} \frac{d^2 \alpha}{\pi } 
[\hbox{Re}(\alpha e^{-i \varphi})]^2  Q(\alpha ,\alpha ^* ) 
- \langle X_\varphi \rangle ^2 =
\langle \Delta X^2 _\varphi \rangle +\frac {1}{4}
\;, 
\end{eqnarray}
where $\langle \Delta X^2 _\varphi \rangle $ is the intrinsic noise,
and the additional term is usually referred to as ``the additional 
3dB noise due to the joint measure'' \cite{Arthurs,yuen82,goodman}.
\par The effect of non-unit quantum efficiency can be taken into account
in analogous way as in Sec. \ref{homsec} for homodyne detection. 
The heterodyne photocurrent is rescaled by an additional factor $\eta
^{1/2}$, and vacuum modes $u$ and $v$ are introduced, thus giving \cite{opti} 
\begin{eqnarray}
Z_\eta = a- b^\dag + \sqrt {\frac {1-\eta }{\eta }} u -{\sqrt \frac
{1-\eta}{\eta }} v^\dag 
\;. 
\end{eqnarray}
Upon tracing over modes $u$ and $v$, one obtain the POVM 
\begin{eqnarray}
\Pi _\eta (z, z^*)&=&\int _{\mathbb C}\frac{d^2 \gamma }{\pi ^2} 
{}_u\langle 0 |{}_v\langle 0 |e^{\gamma (Z^\dag _\eta -z^*)-
\gamma ^*(Z _\eta -z)} |
0 \rangle _u |0 \rangle _v   \\&= &
\int _{\mathbb C}\frac{d^2 \gamma }{\pi ^2} 
e^{\gamma (Z^\dag -z^*)-
\gamma ^*(Z -z)} \, e^{-\frac{1-\eta }{\eta }|\gamma |^2}\nonumber
\\&= & \frac{\eta }{\pi (1-\eta )}e^{-\frac {\eta }{1-\eta }|Z-z|^2 }=
\int _{\mathbb C} \frac{d^2 z'}
{\pi \Delta ^2_\eta }e^{-\frac {|z'-z|^2}{\Delta ^2_\eta } }
|D(z') {\rangle\!\rangle} {\langle\!\langle}D(z')|\;. 
\nonumber 
\end{eqnarray}
The probability distribution is then a Gaussian convolution on the
complex plane of the
ideal probability with rms $\Delta^2_{\eta}=(1-\eta )/\eta $. 
\par  Analogously, the coherent-state POVM for conventional heterodyne
detection with vacuum image band mode is replaced with
\begin{eqnarray}
\Pi _\eta (z, z^*)= 
\int _{\mathbb C}\frac{d^2 z'}
{\pi \Delta ^2_\eta }e^{-\frac {|z'-z|^2}{\Delta ^2_\eta } }
|z' \rangle \langle z'|\;. 
\label{coeta}
\end{eqnarray}
From Eqs. (\ref{convw}) we can equivalently say that the heterodyne
detection probability density is given by the generalized Wigner
function $W_s(\alpha ,\alpha ^* )$, with $s=1-\frac 2\eta$. Notice
that for $\eta <1$, the average of functions $\alpha ^n \alpha ^{*m}$
is related to the expectation value of a different ordering of field
operators. However, one has the relevant identity \cite{cgl2,qsm}
\begin{eqnarray}
\mbox{{\bf :}} (a^{\dag})^na^m \mbox{{\bf :}} _s  =\sum_{k=0}^{(n,m)}
k !\,{n\choose k}{m\choose k}\left( \frac{t-s}{2}\right)^k\,
\mbox{{\bf :}}  (a^{\dag})^{n-k}a^{m-k} \mbox{{\bf :}} _t 
\;,\label{ex2} 
\end{eqnarray}
where $(n,m)=\min(n,m)$, and then 
\begin{eqnarray}
&&\int _{\mathbb C} d^2\alpha\,W_{1-\frac 2\eta }(\alpha,
\alpha ^*)\,\alpha ^m \alpha ^{*n}
\nonumber \\& & =\sum_{k=0}^{(n,m)}
k !\,{n\choose k}{m\choose k}\left(\frac{1-\eta }{\eta }\right)^k\,
\langle a^{m-k} (a^\dag )^{n-k} \rangle
\;.\label{exhet}
\end{eqnarray}
Notice that the measure of the Q-function (or any smoothed version for
$\eta <1$) does not allow to recover the expectation value of {\em
any} operator through an average over heterodyne outcomes. In fact,
one needs the admissibility of anti-normal ordered expansion
\cite{baltin} and the
convergence of the integral in Eq. (\ref{exhet}). In particular, the
matrix elements of the density operator cannot be recovered. For some
operators in which heterodyne measurement is allowed, 
a comparison with quantum homodyne tomography will be given in
Sec. \ref{compare}. 
\par Finally, it is worth mentioning that all results of this section are 
valid also for an image-band mode with the same frequency of the signal. 
In this case a measurement scheme based on multiport homodyne detection 
should be used \cite{hetyuen2,qsm,dmr,sha,sho,wa3,lai,tri,zuc}.

%% file: cap3.tex
\chapter{General tomographic method\label{generaltom}} 
 In this chapter we review the general tomographic method 
of spanning sets of operators of Ref. \cite{orth}, and re-derive in
this general framework the exact homodyne tomography method of
Ref. \cite{dmp}. In the first section, we first give a brief history
of quantum tomography, starting with the original proposal of Vogel and
Risken \cite{vogel}, that extended the conventional tomographic
imaging to the domain of quantum optics. Here we will briefly sketch the
conventional imaging tomographic method, and show the analogy with the
method of Ref. \cite{vogel}. The latter achieves the quantum state via the
Wigner function, which in turn is obtained by inverse 
Radon transform of the homodyne probability distributions for varying
phase with respect to the LO. As already mentioned, the Radon
transform inversion is affected by uncontrollable bias: such
limitations and the intrinsic  unreliability of this method are
thoroughly explained in the same section. 
\par As opposite to the Radon transform method, the first exact 
method of Ref. \cite{dmp} (and successively refined in Ref. 
\cite{dlp}) allows the reconstruction of the density matrix $\rho$, 
bypassing the step of the Wigner function, and achieving the matrix
elements of $\rho$---or the expectation of any arbitrary operator---by
just averaging the pertaining {\em estimators} (also called {\em
Kernel functions} or {\em pattern functions}), evaluated on 
the experimental homodyne data. This method will be re-derived in 
Subsec. \ref{s:qopt}, as a special case of the general tomographic method
of Ref. \cite{orth} here reviewed in Sect. \ref{gent}, where we
introduce the concept of ``quorum'', which is the complete set of
observables whose measurement provides the expectation value of any
desired operator. Here we also show how some ``orthogonality'' and
``completeness'' relations in the linear algebra of operators are
sufficient to individuate a quorum. As another application of the
general method, in Subsect. \ref{s:spin} the tomography of spin
systems \cite{spinmac} is reviewed, which was originally derived from the group
theoretical methods of Refs. \cite{chicago,paini-rete,deconv}. Another
application is the quantum tomography of a free particle state, given
in Subsect. \ref{s:fre}. 
\par In this same chapter, in Sec. \ref{devel} we include some
further developments to improve the tomographic method, 
such as the deconvolution techniques of Ref. \cite{deconv} to correct the
imperfections of detectors and experimental apparatus with a suitable
data processing, and the adaptive tomography of Ref. \cite{adapt} to reduce
the statistical fluctuations of tomographic estimators, by adapting
the averaged estimators to the given sample of experimental data.
\par The other relevant topics of homodyning observables, multimode
tomography, and tomography of quantum operations will be given a
separate treatment in the following chapters of the Review.
\section{Brief historical excursus\label{history}}
The problem of quantum state determination through repeated
measurements on identically prepared systems was originally stated by
Fano in 1957 \cite{fano}, who first recognized the need of measuring
more that two noncommuting observables to achieve such
purpose. However, it was only with the proposal by Vogel and Risken
\cite{vogel} that quantum tomography was born. The first experiments, which already showed
reconstructions of coherent and squeezed states were performed by Michael
Raymer's and his group at the University of Oregon \cite{raymer}. The
main idea at the basis of the first proposal is that it is possible to
extend to the quantum domain the algorithms that are conventionally
used in medical imaging to recover two dimensional (mass)
distributions from unidimensional projections in different
directions. This first tomographic method, however, was unreliable for
the reconstruction of an unknown quantum state, since arbitrary
smoothing parameters were needed in the Radon-transform based imaging
procedure. The first exact unbiased tomographic method was proposed in
Ref. \cite{dmp}, and successively simplified in Ref. \cite{dlp}.
Since then, the new exact method has been practically implemented 
in many experiments, such as the measurement of the photon statistics
of a semiconductor laser \cite{raymer95}, and the reconstruction of
the density matrix of a squeezed vacuum \cite{sch}.  The success of
optical homodyne tomography has then stimulated the development of
state-reconstruction procedures in other quantum harmonic oscillator
systems, such as for atomic beams \cite{atobeams}, and the
vibrational state of a molecule \cite{dunn}, of an ensemble of helium
atoms \cite{exper-freeatoms}, and of a single ion in a Paul trap
\cite{leib}. 
\par After the original exact method, quantum tomography has been
generalized to the estimation of arbitrary observable of the field
\cite{tokio}, to any number of modes \cite{homtom}, and, finally, to
arbitrary quantum systems via group theory
\cite{chicago,paini-rete,deconv,math}, with further improvements such as
noise deconvolution \cite{deconv}, adaptive tomographic methods
\cite{adapt}, and the use of max-likelihood strategies \cite{maxlik},
which has made possible to reduce dramatically the number of
experimental data, up to a factor $10^3\div 10^5$, with negligible
bias for most practical cases of interest. Finally, more recently, a
method for tomographic estimation of the unknown quantum operation 
of a quantum device has been proposed \cite{cptomo}, where a fixed input
entangled state plays the role of all input states in a sort of
quantum parallel fashion. Moreover, as another manifestation of such
a quantum parallelism, one can also estimate the ensemble average
of all operators by measuring only one fixed "universal" observable on an
extended Hilbert space in a sort of {\em quantum hologram} \cite{holo}.
This latest development is based on the general tomographic method 
of Ref. \cite{orth}, where the tomographic reconstruction is based on
the existence of spanning sets of operators, of which the irreducible unitary
group representations of the group-methods of
Refs. \cite{chicago,paini-rete,deconv,math} are just a special case.
\section{Conventional tomographic imaging}\label{convim}
In conventional medical tomography, one collects data in the form of
marginal distributions of the mass function $m(x,y)$.  In the complex
plane the marginal $r(x,\varphi)$ is a projection of the complex
function $m(x,y)$  on the direction indicated by the
angle $\varphi\in[0,\pi]$, namely
\begin{eqnarray}
r(x,\varphi)= \int_{-\infty}^{+\infty}\frac{dy}\pi
m\left((x+iy)e^{i\varphi}, (x-iy)e^{-i\varphi}\right).
\;\label{defmarginale}
\end{eqnarray}
The collection of marginals for different $\varphi$ is called ``Radon
transform''. The tomography process essentially consists in the
inversion of the Radon transform (\ref{defmarginale}), in order to
recover the mass function $m(x,y)$ from the marginals
$r(x,\varphi)$. \par Here we derive inversion of
Eq. (\ref{defmarginale}). Consider the identity
\begin{eqnarray} m(\alpha,\alpha ^*)=\int _{\mathbb C} d^2\beta\;
\delta^{(2)}(\alpha-\beta)\;m(\beta,\beta ^*) 
\;,\label{idenaa1}
\end{eqnarray}
where $\delta ^{(2)}(\alpha )$ denotes the Dirac delta function of
Eq. (\ref{d2}), and $m(\alpha,\alpha ^*)\equiv m(x,y)$, with $\alpha
=x+iy$ and $\alpha ^* =x-iy$. It is convenient to rewrite
Eq. (\ref{d2}) as follows
\begin{eqnarray}
\delta^{(2)}(\alpha)=\int_0^{+\infty}\frac{dk}4k\int_0^{2\pi}
\frac{d\varphi}{\pi^2} e^{-ik\alpha_\varphi}=
\int_{-\infty}^{+\infty}\frac{dk}4|k|\int_0^{\pi}
\frac{d\varphi}{\pi^2} e^{-ik\alpha_\varphi},
\;\label{idenaa2}
\end{eqnarray}
with $\alpha_\varphi\equiv \mbox{Re}(\alpha\;e^{-i\varphi})
=-\alpha_{\varphi+\pi}$. Then, from Eqs. (\ref{idenaa1})
and (\ref{idenaa2}) the inverse Radon transform is
obtained as follows 
\begin{eqnarray}
m(x,y)=\int_0^\pi
\frac{d\varphi}\pi \int_{-\infty}^{+\infty} dx'\;
r(x',\varphi)\int_{-\infty}^{+\infty} \frac{dk}4|k|
\;e^{ik(x'-\alpha_\varphi)}
\;.\label{invradtr}
\end{eqnarray}
Eq. (\ref{invradtr}) is conventionally written as 
\begin{eqnarray}
m(x,y)=\int_0^\pi
\frac{d\varphi}\pi \int_{-\infty}^{+\infty} dx'\;
r(x',\varphi)\;K(x'-\alpha_\varphi),
\;\label{invradtr1}
\end{eqnarray}
where $K(x)$ is given by
\begin{eqnarray}
K(x) \equiv \int_{-\infty}^{+\infty}\frac{dk}4|k|e^{ikx}=
\frac 12\mbox{Re}\int_0^{+\infty} dk\; k e^{ikx}=
 -\frac 12{\cal P}\frac 1{x^2}\;,
\;\label{kernelaa}
\end{eqnarray}
with $\cal P$ denoting the Cauchy principal value. Integrating by
parts Eq. (\ref{invradtr1}) one obtains the tomographic formula that
is usually found in medical imaging, {\it i.e.} \begin{eqnarray}
m(x,y)=\frac 1{2\pi}\int_0^\pi d\varphi\;\; {\cal
P}\int_{-\infty}^{+\infty} dx'\; \frac 1{x'-\alpha_\varphi}\;
\frac{\partial }{\partial x'}\; r(x',\varphi)\;,
\;\label{medicaltom}
\end{eqnarray}
which allows the reconstruction of the mass distribution $m(x,y)$ from
its projections along different directions $r(x,\varphi)$.

\subsection{Extension to the quantum domain}
 In the ``quantum imaging'' process the goal is to reconstruct a quantum
state in the form of its Wigner function starting from its
marginal probability distributions. As shown in Sec. \ref{wigsec}, the
Wigner function is a real normalized function that is in one-to-one
correspondence with the state density operator $\rho$. As noticed in
Eq. (\ref{margw}), the probability distributions of the quadrature
operators $X_\varphi=(a^\dag e^{i\varphi}+ae^{-i\varphi})/2$ are the
marginal probabilities of the Wigner function for the state $\rho $.
Thus, by applying the same procedure outlined in the previous
subsection, Vogel and Risken \cite{vogel} proposed a method to recover the
Wigner function {\it via} an inverse Radon transform from the
quadrature probability distributions $p(x,\varphi)$, namely
\begin{eqnarray} 
W(x,y)= \int_0^\pi \frac{d\varphi}\pi \int_{-\infty}^{+\infty} dx'
\; p(x',\varphi)  \int_{-\infty}^{+\infty} \frac{dk}4\;|k|
\;e^{ik(x'-x\cos\varphi-y\sin\varphi)}
\;\label{inverseradon}.
\end{eqnarray}
[Surprisingly, in the original paper \cite{vogel} the
connection to the tomographic imaging method was never mentioned]. 
As shown in Sec. \ref{homsec} the experimental measurement 
of the quadratures of the field is obtained using the homodyne
detector. The method proposed by Vogel and Risken, namely
the inversion of the Radon transform, was the one which has been used
in the first experiments \cite{raymer}. \par

\par This first method is, however, not reliable for the reconstruction
of an unknown quantum state, due to the intrinsic unavoidable
systematic error related to the fact that the integral on $k$ in 
Eq. (\ref{inverseradon}) is unbounded. In fact, in order to evaluate the inverse
Radon transform, one would need the analytical form of the marginal
distribution of the quadrature $p(x,\varphi)$, which, in turn, can
only be obtained by collecting the experimental data into histograms,
and thence ``spline-ing'' them. This, of course, is not an unbiased
procedure since the degree of spline-ing, the width and the
number of the histogram bins, and finally the number of different
phases used to collect the experimental data sample introduce
systematic errors if they are not set above some minimal values, which
actually depend on the unknown quantum state that one wants to
reconstruct. Typically, an over-spline-ing  will wash-out  the
quantum features of the state, whereas, {\em vice-versa}, an
under-spline-ing will create negative photon
probabilities in the reconstruction (see Ref.  \cite{dmp} for
details).

\par A new exact method was then proposed in Ref. \cite{dmp},
alternative to the Radon-transform technique.  This approach, referred
to as {\em quantum homodyne tomography}, allows to recover the 
quantum state of the field $\rho$---along with any ensemble average of
arbitrary operators---by directly averaging functions of the homodyne
data, abolishing the intermediate step of the Wigner function, which
is the source of all systematic errors. Only statistical errors are present, and
they can be reduced arbitrarily by collecting more experimental data.
This  exact method will be re-derived from the general tomographic
theory in Sec. \ref{s:qopt}.

\section{General method of quantum tomography \label{gent}}
In this section the general method of quantum tomography is explained
in detail.
 First, we give the basics of Monte Carlo integral theory
which are needed to implement the tomographic algorithms in actual
experiments and in numerical simulations. Then, we derive the formulas
on which all schemes of state reconstruction are based.
\subsection{Basic statistics \label{s:stat}}
The aim of quantum tomography is to estimate, for arbitrary quantum
system, the mean value $\langle O\rangle $ of a system operator $O$
using only the results of the measurements on a set of observables
$\{Q_\lambda,\ \lambda\in\Lambda\}$, called ``quorum''. The procedure
by which this can be obtained needs the {\em estimator} or ``Kernel
function'' ${\cal R}[O](x, \lambda )$ which is a function of the
eigenvalues $x$ of the quorum operators. Integrating the estimator
with the probability $p(x, \lambda )$ of having outcome $x $ when
measuring $Q_\lambda$, the mean value of $O$ is obtained as follows
\begin{eqnarray} 
\langle O\rangle =\int_\Lambda d\lambda\int d\mu _\lambda (x)\;
p(x,\lambda)\;{\cal R}[O](x, \lambda )
\;\label{aa1},
\end{eqnarray}
where the first integral is performed on the values of $\lambda$ that
designate all quorum observables, and the second on the eigenvalues of
the quorum observable $Q_\lambda$ determined by the $\lambda$ variable
of the outer integral.  For discrete set $\Lambda$ and/or discrete spectrum of 
the quorum, both integrals in (\ref{aa1}) can suitably
replaced by sums.  \par The algorithm to estimate $\langle O\rangle $ with
Eq. (\ref{aa1}) is the following. One chooses a quorum operator
$Q_\lambda$ by drawing $\lambda$ with uniform probability in
$\Lambda$ and performs a measurement, obtaining the result $x_i$. By
repeating the procedure $N$ times, one collects the set of
experimental data $\{(x_i, \lambda_i)$, with $i=1,\cdots,N\}$, where
$\lambda_i$ identifies the quorum observable used for the $i$th
measurement, and $x_i$ its result. From the same set of data the mean
value of any operator $O$ can be obtained. In fact, one evaluates the
estimator of $\langle O\rangle $ and the quorum $Q_\lambda$, and
then samples the double integral of (\ref{aa1}) using the limit
\begin{eqnarray}
\langle O\rangle =\lim_{N\to\infty}\frac 1N\;\sum_{i=1}^N{\cal
  R}[O](x_i, \lambda _i)
\;\label{aa2}.
\end{eqnarray}
Of course the finite sum
\begin{eqnarray}
F_N=\frac 1N\sum_{i=1}^N{\cal R}[O](x_i, \lambda _i)
\;\label{fin}.
\end{eqnarray}
gives an approximation of $\langle O \rangle $.  To estimate the error
in the approximation one applies the central limit theorem that we
recall here.  

\par {\bf Central limit theorem.} Consider $N$ statistically
uncorrelated random variables $\{z_i,$ $ i=1,\cdots,N\}$, with mean
values $\mu(z_i)$, variances $\sigma^2(z_i)$ and bounded third order
moments. If the variances $\sigma^2(z_i)$ are all of the same order
then the statistical variable ``average'' $y$ defined as
\begin{equation}
y_N=\frac{1}{N}\sum_{i=1}^Nz_i\label{yN}
\end{equation}
has mean and variance  
\begin{equation}
\mu(y_N)=\frac{1}{N}\sum_{i=1}^N\mu(z_i),
\qquad\sigma^2(y_N)=\frac{1}{N^2}\sum_{i=1}^N\sigma^2(z_i)
\label{defsommas}.
\end{equation}
The distribution of $y_N$ approaches asymptotically a Gaussian for
$N\to\infty$. In practical cases, the distribution of $y$ can be
considered Gaussian already for $N$ as low as $N\sim 10$.  \par For
our needs the hypotheses are met if the estimator ${\cal
R}[O](x_i,\lambda _i )$ in Eq. (\ref{fin}) has limited moments up to
the third order, since, even though $x_i$ have different probability densities
depending on $\lambda_i$, nevertheless, since $\lambda_i$ is also
random all $z_i$ here given by
\begin{equation}
z_i={\cal R}[O](x_i,\lambda _i)\label{zi}
\end{equation}
have common  mean
\begin{eqnarray}
\mu(z_i)=\langle O \rangle \;
\end{eqnarray}
and variance
\begin{eqnarray}
\sigma ^2(z_i)=\int_\Lambda d\lambda\int d\mu _\lambda (x)\;
p(x,\lambda){\cal R}^2 [O](x, \lambda ) - \langle O \rangle ^2. 
\end{eqnarray}
Using the central limit theorem, we can conclude that the experimental
average $y\equiv F_N $ in Eq. (\ref{fin}) 
is a statistical variable distributed as a Gaussian with mean value 
$\mu (y_N)\equiv\mu(z_i)$ and variance $\sigma^2 (y_N)\equiv
\frac{1}{N}\sigma^2(z_i)$. Then the tomographic estimation converges
with statistical error that decreases as $N^{-1/2}$. A statistically
precise estimate of the confidence interval is given by 
\begin{equation} 
\epsilon_N=\sqrt{\frac{\sum_{i=1}^N[z_i-y_N]^2}{N(N-1)}}\;, 
\label{esterrorbar}
\end{equation}
with $z_i$ given by Eq. (\ref{zi}) and $y_N$ by Eq. (\ref{yN}). In order to
test that the confidence intervals are estimated correctly, one can
check that the $F_N$ distribution is actually Gaussian.  This can be
done by comparing the histogram of the block data with a Gaussian, or
by using the $\chi^2$ test.

\subsection{Characterization of the quorum}
Different estimation techniques have been proposed tailored to
different quantum systems, such as the radiation field
{\cite{dlp,homtom}}, trapped ions and molecular vibrational states
{\cite{welschrev}}, spin systems {\cite{weigert}}, etc. All the known
quantum estimation techniques can be embodied in the following
approach.

\par The tomographic reconstruction of an operator $O$ is possible
when there exists a resolution of the form
\begin{eqnarray}
O= \int_{\Lambda } d \lambda  \;  \hbox{Tr} \left[O B^\dagger 
(\lambda )\right]C(\lambda )\;\label{gentomo},
\end{eqnarray}
where $\lambda $ is a (possibly multidimensional) parameter living on
a (discrete or continuous) manifold $\Lambda $. The only hypothesis in
(\ref{gentomo}) is the existence of the trace.  If, for example, $O$
is a trace--class operator, then we do not need to require $B(\lambda )$ to 
be of Hilbert-Schmidt class, since it is sufficient to require 
$B(\lambda )$ bounded.  The operators $ C(\lambda )$ are functions of the
quorum of observables measured for the reconstruction, whereas the
operators $ B(\lambda )$ form the {\it dual basis} of the set $
C(\lambda )$. The term
\begin{eqnarray}
{\cal E}[  O] (\lambda ) = \hbox{Tr} \left[   OB^\dagger (\lambda
  )\right]
C(\lambda )
\label{qest}\;
\end{eqnarray}
represents
the quantum estimator for the operator $  O$. The expectation 
value of $  O$ is given by
the ensemble average
\begin{eqnarray} 
\langle   O\rangle \equiv \hbox{Tr}\left[  O \rho\right]=
 \int_{\Lambda } d \lambda   \;  \hbox{Tr} \left[   OB^\dagger (\lambda )\right] 
\hbox{Tr}\left[  C(\lambda ) \rho\right] \equiv \int_{\Lambda } d
 \lambda   
\,\langle{\cal E}[  O] (\lambda ) \rangle
\;\label{expval},
\end{eqnarray}
where $ \rho$ is the density matrix of the quantum system under
investigation.  Notice that the quantity $\hbox{Tr}\left[ C(\lambda )
\rho\right]$ depends only on the quantum state, and it is related to
the probability distribution of the measurement outcomes, whereas the
term $\hbox{Tr} \left[ O B^\dagger (\lambda )\right]$ depends only on
the quantity to be measured. In particular, the tomography of the
quantum state of a system corresponds to writing Eq. (\ref{gentomo})
for the operators $ O=|k\rangle\langle n|$, $\{|n\rangle\}$ being a
given Hilbert space basis. For a given system, the existence of a set
of operators $ C(\lambda )$, together with its dual basis $ B(\lambda
)$ allows universal quantum estimation, i. e. the reconstruction of
any operator. 

\par We now give two characterizations of the sets $ B(\lambda )$ and
$ C(\lambda )$ that are necessary and sufficient conditions for
writing Eq. (\ref{gentomo}).  

\par {\bf Condition 1: bi-orthogonality} \par \noindent Let us
consider a complete orthonormal basis of vectors $|n\rangle\;
(n=0,1,\cdots)$.  Formula (\ref{gentomo}) is equivalent to the
bi-orthogonality condition
\begin{eqnarray}
\int_{\Lambda } d \lambda  \;  \langle q|  B^\dagger (\lambda )|p \rangle \;  
\langle m|  C (\lambda )|l \rangle \;  = \;  \delta_{mp} \delta_{lq}
\label{biort}\;,
\end{eqnarray}
where $\delta_{ij}$ is the Kronecker delta. Eq. (\ref{biort}) can be
straightforwardly generalized to a continuous basis. 

\par {\bf Condition 2: completeness} \par \noindent If the set of
operators $ C (\lambda )$ is {\em complete}, namely if any operator
can be written as a linear combination of the $ C(\lambda )$ as
\begin{eqnarray}
  O = \int_{\Lambda } d \lambda   \;  a(\lambda ) \;    C (\lambda )
\;\label{irred},
\end{eqnarray}
then Eq. (\ref{gentomo}) is also equivalent to the trace condition
\begin{eqnarray}
\hbox{Tr}\left[  B^\dagger (\lambda )\;   C (\mu )\right]\;  = \;
\delta (\lambda ,\mu )
\label{binorm}\;,
\end{eqnarray} 
where $\delta(\lambda ,\mu )$ is a {\em reproducing kernel} for the set $  B
(\lambda )$, namely
it is a function or a tempered distribution which satisfies
\begin{eqnarray}
\int_{\Lambda }d \lambda   \;  B(\lambda )\; \delta(\lambda ,\mu )=
B(\mu  )
\;.\label{deltadef}
\end{eqnarray}
An analogous identity holds for the set of $C(\lambda )$ 
\begin{eqnarray}
\int_{\Lambda }d \lambda  \;  C(\lambda )\; \delta(\lambda ,\mu )=
C(\mu )
\;\label{delta2}.
\end{eqnarray}
The proofs are straightforward.  The completeness condition on the
operators $ C (\lambda )$ is essential for the equivalence of
(\ref{gentomo}) and (\ref{binorm}). A simple counterexample is
provided by the set of projectors $ P (\lambda) =
|\lambda\rangle\langle\lambda |$ over
the eigenstates of a self-adjoint operator $L$. In fact,
Eq. (\ref{binorm}) is satisfied by $C(\lambda)=B(\lambda)\equiv
P(\lambda)$. However, since they do not 
form a complete set in the sense of Eq. (\ref{irred}), it is not possible to express a generic
operator in the form $X=\int_\Lambda d\lambda \; \langle\lambda | O |\lambda\rangle \;
|\lambda\rangle\langle\lambda |$.  \par\noindent If either the set $ B(\lambda )$
or the set $ C(\lambda )$ satisfy the additional trace condition
\begin{eqnarray}
&&\hbox{Tr}\left[  B^\dag (\mu )   B (\lambda )\right]  =  \delta
(\lambda ,\mu )\;,
\label{bbd1} \\
&&\hbox{Tr}\left[  C^\dag (\mu )C (\lambda )\right]  =  \delta (\lambda
,\mu )
\label{bbd}\;,
\end{eqnarray}
then we have $ C (\lambda ) = B (\lambda )$ (notice that neither $
B(\lambda )$ nor $ C(\lambda )$ need to be unitary). In this case,
Eq. (\ref{gentomo}) can be rewritten as
\begin{eqnarray}
  O = \int_{\Lambda } d \lambda  \;  \hbox{Tr} \left[   O C^\dagger
(\lambda )\right]  
  C (\lambda )
\label{gentomo2}\;.
\end{eqnarray}
\par A certain number of observables $ Q_\lambda $ constitute a quorum
when there are functions $f_\lambda ( Q_\lambda )= C(\lambda )$ such
that $C(\lambda )$ form an irreducible set. The quantum estimator for
$O$ in Eq. (\ref{qest}) then writes as a function of the quorum
operators
\begin{eqnarray}
{\cal E}[O](\lambda )\equiv {\cal E}_\lambda [O](Q_\lambda) \;.
\end{eqnarray}
Notice that if a set of observables $ Q_\lambda $ constitutes a
quorum, than the set of projectors $|q\rangle_\lambda {}_\lambda
\langle q|$ over their eigenvectors provides a quorum too, with the
measure $d \lambda $ in Eq. (\ref{gentomo}) including the measure
$d\mu _\lambda (q)$.  
Notice also that, even once the quorum has been fixed, the
unbiased estimator for an operator $O$ will not in general be unique,
since there can exist functions ${\cal N}(Q_\lambda)$ that satisfies
\cite{adapt}
\begin{eqnarray}
\int _\Lambda d\lambda \,{\cal N}(Q_\lambda)=0
\;\label{nullestim},
\end{eqnarray}
and that will be called `null estimators'.  Two unbiased estimators
that differ by a null estimator yield the same results when estimating
the operator mean value. We will see in Sec. \ref{s:adapt} how the
null estimators can be used to reduce the statistical noise.  

\par In terms of the quorum observables $Q_\lambda $
Eq. (\ref{expval}) rewrites
\begin{eqnarray}
\langle O \rangle &=&\int_{\Lambda }d \lambda 
\;  \hbox{Tr} \left[   OB^\dagger (\lambda )\right] 
\hbox{Tr}\left[  f_ \lambda (Q_\lambda ) \rho\right] 
\nonumber \\&= &
\int_{\Lambda } d \lambda \int d\mu _{\lambda }(q) \,p(q,\lambda )\, 
\hbox{Tr} \left[   OB^\dagger (\lambda )\right]\,f_\lambda (q)   
\;,\label{due}
\end{eqnarray}
where $p(q,\lambda )= {}_\lambda \langle q|\rho |q \rangle
_\lambda $ is the probability density of getting the outcome
$q$ from the measurement of $Q_\lambda $ on the state $\rho
$. Eq. (\ref{due}) is equivalent to the expression (\ref{aa1}), with
estimator
\begin{eqnarray}
{\cal R}[O](q,\lambda )= \hbox{Tr}
\left[ OB^\dagger (\lambda )\right]\,f_\lambda (q)
\;.
\end{eqnarray}  
\par Of course it is of interest to connect a quorum of observables to
a resolution of the form (\ref{gentomo}), since only in this case
there can be a feasible reconstruction scheme. If a resolution formula
is written in terms of a set of selfadjoint operators, the set itself
constitutes the desired quorum. However, in general a quorum of
observables is functionally connected to the corresponding resolution
formula. If the operators $ C(\lambda )$ are unitary, then they can
always be taken as the exponential map of a set of selfadjoint
operators, which then are identified with our quorum $ Q_\lambda
$. The quantity $\hbox{Tr}\left[ C(\lambda ) 
\rho\right]$ is thus connected with the moment generating function of
the set $ Q_\lambda $, and hence to the probability density 
$p(q,\lambda )$ of the measurement outcomes, which play the role of the
Radon transform in the quantum tomography of the harmonic oscillator.
In general, the operators $ C(\lambda )$ can be any function (neither
self-adjoint nor unitary) of observables and, even more generally,
they may be connected to POVMs rather than observables.
\par The dual set $B(\lambda )$ can be obtained from the set
$C(\lambda )$ by solving Eq. (\ref{binorm}). For finite quorums, this
resorts to a matrix inversion. An alternative procedure uses the
Gram-Schmidt orthogonalization procedure \cite{orth}.  No such a
general procedure exists for a continuous spanning set. Many cases,
however, satisfy conditions (\ref{bbd1}) and (\ref{bbd}), and thus 
we can write $B(\lambda )= C(\lambda )^\dag$.

\subsection{Quantum estimation for harmonic-oscillator
  systems}\label{s:qopt}
The harmonic oscillator models several systems of interest in quantum
mechanics, as the vibrational states of molecules, the motion of an
ion in a Paul trap, and a single mode radiation field. Different
proposals have been suggested in order to reconstruct the quantum
state of a harmonic system, which all fit the framework of the
previous section, which is also useful for devising novel 
estimation techniques.  Here, the basic resolution formula involves
the set of displacement operators $ D (\alpha) = \exp (\alpha a^\dag -
\alpha ^* a)$, which can be viewed as exponentials of the
field-quadrature operators $ X_{\varphi }=(a^{\dag} e^{i\varphi
}+ae^{-i\varphi })/2$.  We have shown in Sec. \ref{homsec} that for a
single-mode radiation field $X_\varphi $ is measured through homodyne
detection.  For the vibrational tomography of a molecule or a trapped
ion $X_\varphi $ corresponds to a time-evolved position or momentum.
The set of displacement operators satisfies Eqs. (\ref{binorm}) and
(\ref{bbd}), since
\begin{eqnarray}
\hbox{Tr}[   D(\alpha)  D^\dag (\beta)] 
= \pi\delta^{(2)} (\alpha -\beta)\; ,\;\label{biorthom}
\end{eqnarray}
whereas Eq. (\ref{gentomo2}) reduces to the Glauber formula 
\begin{eqnarray}
  O = \int_{\mathbb C} \frac{d^2\alpha}{\pi} \; \hbox{Tr} \left[ O
  D^\dag (\alpha )\right] D (\alpha ) \label{glauber}\;.
\end{eqnarray}
Changing to polar variables $\alpha = (-i/2)k
e^{i\varphi}$, Eq. (\ref{glauber}) becomes 
\begin{equation}
  O = \int^{\pi}_0\frac{d\varphi}{\pi}\int^{+\infty}_{-\infty} 
\frac{d k\, |k|}{4}\,\hbox{Tr} [  O\;  e^{ik X_{\varphi}}]\, 
e^{-ik X_{\varphi}}\; , \label{op}
\end{equation}
which shows explicitly the dependence on the quorum $ X_\varphi $.
Taking the ensemble average of both members and evaluating the trace
over the set of eigenvectors of $ X_{\varphi}$, one obtains
\begin{eqnarray}
\langle   O  \rangle = 
\int^{\pi}_0\frac{d\varphi }{\pi}  \int^{+\infty}_{-\infty}\!\! dx\;
p(x,\varphi) 
\;  {\cal R}[O] (x,\varphi) \label{qht1}\;,
\end{eqnarray}
where $p(x;\varphi)={}_\varphi\langle x| \rho|x\rangle_\varphi$ is the
probability distribution of quadratures outcome.  The estimator of the
operator ensemble average $\langle O\rangle$ is given by
\begin{eqnarray}
{\cal R}[  O](x,\varphi) =
\hbox{Tr} [  O K(X_{\varphi}-x)]\;,\label{tomke}
\end{eqnarray}
where $K(x)$ is the same as in Eq. (\ref{kernelaa}).

\par Eq. (\ref{qht1}) is the basis of quantum homodyne
tomography. Notice that even though $K(x) $ is unbounded,
however, the matrix element $\langle \psi |K(X_\varphi
-x)|\phi\rangle$ can be bounded, whence it can be used to sample the
matrix element $\langle \psi|\rho |\phi \rangle$ of the state $\rho$, which,
according to Sec. \ref{s:stat}, is directly obtained by averaging the
estimator (\ref{tomke}) over homodyne
experimental values. In fact, for bounded $\langle \psi|K(X_\varphi
-x)|\phi\rangle $, the central limit theorem guarantees that
\begin{eqnarray}
\langle\psi|\rho|\phi\rangle&=&\int_0^\pi
\frac{d\varphi}\pi\int_{-\infty}^{+\infty} dx\ p(x,\varphi
)\;\;\langle \psi|K(X_\varphi -x)|\phi\rangle\label{mcsample0}\\&=& \lim_{N\to\infty}
\frac 1N \sum_{n=0}^N \langle \psi|K(x_{\varphi_n}-x_n))|\phi\rangle 
\;,\label{mcsample1}
\end{eqnarray} 
where $x_n$ is the homodyne outcome measured at phase $\varphi_n$ and
distributed with probability $p(x,\varphi)$.  Systematic errors are
eliminated by choosing randomly each phase $\varphi_n$ at which
homodyne measurement is performed.  As shown in Sec. \ref{s:stat}, for
finite number of measurements $N$, the estimate (\ref{mcsample1}) of
the integral in Eq. (\ref{mcsample0}) is Gaussian distributed around the true value $\langle
\psi |\rho |\phi \rangle $, with statistical error decreasing as
$N^{-1/2}$.  Notice that the measurability of the density operator
matrix element depends only on the boundedness of the matrix element
of the estimator, and that no adjustable parameters are needed
in the procedure, which thus is unbiased.  

\par The general procedure for noise deconvolution is presented in
Sec. \ref{noise}.  However, we give here the main result for the
density matrix reconstruction.  As shown in Sec. \ref{homsec}, the
effect of the efficiency in homodyne detectors is a Gaussian
convolution of the ideal probability $p(x,\varphi)$, as
\begin{eqnarray}
p_\eta(x,\varphi)=\sqrt{\frac{2\eta}{\pi(1-\eta)}}\;
\int_{-\infty}^{+\infty} dx'\ 
e^{-\frac{2\eta}{1-\eta}(x-x')^2}p(x',\varphi) 
\;\label{peta1}.
\end{eqnarray}
The tomographic reconstruction procedure still holds upon replacing 
$p(x,\varphi)$ with
$p_\eta(x,\varphi)$, so that 
\begin{eqnarray}
\rho=\int_0^\pi\frac{d\varphi}\pi\int_{-\infty}^{+\infty}
dx\; p_\eta(x,\varphi)\;K_\eta(X_\varphi -x),
\;\label{formtomhometa}
\end{eqnarray}
where now the estimator is 
\begin{eqnarray}
K_\eta(x)=\frac 12\hbox{Re}\int_{0}^{+\infty}
k\;dk\;e^{\frac{1-\eta}{8\eta}k^2+ikx}
\;\label{kerneleta1}.
\end{eqnarray}
In fact, by taking the Fourier transform of both members of
Eq. (\ref{peta1}), one can easily check that 
\begin{eqnarray}
\rho &=&\int_0^\pi\frac{d\varphi}\pi\int_{-\infty}^{+\infty} dx\;
p_\eta(x,\varphi)\;K_\eta(X_\varphi -x) \nonumber \\&= &
\int_0^\pi\frac{d\varphi}\pi\int_{-\infty}^{+\infty} dx\;
p(x,\varphi)\;K(X_\varphi -x)\;.\label{etaderc}
\end{eqnarray}
Notice that the anti-Gaussian in Eq. (\ref{kerneleta1}) causes a much
slower convergence of the Monte Carlo integral (\ref{formtomhometa}):
the statistical fluctuation will increase exponentially for decreasing
detector efficiency $\eta$. In order to achieve good reconstructions
with non-ideal detectors, then one has to collect a larger number of
data.

\par It is clear from Eq. (\ref{mcsample1}) that the measurability of
the density matrix depends on the chosen representation and on the
quantum efficiency of the detectors. For example, for the
reconstruction of the density matrix in the Fock basis the estimators are given by
\begin{eqnarray}
&&{\cal R}_\eta[|n\rangle \langle  n+d|](x,\varphi)= \int_{-\infty}^{+\infty}\frac
{dk\,|k|}{4}\, e^{\frac{1-\eta}{8\eta}k^2-ikx}\langle  n+d| e^{ik
X_{\varphi}}| n\rangle  \label{estimat}\\
&&=
e^{id(\varphi+\frac{\pi}{2})}\sqrt{\frac{n!}{(n+d)!}}\int_{-\infty}^{+\infty}
dk\,|k| e^{\frac{1-2\eta}{2\eta}k^2-i2kx} k^d L_n^d(k^2)\;,\nonumber
\end{eqnarray}
where $L_n^d(x)$ denotes the generalized Laguerre polynomials. Notice
that the estimator is bounded only for $\eta >1/2$, and below
the method would give unbounded statistical errors. However, this
bound is well below the values that are reasonably achieved in the
lab, where actual homodyne detectors have efficiency ranging between
$70\%$ and $90\%$ \cite{sch,kumar}. Moreover, a more efficient
algorithm is available, that uses the factorization formulas that hold
for $\eta=1$ \cite{Richter,factulf} 
\begin{eqnarray}
&&{\cal R}[|n\rangle \langle n+d|](x,\varphi)=e^{id\varphi }[4x
u_n(x)v_{n+d}(x) \\& &- 2\sqrt{n+1}u_{n+1}(x)v_{n+d}(x)-
2\sqrt{n+d+1}u_{n}(x)v_{n+d+1}(x)]\nonumber \;,
\end{eqnarray}
where $u_j(x)$ and $v_j(x)$ are the normalizable and unnormalizable
eigenfunctions of the harmonic oscillator with eigenvalue $j$,
respectively. The noise from quantum efficiency can be unbiased via
the inversion of the Bernoulli convolution, which holds for $\eta
>1/2$ \cite{invb}.

\par The use of Eq. (\ref{qht1}) to estimate arbitrary operators
through homodyne tomography will be the subject of the following
Chapter.  Notice that Eq. (\ref{glauber}) cannot be used for unbounded 
operators, however the estimators also for some unbounded operators
will be derived in Sec. \ref{s:ob}. 
\subsection{Some generalizations} 
Using condition (\ref{binorm}) one can see that the Glauber formula
can be generalized to
\begin{eqnarray}
  O = \int_{\mathbb C} \frac{d^2\alpha}{\pi} \; \hbox{Tr} \left[ O F_1
  D (\alpha ) F_2 \right] F_2^{-1} D^\dag (\alpha ) F_1^{-1}
\label{genglauber}\;, 
\end{eqnarray}
where $  F_1$ and $  F_2$ are two generic invertible operators. 
By choosing $  F^\dag_1 =   F_2 =   S (\zeta )$, where $S (\zeta )$  
is the squeezing operator
\begin{eqnarray}
  S (\zeta) = \exp\left[\frac12 \left( \zeta^2 a^{\dag 2} - {\zeta ^*}^2
a^2\right)\right]\;,\qquad \zeta \in {\mathbb C}\label{rotsque}\;,
\end{eqnarray}
we obtain the tomographic resolution
\begin{eqnarray}
\langle   O  \rangle = 
\int^{\pi}_0\frac{d\varphi  }{\pi}  \int^{+\infty}_{-\infty}\!\! dx 
\;  p_{\zeta}(x,\varphi  ) \;  \hbox{Tr} \left[  O  K(X_{\varphi  \zeta} -x)
\right] \label{qht5}\;,
\end{eqnarray}
in terms of the probability distribution of the generalized squeezed 
quadrature operators 
\begin{eqnarray}
  X_{\varphi  \zeta} =   S^\dag (\zeta)  
X_\varphi     S (\zeta) =
\frac12 \left[ (\mu e^{i\varphi  }+\nu e^{-i\varphi  }) a^\dag+
(\mu e^{-i\varphi  }+\nu ^*e^{i\varphi  }) a \right]
\label{rotquad}\;,
\end{eqnarray}
with $\mu = \cosh |\zeta|$ and $\nu = \sinh |\zeta| \exp[2i
\arg (\zeta )]$.  Such an estimation technique has been investigated in
detail in Ref. {\cite{cometipare}}.

\par A different estimation technique can be obtained by choosing in
Eq.  (\ref{genglauber}) $ F_1= I$, the identity operator, and $
F_2=(-)^{a^\dag a}$, the parity operator. In this case one gets
\begin{eqnarray}
  O = \int_{\mathbb C} \frac{d^2\alpha}{\pi} \; \hbox{Tr} \left[ O
  D^\dag ( \alpha) (-)^{a^\dag a}\right]\; (-)^{a^\dag a} D (\alpha
  )\label{phspa0}\; .
\end{eqnarray} 
Changing variable to $\alpha =2 \beta $ and using the relation
\begin{eqnarray}
(-)^{a^\dag a}   D (2\beta ) =   D^\dag (\beta )(-)^{a^\dag a}   D
  (\beta )\;
\end{eqnarray}
it follows  
\begin{eqnarray}
\langle O \rangle = \int_{\mathbb C} \frac{d^2\beta}{\pi} \; \hbox{Tr}
\left[ O 4 D^\dag (\beta )(-)^{a^\dag a} D (\beta)\right]\;
\hbox{Tr}\left[ D (\beta ) \rho D^\dag (\beta ) (-)^{a^\dag a}
\right]\label{phspa}\;.
\end{eqnarray}
Hence, it is possible to estimate $ \langle O \rangle $ by repeated
measurement of the parity operator on displaced versions of the state
under investigation. An approximated implementation of this technique
for a single mode radiation field has been suggested in
Refs. \cite{wodb,opa} through the measurement of the photon number
probability on states  displaced by  a beam splitter. A
similar scheme has been used for the experimental determination of
the motional quantum state of a trapped atom \cite{leib}.  In
comparison with the approximated methods, Eq. (\ref{phspa}) allows to
directly obtain the estimator ${\cal R}[ O](\alpha )$ for any operator $
O$ for which the trace exists.  For instance, the reconstruction of
the density matrix in the Fock representation is obtained by averaging
the estimator
\begin{eqnarray} 
{\cal R}[|n\rangle\langle n+d||](\alpha ) &=& 4 \langle n+d | D^\dag
  ( \alpha) (-)^{a^\dag a} D (\alpha )| n \rangle \\ &=& 4\;
  (-)^{n+d}\; \sqrt{\frac{n!}{(n+d)!}} \; (2 \alpha )^d \; e^{- 2
  |\alpha |^2}\, L_n^d (4 |\alpha |^2) \nonumber
\label{phspa1}\;, 
\end{eqnarray}
without the need of artificial cut-off in the Fock space \cite{leib}.
\subsection{Quantum estimation for spin systems}\label{s:spin}
 The spin tomographic methods of Refs. \cite{deconv,weigert,spinmac}
allow the reconstruction of the quantum state of a spin system.  These
methods utilize measurements of the spin in different directions, {\it i.e.} the
quorum is the set of operators of the form $\vec S\cdot\vec n$, where $\vec S$ is
the spin operator and $\vec n \equiv (\cos\varphi\sin\vartheta
,\sin\varphi\sin\vartheta,\cos\vartheta)$ is a varying unit vector.
Different quorums can be used, that exploit different sets of
directions.
\par The easiest choice for the set of directions $\vec n$ is to consider all possible
directions. The procedure to derive the tomographic formulas for this
quorum is analogous to the one employed in Sec. \ref{s:qopt} for
homodyne tomography. The reconstruction formula for spin tomography
for the estimation of an arbitrary operator $ O$ writes
\begin{eqnarray}
\langle  O\rangle=\sum_{m=-s}^s\int_{\Omega}\frac{d\vec
n}{4\pi}\;p(m,\vec n)\;{\cal R}[  O](m,\vec n)\;\label{spintom},
\end{eqnarray}
where $p(m,\vec n)$ is the probability of obtaining the eigenvalue $m$
when measuring the spin along direction $\vec n$, $ {\cal R}[
O](m,\vec n)$ is the tomographic estimator for the operator $ O$, and
$\Omega$ is the unit sphere. In this case the operators $ C(\lambda )$
of Eq. ({\ref{gentomo}) are given by the set of projectors over the
eigenstates $|m,\vec n\rangle$ of the operators $\vec S\cdot\vec n$.
 Notice that this is a complete set of operators on the system Hilbert
space $\cal H$.  In order to find the dual basis $ B$, one must
consider the unitary operators obtained by exponentiating the quorum,
{\it i.e.} $ D(\psi,\vec n)=\exp(i\psi\vec S\cdot\vec n)$, which
satisfy the bi-orthogonality condition (\ref{biort}). In fact, $
D(\psi,\vec n)$ constitutes a unitary irreducible representation of
 the group ${\mathbb G}=SU(2)$, and the bi-orthogonality condition is just the
orthogonality relations between the matrix elements of the group
representation \cite{murna}, {\it i.e.} 
\begin{eqnarray} \int_{{\mathbb G}}
dg\;{ D}_{jr}(g){ D}^\dagger_{tk}(g)=\frac Vd\delta_{jk}\delta_{tr}
\;\label{murnagh},
\end{eqnarray}
where $ D$ is a unitary irreducible representation of dimension $d$,
$dg$ is the group Haar invariant measure, and $V=\int_{\mathbb G}
dg$. For ${\mathbb G}=SU(2)$, with the $2s+1$-dimensional unitary
irreducible representation $D(\psi,\vec n)$ ($\vec n\in S^2$ unit
vector on the sphere, and $\psi\in[0,4\pi]$ the rotation angle around
$\vec n$) the Haar's invariant measure is $\sin^2\frac\psi2\sin\vartheta\;
d\vartheta\; d\varphi\; d\psi$, and  $\frac
Vd=\frac{8\pi^2}{2s+1}$. We need however to integrate only for
$\psi\in[0,2\pi]$ (the change of sign for $2\pi$ rotation is
irrelevant), whence the bi-orthogonality condition writes  
\begin{eqnarray}
\frac{2s+1}{4\pi^2}\int_\Omega d\vec n
\int_0^{2\pi}d\psi\;\sin^2\frac\psi2\; \langle j|e^{i\psi\vec n\cdot\vec
S}|r\rangle\langle t|e^{-i\psi\vec n\cdot\vec
S}|k\rangle=\delta_{jk}\delta_{tr}
\;\label{murmaghsu2}, 
\end{eqnarray}
and hence the spin tomography identity is given by 
\begin{eqnarray}
  O=\frac{2s+1}{4\pi^2}\int_\Omega d\vec n
\int_0^{2\pi}d\psi\;\sin^2\frac\psi2 \; \hbox{Tr} \left[ O
D^\dagger(\psi,\vec n)\right] D(\psi,\vec n) \;\label{spinglaub}.
\end{eqnarray}
Notice the analogy between Eq. (\ref{spinglaub}) and Glauber's formula
(\ref{glauber}). In fact, both homodyne and spin tomography can be
derived using the method of Group Tomography
{\cite{deconv}}, and the underlying groups are the Weyl-Heisenberg group and
the $SU(2)$ group, respectively. Formula (\ref{spintom}) is obtained
from Eq. (\ref{spinglaub}) through the expectation value calculated on
the eigenstates of $\vec S\cdot\vec n$. Thus, the explicit form of the
tomographic estimator is obtained as
\begin{eqnarray}
{\cal R}[ O](m,\vec
n)=\frac{2s+1}\pi\int_0^{2\pi}d\psi\;\sin^2\frac\psi2 \;  \hbox{Tr}
\left[  O\, e^{-i\psi \vec S\cdot\vec n}\right]\,e^{i\psi m}
\;\label{spinkern}.
\end{eqnarray}
\par As already noticed, there are other possible quorums for spin
tomography. For example, for spin $s=\frac 12$ systems, a self-dual
basis for the operator space is given by the identity and the Pauli
matrices. In fact, from the properties $\hbox{Tr}[\sigma _\alpha ]=0$
and $\sigma _\alpha \sigma _\beta =i\sum _{\gamma }\epsilon _{\alpha
  \beta \gamma }\sigma _\alpha \sigma _\beta $ ($\alpha,\beta ,\gamma
=x,y,z$), both the
bi-orthogonality relation (\ref{biort}) and the trace condition
(\ref{binorm}) follow. In this case the reconstruction formula writes
\begin{eqnarray}
\langle  O\rangle=\frac 12 \hbox{Tr}
\left[  O\right] +\frac 12 
\sum_{\alpha=x,y,z} \sum_{m=\pm\frac 12} m\,p(m,\vec n_\alpha) \;  \hbox{Tr}
\left[  O \sigma_\alpha\right]
\;\label{spintom12}.
\end{eqnarray}
\par In the case of generic $s$ spin system, Weigert has also shown
{\cite{weigert}} that by choosing $(2s+1)^2$ arbitrary directions for
$\vec n$, it is possible to obtain (in almost all cases) a quorum of
projectors $|s,\vec n_j\rangle\langle s,\vec n_j|$
($j=1,\cdots,(2s+1)^2$), where $|s,\vec n_j\rangle$ is the eigenstate
pertaining to the maximum eigenvalue $s$ of $\vec S\cdot\vec n_j$.
\subsection{Quantum estimation for a free particle}\label{s:fre}
The state of a moving packet can be inferred from position measurement
at different times \cite{wpack}. Assuming a particle with unit mass
and using normalized unit $\hbar/2 =1$, the free Hamiltonian is given
by the square of momentum operator $ H_F = p^2$.  In terms of the
eigenvectors $|x\rangle$ of the position operator and of the
selfadjoint operator
\begin{eqnarray}
  R(x,\tau) = e^{- i  p^2 \tau}\;  |x\rangle\langle x|\; 
e^{i  p^2 \tau}
\label{free1}\;,
\end{eqnarray}
the probability density of the position of the free particle at time
$\tau$ is obtained as $p(x,\tau)= \hbox{Tr}[ \rho\, R(x,\tau)]$. The
operators $ R(x,\tau)$ provide a self-dual basis, and an arbitrary
particle state can be written as
\begin{eqnarray}
  \rho = \int_{\mathbb R}\int_{\mathbb R} dx \; d\tau\;  p(x,\tau) \;  
  R(x,\tau)\label{free3}\;.
\end{eqnarray}
\section{Noise deconvolution and adaptive tomography \label{devel}}
In this section we will analyze: 1) the noise deconvolution scheme of
Refs. {\cite{deconv,maurogtslov}}, that allows to eliminate the experimental
noise that arises from imperfect detection and lossy devices; 2) the
adaptive tomography technique of Ref. {\cite{adapt}} that allows to tune the
unbiased tomographic estimators to a specific sample of experimental data, in
order to reduce the statistical noise.
\subsection{Noise deconvolution \label{noise}}
In short, it is possible to eliminate detection noise when it is
possible to invert the noise map. A noise process is described by a
trace preserving completely positive map $\Gamma $.  The noise can be
deconvolved at the data analysis if
\begin{itemize}
\item the inverse of $\Gamma$ exists, namely $\Gamma^{-1}:{\cal
L}({\cal H})\to{\cal L}({\cal H})$, with
$\Gamma^{-1}\left[\Gamma\left[O\right]\right]=O$, for $\forall
O\in{\cal L}({\cal H})$. 
\item the estimator ${\cal E}_\lambda[O](Q_\lambda)$ is in the domain of
$\Gamma^{-1}$ 
\item the map $\Gamma^{-1}\left[{\cal E}_\lambda[O](Q_\lambda)\right]$ is a
function of $Q_\lambda$. 
\end{itemize} 
If the above conditions are met, we can recover the ``ideal''
expectation value $\langle O\rangle $ that we would get without noise. This is
achieved by replacing ${\cal E}_\lambda [O](Q_\lambda )$ with $\Gamma
^{-1}[{\cal E}_\lambda [O](Q_\lambda ) ]$, and evaluating the ensemble
average with the state $\Gamma ^\tau (\rho )$, namely the state
affected by the noise ($\Gamma ^\tau $ represents the dual map, that
provides the evolution in the Schroedinger picture). Hence, one has
\begin{eqnarray}
 \langle O \rangle = \int _{\Lambda }d \lambda \, \hbox{Tr} [\Gamma
^{-1}[{\cal E}_\lambda [O](Q_\lambda )] \Gamma ^\tau (\rho )]\equiv
\int _{\Lambda }d \lambda \, \langle \Gamma ^{-1}[{\cal E}_\lambda
[O](Q_\lambda )] \rangle _{\Gamma }\;.
\end{eqnarray}
\par Consider for example the noise arising from non-unity quantum
efficiency $\eta$ of homodyne detectors. Recall that the ideal
probability density is replaced by a Gaussian convolution with rms
$\Delta _\eta ^2=(1-\eta)/(4\eta )$.  Then, the map $\Gamma _\eta $
acts on the quorum as follows
\begin{eqnarray}
\Gamma _\eta [e^{ikX_\varphi }] &=&
\int _{-\infty }^{+\infty} dx 
\, e^{ikx} \,\Gamma _\eta [|x\rangle \langle x|] \\&= & 
\int _{-\infty }^{+\infty} dx 
\int _{-\infty }^{+\infty} dx' 
\, e^{ikx}\,e^{-\frac {(x-x')^2}{2\Delta ^2}} 
[|x'\rangle \langle x'|]=
e^{-\frac 12 \Delta ^2 k^2}\,
e^{ikX_\varphi}\nonumber \;.
\end{eqnarray}
Of course one has 
\begin{eqnarray}
\Gamma ^{-1}_\eta [e^{ikX_\varphi }]= e^{ \frac 12 \Delta ^2
k^2}\,e^{ikX_\varphi}\;.
\end{eqnarray}
In terms of the Fourier transform of the estimator
\begin{eqnarray}
\tilde {\cal R}[O](y,\varphi)= \int _{-\infty }^{+\infty}
\frac{dx}{2\pi }\, 
e^{ixy}\,{\cal R}[O](x, \varphi)
\; ,
\end{eqnarray}
one has 
\begin{eqnarray}
\tilde {\cal R}_
\eta [O](y,\varphi)= e^{\frac 12 \Delta ^2 y^2}
\tilde {\cal R}[O](y,\varphi)
\;.\label{rtra}
\end{eqnarray}
We applied the above result in Sec. \ref{s:qopt}, where the effect of
non-unity quantum efficiency for reconstructing the density matrix
elements was discussed. The use of the estimator in
Eq. (\ref{kerneleta1}) and the origin of the bound $\eta > 1/2$ is now
more clear.  \par Another simple example of noise deconvolution is
given here for a spin $1/2$ system. Consider the map that describes
the ``depolarizing channel''
\begin{eqnarray}
\Gamma_p[O]=(1-p)O+\frac p2\,\hbox{Tr}
[O]\,I\;,\qquad 0\leq p\leq
1
\;\label{depolarchann}.
\end{eqnarray}
This map  can be inverted for $p \neq 1 $ as follows
\begin{eqnarray}
\Gamma^{-1}_p[O]=\frac {1}{1-p}\left (O -\frac p2\,\hbox{Tr}[O]\,I\right )
\;.
\end{eqnarray}
Then Eq. (\ref{spintom12}) can be replaced with 
\begin{eqnarray}
\langle  O\rangle=\frac 12 \hbox{Tr}
\left[  O\right] +\frac {1}{2(1-p)} \sum_{m=\pm\frac 12}
\sum_{\alpha=x,y,z} m\,p_p(m,\vec n_\alpha) \;  \hbox{Tr}
\left[  O \sigma_\alpha\right]\;, 
\end{eqnarray}
where now $p_p(m,\vec n_\alpha)$ represents the probability of outcome
$m$ when measuring $\sigma _\alpha $ on the noisy state $\Gamma
_p[\rho ]$. 
\subsection{Adaptive tomography\label{s:adapt}}
 The idea of adaptive tomography is that the tomographic null
estimators of Eq. (\ref{nullestim}) can be used to reduce 
statistical errors. In fact, the addition of a null estimator in the
ideal case of infinite statistics does not change the average since
its mean value is zero, but can change the variance.
Thus, one can look for a procedure to reduce the variance by adding
suitable null functions. Consider the class of equivalent estimators for $O$
\begin{eqnarray}
{\cal E}'_\lambda[O](Q_\lambda) =
{\cal E}_\lambda[O](Q_\lambda)+\sum_{i=1}^M\nu_i{\cal N}_i(Q_\lambda)
\;\label{nullproced}.
\end{eqnarray}
Each estimator in the class ${\cal E}'$ is identified by the coefficient
vector $\vec\nu$. The variance of the tomographic averages can be
evaluated as
\begin{eqnarray}
\overline{\Delta^2{\cal E}'[O]}=\overline{\Delta^2{\cal E}[O]}+
 2\sum_{i=1}^M\nu_i\overline{{\cal
N}_i{\cal E}[O]}\;+\;\sum_{i,j=1}^M\nu_i\nu_j\overline{{\cal N}_i{\cal N}_j}
\;\label{nullvar},\end{eqnarray}
where $\overline{F}\equiv \left\langle \int _\Lambda d\lambda \;
F(Q_\lambda)\right\rangle $, and 
\begin{eqnarray}
\overline{\Delta^2{\cal E}[O]}=\overline{{\cal E} ^2[O]}- \overline{{\cal E}[O]} ^2
\;\label{varianm}.
\end{eqnarray}
Minimizing $\overline{\Delta^2{\cal E}'[O]}$ with respect to the
coefficients $\nu_i$, one obtains the equation 
\begin{eqnarray}
\sum_{j=1}^M\nu_j\overline{{\cal N}_i{\cal N}_j}=-\overline{{\cal E}[O]{\cal
N}_i}
\;\label{minimnu}, 
\end{eqnarray}
which can be solved starting from the estimated mean
values, with the vector $\vec\nu$ as unknown. 
Notice that the obtained vector $\vec\nu$ will depend on
the experimental data, and has to be calculated with the above
procedure for any new set of data. 
\par In this way we obtain an adaptive tomographic
algorithm, which consists in the following steps:
\begin{itemize}\item Find the null estimators ${\cal
N}_i(Q_\lambda)\ (i=1,\cdots,M)$ for the quorum which is being used in
the experiment.\item Execute the experiment and collect the input
data.\item Calculate, using the obtained data, the mean values
$\overline{{\cal N}_i{\cal N}_j}$ and $\overline{{\cal E}[O]{\cal N}_i}$,
and solve the linear system (\ref{minimnu}), to obtain $\vec\nu$.\item
Use the vector $\vec\nu$ obtained in the previous step to build the
`optimized estimator' ${\cal E}'[O](Q_\lambda )={\cal E}[O](Q_\lambda )
+\sum_i\nu_i\,{\cal N}_i(Q_\lambda )$. Using
the data collected in the first step, the mean value $\langle O
\rangle $ is now evaluated as
\begin{eqnarray}
\langle O\rangle =\int_\Lambda d\lambda \;
\langle {\cal E'}_\lambda[O](Q_\lambda)\rangle 
\;\label{defestimotti},
\end{eqnarray}
where the optimized
estimator has been used. 
\item For each new set of data the whole procedure must be repeated, 
as $\vec\nu$ is dependent on the data.
\end{itemize}
Notice that also the experimental mean values are slightly modified in
the adaptive tomographic process, since null estimators do not change
mean values only in the limiting case of infinite
statistics. 
Examples of simulations of the adaptive technique that efficiently
reduce statistical noise of homodyne tomographic reconstructions can
be found in Ref. {\cite{adapt}}. In homodyne tomography null
estimators are obtained as linear combinations of the following
functions 
\begin{eqnarray}
{\cal N}_{k,n}(X_\varphi)=X_\varphi ^k\,e^{\pm
  i(k+2+2n)\varphi}\;,\qquad k,n\geq 0\;.
\end{eqnarray}
One can easily check that such functions have zero average over
$\varphi$, independently on $\rho$. Hence, for every operator $O$ one
actually has an equivalence class of infinitely many unbiased
estimators, which differ by a linear combination of functions ${\cal
N}_{k,n}(X_\varphi )$. It is then possible to minimize the rms error
in the equivalence class by the least-squares method, obtaining in
this way an optimal estimator that is {\em adapted} to the particular
set of experimental data.

%% file: cap4.tex
\chapter{Universal homodyning}
As shown in Ref. \cite{tokio}, homodyne tomography can be used as a
kind of universal detector, for measuring generic field operators,
however at expense of some additional noise. In this chapter, the
general class of field operators that can be measured in this way
is reviewed, which includes also operators that are 
inaccessible to heterodyne detection.  In Ref. \cite{added} the most
relevant observables were analyzed---such as the intensity, the real,
the complex field, and the phase---showing how their tomographic
measurements are affected by noise that is always larger than the
intrinsic noise of the direct detection of the considered
observables. On the other hand, by comparing the noise from homodyne
tomography with that from heterodyning (for those operators that can
be measured in both ways), in Ref. \cite{added} it was shown that for
some operators homodyning is better than heterodyning when the mean
photon number is sufficiently small, i.e. in the quantum regime, and
in this chapter such comparison will be also reviewed. 

\section{Homodyning observables}\label{s:ob} 
Homodyne tomography provides the maximum achievable information on the
quantum state of a single-mode radiation field through the use of the
estimators in Sec. \ref{s:qopt}. In principle, the knowledge of the
density matrix should allow one to calculate the expectation value
also for unbounded operators.  However, this is generally true only
when one has an analytic knowledge of the density matrix, but it is
not true when the matrix has been obtained experimentally.  In fact,
the Hilbert space is actually infinite dimensional, whereas
experimentally one can achieve only a finite matrix, each element
being affected by an experimental error. Notice that, even though the
method allows one to extract {\em any} matrix element in the Hilbert
space from the same bunch of experimental data, it is the way in which
errors converge in the Hilbert space that determines the actual
possibility of estimating the trace $\langle O\rangle =\mbox{Tr}[O
\rho]$ for an arbitrary operator $O$.  This issue has been debated in
the set of papers of Ref. \cite{invb}.  Consider for example the
number representation, and suppose that we want to estimate the
average photon number $\langle a^{\dag}a\rangle $.  In Ref.
\cite{nico} it has been shown that for non-unit quantum efficiency the
statistical error for the diagonal matrix element $\langle
n|\rho|n\rangle $ diverges faster than exponentially versus $n$,
whereas for $\eta=1$ the error saturates for large $n$ to the
universal value $\varepsilon_n=\sqrt{2/N}$ that depends only on the
number $N$ of experimental data, but is independent on both $n$ and on
the quantum state. Even for the unrealistic case $\eta=1$, one can see
immediately that the estimated expectation value $\langle
a^{\dag}a\rangle =\sum_{n=0}^{H-1}n\rho_{nn}$ based on the measured
matrix elements $\rho_{nn}$, will exhibit an unbounded error versus
the truncated-space dimension $H$, because the non-vanishing error of
$\rho_{nn}$ versus $n$ multiplies the increasing eigenvalue $n$.  \par
Here, we report the estimators valid for any operator that admits a
normal ordered expansion, giving the general class of operators that
can be measured in this way, also as a function of the quantum
efficiency $\eta$.  Hence, from the same tomographic experiment, one
can obtain not only the density matrix, but also the expectation value
of various field operators, also unbounded, and including some
operators that are inaccessible to heterodyne detection. However, the
price to pay for such detection flexibility is that all measured
quantities will be affected by noise.  If one compares this noise with
that from heterodyning (for those operators that can be measured in
both ways), it turns out that for some operators homodyning is anyway
less noisy than heterodyning, at least for small mean photon
numbers. The procedure for estimating the expectation $\langle
O\rangle $ will be referred to as {\em homodyning the observable} $O$.
\par By homodyning the observable $O$ we mean averaging an appropriate
estimator ${\cal R}[O](x,\varphi)$, independent on the state $\rho$,
over the experimental homodyne data, achieving in this way the
expectation value $\langle O\rangle $ for every state $\rho$, as in
Eq. (\ref{qht1}). For unbounded operators one can obtain the explicit
form of the estimator ${\cal R}[O](x,\varphi)$ in a different way.
Starting from the identity involving trilinear products of Hermite
polynomials \cite{gradshtein}
\begin{eqnarray}
\int^{+\infty}_{-\infty}\, d x\,e^{-x^2}\,H_k(x)\,H_m(x)\,H_n(x)=\frac{
2^{\frac{m+n+k}{2}}\pi^{{\frac 1 2}}k!m!n!}{(s-k)!(s-m)!(s-n)!}\;,
\end{eqnarray}
for $k+m+n=2s$ even, 
Richter proved  the following  nontrivial formula for  the expectation
value of the normally ordered field operators \cite{Rich}
\begin{eqnarray}
\langle a^{\dag}{}^n a^m\rangle =\int^{\pi}_0 \frac{d\varphi}{\pi}
\int^{+\infty}_{-\infty} \, d x\, p(x,\varphi)
e^{i(m-n)\varphi}\frac{H_{n+m}(\sqrt{2}x)}{ \sqrt{2^{n+m}}{{n+m}\choose
n}}\;,
\end{eqnarray}
which corresponds to the estimator
\begin{eqnarray}
{\cal R}[a^{\dag}{}^n a^m](x,\varphi)=e^{i(m-n)\varphi}
\frac{H_{n+m}(\sqrt{2}x)}{\sqrt{2^{n+m}}{{n+m}\choose n}}\;.
\end{eqnarray}
This result can be easily extended to the case of non-unit quantum 
efficiency $\eta< 1$. Using Eq. (\ref{rtra}) one obtains 
\begin{eqnarray}
{\cal R}_{\eta}[a^{\dag}{}^n a^m](x,\varphi)=e^{i(m-n)\varphi}
\frac{H_{n+m}(\sqrt{2\eta }x)}{\sqrt{(2\eta)^{n+m}}{{n+m}\choose n}}\;.\label{Rnm}
\end{eqnarray}
From Eq. (\ref{Rnm}) by linearity one can obtain
the estimator ${\cal R}_{\eta}[f](x,\varphi)$ for any operator function
$f$ that has normal ordered expansion
\begin{eqnarray}
f\equiv f(a,a^{\dag})=\sum_{nm=0}^{\infty}f^{(N)}_{nm}a^{\dag}{}^n a^m\;.
\end{eqnarray}
From Eq. (\ref{Rnm}) one obtains
\begin{eqnarray}
{\cal R}_{\eta}[f](x,\varphi)&=&\sum_{s=0}^{\infty}\frac{H_s(\sqrt{2 \eta
  }x)}{s!
(2\eta)^{s/2}}\sum_{nm=0}^{\infty}f^{(N)}_{nm}e^{i(m-n)\varphi}n!m!\delta_{n+m,s}
\nonumber \\&= & 
\sum_{s=0}^{\infty}\frac{H_s(\sqrt{2 \eta }x)i^s}{s!
(2\eta)^{s/2}}\frac{d^s}{dv^s}\Bigg|_{v=0}\!\!\!\!
{\cal F}[f](v,\varphi),\label{123}
\end{eqnarray}
where
\begin{eqnarray}
{\cal F}[f](v,\varphi)=\sum_{nm=0}^{\infty}f^{(N)}_{nm}
{{n+m}\choose m}^{-1}(-iv)^{n+m}e^{i(m-n)\varphi}\;.\label{ff}
\end{eqnarray}
Continuing from Eq. (\ref{123}) one has
\begin{eqnarray}
{\cal R}_{\eta}[f](x,\varphi)=\exp\left(\frac{1}{2\eta}\frac{d^2}{dv^2}+
\frac{2ix}{\sqrt{\eta}}\frac{d}{dv}\right)\Bigg|_{v=0}{\cal F}[f](v,\varphi)\;,
\end{eqnarray}
and finally  
\begin{eqnarray}
{\cal R}_{\eta}[f](x,\varphi)=\int_{-\infty}^{+\infty}\frac{dw}{\sqrt{2\pi\eta^{-1}}}
e^{-\frac{\eta}{2}w^2}{\cal F}[f](w+2ix/\sqrt{\eta},\varphi)\;.
\label{generalf}
\end{eqnarray}
Hence one concludes that the operator $f$ can be measured by homodyne
tomography if the function ${\cal F}[f](v,\varphi)$ in Eq.  (\ref{ff})
grows slower than $\exp (-\eta v^2/2)$ for $v\to\infty$, and the
integral in Eq. (\ref{generalf}) grows at most exponentially for
$x\to\infty$ (assuming $p(x,\varphi)$ goes to zero faster than
exponentially at $x\to\infty$).  
\par The robustness to additive phase-insensitive noise of this method
of homodyning observables has also been analyzed in
Ref. \cite{tokio}, where it was shown that just half photon of thermal
noise would spoil completely the measurement of the density matrix
elements in the Fock representation.
\par In Table \ref{td1} we report the estimator ${\cal
R}_{\eta}[O](x,\varphi)$ for some operators $O$. The operator $\hat W_s$
gives the generalized Wigner function $W_s (\alpha,\alpha ^*)$ for
ordering parameter $s$ through the relation in Eq. (\ref{Ws2}). From
the expression of ${\cal R}_{\eta}[\hat W_s](x,\varphi)$ it follows that
by homodyning with quantum efficiency $\eta$ one can measure the
generalized Wigner function only for $s<1-\eta^{-1}$: in particular
the usual Wigner function for $s=0$ cannot be measured for any quantum
efficiency. 
\begin{table}[hbt]
\begin{center}
\begin{tabular}{|c|c|}
\hline \hline
 $O$ & ${\cal R}_{\eta}[O](x,\varphi)$ \\ \hline \hline 
 $a^{\dag}{}^n a^m$ & ${\displaystyle e^{i(m-n)\varphi}
\frac{H_{n+m}(\sqrt{2\eta }x)}{\sqrt{(2\eta )^{n+m}}{{n+m}\choose n}} }$ \\ \hline
$a$ & $2e^{i\varphi}x$ \\ \hline
 $a^2$ & $e^{2i\varphi}(4x^2-1/\eta)$ \\ \hline
 $a^{\dag}a$ & $2x^2-{\frac {1}{2\eta }}$ \\ \hline
 $(a^{\dag}a)^2$ & ${\frac 8 3}x^4-\frac{4-2\eta }{\eta
 }x^2+\frac{1-\eta }{2\eta ^2}$ \\ \hline
 $\hat W_s=\frac{2}{\pi(1-s)}\left(\frac{s+1}{s-1}\right)^{a^{\dag}a}$ & 
${\displaystyle \int_0^{\infty}dt\frac{2e^{-t}}{\pi(1-s)-{\frac 1\eta}}
\cos\left(2\sqrt{\frac{2t}{(1-s)-{\frac 1 \eta}}}x\right)}$ \\ \hline
 $|n\rangle \langle  n +d |$ & ${\cal R}_\eta [|n \rangle \langle n+d|] (x,\varphi
 )$ in Eq. (\protect\ref{estimat}) \\
\hline \hline
\end{tabular} 
\end{center}
\caption{\footnotesize Estimator ${\cal R}_{\eta}[O](x,\varphi)$ for 
some operators $O$ (From Ref.\cite{tokio}).}\label{td1}
\end{table}
\section{Noise in tomographic measurements}\label{s:add}
In this section we will review the analysis of Ref. \cite{added},
where the tomographic measurement of following four relevant field
quantities have been studied: the field intensity, the real field or 
quadrature, the complex field, and the phase. For all these quantities the
conditions given after Eq. (\ref{generalf}) are fulfilled.  
\par The tomographic measurement of the observable $O$ is provided in
terms of the average 
$\overline{w_{\eta}}$ of the estimator $w_{\eta}\equiv {\cal R}_{\eta}[O]
(x,\varphi )$ over the homodyne data. The precision of the measurement is
given by the confidence interval $\sqrt{\overline{\Delta w^2_{\eta}}}$. When  
$w_{\eta}$ is a  real quantity, one has
\begin{eqnarray}
\overline{\Delta w^2_{\eta}}= \overline{w ^2_{\eta}}
-\overline{w_{\eta }}\,^2  \;,
\end{eqnarray}
where
\begin{eqnarray}
\overline{w^2_{\eta}} \equiv \overline{{\cal R}_{\eta}^2[O](x,\varphi)}=
\int^{\pi}_{0} \frac{d\varphi}{\pi}\int_{-\infty}^{\infty} \!\!
dx\;p_{\eta} (x,\varphi)\; {\cal R}^2_{\eta}[O](x,\varphi)\;.
\label{ysquared}
\end{eqnarray}
When $w_\eta $ is complex, one has to consider the eigenvalues of the
covariance matrix, namely
\begin{eqnarray}
\overline{\Delta w_{\eta}^2} =
\frac{1}{2}\left[\overline{|w|_{\eta}^2} -
|\overline{w_{\eta}}|^2\pm |
\overline{w^2_{\eta}}-
\overline{w_\eta }^2
|\right]
\label{amp1}\;.
\end{eqnarray}
When the observable $O$ can also be directly measured by a specific
setup we can compare the tomographic precision $\overline{\Delta w^2}$
with $\langle \Delta O ^2 \rangle =\langle O^ 2 \rangle -\langle O
\rangle ^2$.  Notice that, when we deal with $\eta <1 $ the noise
$\langle \Delta O ^2_\eta \rangle $ is larger that the quantum
fluctuations due to smearing effect of non-unit quantum efficiency.  As
we will see, the tomographic measurement is always more noisy than the
corresponding direct measurement for any observable at any quantum
efficiency $\eta$.  This is not surprising, in view of the larger
amount of information retrieved in the tomographic measurement 
as compared to the direct measurement of a single quantity.  \par
According to Eq. (\ref{ysquared}), the evaluation of the added noise
requires the average of the squared estimator. For the estimators in
Eq. (\ref{Rnm}) it is very useful the following identity for the
Hermite polynomials \cite{wun}
\begin{eqnarray}
H^2_{n}(x)= 2^n n!^2 \sum_{k=0}^{n}\frac{H_{2k}(x)}{k!^2\;2^k\;(n-k)!}
\label{er2}\;, 
\end{eqnarray}
that allows to write 
\begin{eqnarray}
 &&{\cal R}^2_{\eta} [a^{\dag n} a^m](x,\varphi) = \nonumber \\& & e^{2i\varphi(m-n)}
\frac{n!^2 m!^2}{\eta^{m+n}} 
\sum_{k=0}^{m+n}\frac{(2k)!\eta^k}{k!^4
(n+m-k)!}{\cal R}_{\eta}[a^{\dag k} a^k](x,\varphi)
\label{er3}\;, 
\end{eqnarray}
namely the squared estimator
${\cal R}^2_{\eta} [a^{\dag n} a^m](x,\varphi)$ can be written just 
in terms of "diagonal" estimators 
${\cal R}_{\eta}[a^{\dag k} a^k](x,\varphi)$. 
\subsection{Field-Intensity}
Photodetection is the direct measurement of the field-intensity. For
non-unit quantum efficiency $\eta$, the probability of detecting $m$
photons is given by the Bernoulli convolution in
Eq. (\ref{conv_n}). Let us consider the rescaled photocurrent
\begin{eqnarray}
I_{\eta} = \frac{1}{\eta} a^{\dag}a
\label{pho2}\;,
\end{eqnarray}
which traces the photon number, namely 
\begin{eqnarray}
\langle I_{\eta} \rangle = \frac{1}{\eta}
\sum_{m=0}^{\infty} m\;p_\eta (m) = \langle a^\dag a \rangle \equiv \bar{n}
\label{pho3}\;.
\end{eqnarray}
The variance of $I_{\eta}$ is
given by 
\begin{eqnarray}
\langle \Delta I^2 _\eta \rangle  =\frac{1}{\eta^2}
\sum_{m=0}^{\infty} m^2 p(m) - \bar n ^2= \langle\Delta n^2\rangle  +
\bar{n} \left(\frac{1}{\eta}-1\right)
\label{pho4}\;, 
\end{eqnarray}
where $\langle\Delta n^2\rangle$ denotes the intrinsic photon number
variance, and $\bar{n} (\eta^{-1} -1)$ represents the noise introduced
by inefficient detection.  The tomo\-graphic estimator that traces the
pho\-ton number is given by the pha\-se-inde\-pendent function
$w_{\eta}\equiv 2x^2 - (2\eta)^{-1}$.  Using Eq. (\ref{er3}) we can
evaluate its variance as follows
\begin{eqnarray}
\overline{\Delta w_{\eta}^2} = \langle\Delta n^2 \rangle
+ \frac{1}{2}\langle n^2 \rangle + \bar{n} \left(\frac{2}{\eta}
-\frac{3}{2}\right) + \frac{1}{2\eta^2} 
\label{pho5}\;.
\end{eqnarray}
The noise 
$N[n]$ added by tomography in the measurement of the field 
intensity $n$ is then given by 
\begin{eqnarray}
N[n] = 
\overline{\Delta w_{\eta}^2} -
\langle \Delta I^2 \rangle_{\eta}=
\frac{1}{2}\left[ \langle n^2 \rangle+
\bar{n}\left(\frac{2}{\eta}-1 \right) + \frac{1}{\eta^2}\right]
\label{pho6}\;.
\end{eqnarray}
Notice that $N[ n]$ is always positive, and largely 
depends on state under examination. 
For coherent states we have the noise-ratio
\begin{eqnarray}
\delta n _\eta = \sqrt{\frac{\overline{\Delta w_{\eta}^2}}{\langle
\Delta I^2\rangle_{\eta}}}=\left[2+\frac{1}{2}\left(\eta\bar{n}+
\frac{1}{\eta\bar{n}}\right)\right ]^{1/2}        
\label{pho7}\;,
\end{eqnarray}                                   
which is minimum for $\bar{n}=\eta^{-1}$.
\subsection{Real Field}
For single mode radiation the electric field is proportional to a
quadrature $X =(a+a^{\dag})/2$, which is just traced by homodyne
detection at fixed zero-phase with respect to the local oscillator.
The tomographic estimator is given by $w_{\eta}\equiv {\cal R}_{\eta}
[X](x,\varphi) = 2 x \cos\varphi$, independently on $\eta$, whereas the
squared estimator ${\cal R}^2_\eta[ X ]$ can be written as 
\begin{eqnarray}
w_{\eta}^2 =\frac{1}{4}\left[
{\cal R}_{\eta }[a^2](x,\varphi) + {\cal R}_\eta [a^{\dag 2}](x,\varphi)\right]
+{\cal R}_\eta [a^{\dag} a](x,\varphi) +
\frac{1}{2\eta}+\frac{\cos(2\varphi )}{2}
\label{fie1}\;.
\end{eqnarray}    
Then one has 
\begin{eqnarray}
\overline{\Delta w_{\eta }^2} &=&
\frac{1}{4}\left[\langle a^{\dag 2} \rangle
+ \langle a^2\rangle\right] + \bar{n} + \frac{1}{2\eta} - \frac 14
\left\langle 
a+a^{\dag}\right\rangle^2  \nonumber \\&= & 
 \langle \Delta X^2
\rangle + \frac{1}{2}\bar{n} +\frac{2-\eta}{4\eta}
\label{fie11}\;, 
\end{eqnarray}
where $\langle \Delta X^2 \rangle$ represents the intrinsic quadrature
fluctuations.  The tomographic noise in Eq. (\ref{fie11}) can be
compared with the rms variance of direct homodyne detection (see
Sec. \ref{homsec})
\begin{eqnarray}
\langle\Delta X^2 \rangle_{\eta} =
\langle \Delta X^2 \rangle +\frac{1-\eta}{4\eta}
\label{fie20}\;.
\end{eqnarray}    
Then the added noise reads
\begin{eqnarray}
 N[X]= \frac{\bar n}{2}+\frac{1}{4\eta}
\label{fie2}\;. 
\end{eqnarray}
For coherent states $\langle \Delta X^2 \rangle =1/4$, and  
one has the noise-ratio
\begin{eqnarray}
\delta x _\eta = \sqrt{\frac{\overline{\Delta w_{\eta}^2}}{\langle
\Delta X^2 \rangle_{\eta}}} =\sqrt{2\eta\bar {n} +2} 
\label{fie3}\;.
\end{eqnarray}        
\subsection{Field amplitude}\label{fampl}
The detection of the complex field amplitude of a single-mode
light-beam is represented by the generalized measurement of the
annihilation operator $a$. The tomographic estimator for $a$ is given by
the complex function $w_{\eta}\equiv{\cal R}_\eta [a](x,\varphi) = 2x
\exp(i\varphi)$, and the precision of the measurement is evaluated as in
Eq. (\ref{amp1}).  From Eq. (\ref{er3}) one obtains
\begin{eqnarray}
w_{\eta}^2 \equiv {\cal R}^2_{\eta}[a](x,\varphi) = e^{i2\varphi}
\left[\frac{1}{\eta}+2{\cal R}_{\eta}[a^{\dag}a ](x,\varphi) \right] =
\frac{e^{i2\varphi}}{\eta} + {\cal R}_{\eta}[a^2 ](x,\varphi) 
\label{amp2}\;,
\end{eqnarray}    
and 
\begin{eqnarray}
\left|w_{\eta}\right|^2 \equiv  \left|{\cal R}_{\eta}[a](x,\varphi)\right|^2 =
\frac{1}{\eta} \left[1+2\eta {\cal R}_{\eta}[a^{\dag}a ](x,\varphi)\right]
\label{amp3}\;, 
\end{eqnarray}
and hence 
\begin{eqnarray}
\overline{\Delta w_{\eta}^2} = \frac{1}{2}
\left[\frac{1}{\eta}+ 2 \bar{n} - \left|\langle a \rangle \right|^2
\pm\left| \langle a^2 \rangle- \langle a \rangle^2  \right|
\right]
\label{amp4}\;.
\end{eqnarray}
The optimal measurement of the complex field $a$ is obtained through
heterodyne detection. As notice in Sec. \ref{hetsec}, the probability
distribution is given by the generalized Wigner function $W_{s}
(\alpha,\alpha ^*)$, with $s=1-\frac 2 \eta $. Using
Eq. (\ref{exhet}) the precision of the measurement is easily evaluated
as follows
\begin{eqnarray}
\langle \Delta a^2\rangle _{\eta}&=& 
\frac{1}{2}\left[\overline{|\alpha|^2} - |\overline{\alpha}|^2
\pm \left|\overline{\alpha^2} -\overline{\alpha}^2\right|\right]     
\nonumber \\&= & 
\frac{1}{2}\left[\bar{n}+\frac{1}{\eta} - |\langle a\rangle|^2
\pm\left|\langle a^2\rangle -\langle a\rangle^2 \right|\right]     
\label{amp51}\;.
\end{eqnarray}
The noise added by quantum tomography then reads
\begin{eqnarray}
N[a] = \frac{1}{2}\bar{n} 
\label{amp6}\;,  
\end{eqnarray}           
which is independent on quantum efficiency. For a coherent state we
have
\begin{eqnarray}
\overline{\Delta w_{\eta}^2} = \frac{1}{2} \left[\bar{n} +
\frac{1}{\eta} \right]\;,  \qquad 
\langle \Delta a^2\rangle _{\eta}= \frac{1}{2\eta}
\label{amp7}\;,  
\end{eqnarray}           
and the noise ratio then writes 
\begin{eqnarray}
\delta a _\eta = \sqrt{\frac{\overline{\Delta w_{\eta}^2}}{\langle
\Delta a^2 \rangle_{\eta}}} =\sqrt{1+\eta\bar{n}}
\label{amp8}\;.
\end{eqnarray}         
\subsection{Phase}
\begin{figure}[h]
\begin{center}
\epsfxsize=.4\hsize\leavevmode\epsffile{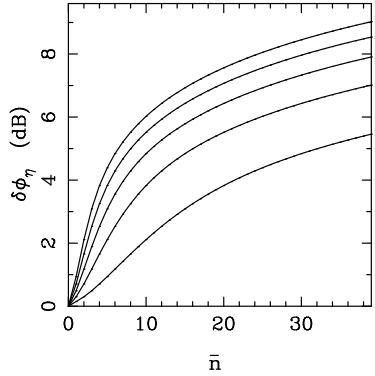}
\end{center}       
\caption{Ratio between tomographic and heterodyne noise in the
measurement of the phase for low excited coherent states, The noise
ratio is reported versus the mean photon number $\bar{n}$ for some
values of the quantum efficiency. From bottom to top we have
$\eta=0.2,0.4,0.6,0.8,1.0$ (From Ref. \cite{added}). }\label{f:pha}
\end{figure}

The canonical description of the quantum optical phase is given by the 
probability operator measure \cite{hol,hel}
\begin{eqnarray}
d\mu(\varphi) = \frac{d\varphi}{2\pi} \sum_{n,m=0}^{\infty}
\exp [i(m-n)\varphi] |n\rangle\langle m|
\label{pha1}\;.
\end{eqnarray}
However, no feasible setup is known that achieves 
the optimal measurement (\ref{pha1}). 
For this reason, here we consider the heterodyne measurement of the
phase, and compare it with the phase of the tomographic estimator for the
corresponding field operator $a$, i.e. $w_{\eta}=\arg(2xe^{i\varphi})$.
Notice that the phase $w_{\eta}$ does not coincide with the local oscillator
phase $\varphi $, because $x$ has varying sign. The 
probability distribution of $w_\eta $ can be obtained by the following identity
\begin{eqnarray}
\int_{0}^{\pi} \frac{d\varphi}{\pi}\int_{-\infty}^{\infty} dx\; 
p_{\eta} (x,\varphi)= 1 = \int_{-\pi}^{\pi} 
\frac{dw_{\eta}}{\pi}\int_{0}^{\infty}  dx\; p_{\eta} (x,w_{\eta})
\label{pha4}\;, 
\end{eqnarray}
which implies
\begin{eqnarray}
p_{\eta} (w_{\eta}) = \frac{1}{\pi} \int_{0}^{\infty} \!\! dx \;
p_{\eta} (x,w_{\eta})
\label{pha5}\;. 
\end{eqnarray}
The precision in the tomographic phase measurement is given by the rms
variance $\overline{\Delta w_{\eta}^2}$ of the probability (\ref{pha5}).
In the case of a coherent state with positive amplitude 
$|\beta\rangle \equiv |
|\beta|\rangle$, Eq. (\ref{pha5}) gives
\begin{eqnarray}
p_{\eta} (w_{\eta}) = \frac{1}{2\pi} \left[1+\hbox{\rm Erf}
\left(\frac{\sqrt{2}|\beta| \cos w_{\eta}}{\sqrt{\eta}}\right)\right]
\label{pha6}\;, 
\end{eqnarray}
which approaches a "boxed" distribution in $[-\pi/2,\pi/2]$ for 
large intensity $|\beta  |\gg 1$. 
We compare the tomographic phase measurement with  heterodyne detection,
namely the phase of the direct-detected complex field $a$.
The outcome probability distribution is the marginal distribution
of the generalized Wigner function $W_{s} (\alpha,\alpha ^*)$
($s=1-\frac 2\eta$) integrated over the radius 
\begin{eqnarray}
p_{\eta}(\varphi )= \int_0^{\infty} \! \rho\; d\rho\; W_{s}(\rho e^{i\varphi},
\rho e^{-i\varphi}) \label{pha3}\;, 
\end{eqnarray}                           
whereas the precision in the phase measurement is given by its rms 
variance $\overline{\Delta \varphi _{\eta}^2}$.
We are not able to give a closed formula for the added noise
$N[\varphi]=\overline{\Delta w_{\eta}^2}-\overline{\Delta\varphi_{\eta}^2}$.
However, for high excited coherent states $|\beta\rangle \equiv |
|\beta|\rangle$ (zero mean phase) one has $\overline{\Delta w_{\eta }^2}
= \pi^2/12$ and $\overline{\Delta \varphi _{\eta}^2}= (2\eta\bar{n})^{-1}$.
The asymptotic noise-ratio is thus given by
\begin{eqnarray}
\delta \varphi _\eta = \sqrt{\frac{\overline{\Delta y _{\eta } ^2}}
{\overline{ \Delta \varphi _{\eta} ^2 }}} = \pi \sqrt{\frac{\eta\bar{n}}{6}} \;,    
\qquad \bar{n} \gg 1
\label{pha7}\;.
\end{eqnarray}         
A comparison for low excited coherent states can be performed
numerically. The noise ratio $\delta\varphi_{\eta}$ (expressed in dB) 
is  shown in Fig. \ref{f:pha} for some values of the quantum
efficiency $\eta$. 
\par\noindent It is apparent that the tomographic determination of the
phase is more noisy than heterodyning also in this low-intensity
regime.  

\par In Table \ref{t:two} a synthesis of the results of this section
is reported.  We have considered the ratio between the tomo\-graphic
and the di\-rect-measu\-rement noise. This is an increasing function
of the mean photon number $\bar{n}$, however scaled by the quantum
efficiency $\eta$. Therefore homodyne tomography turns out to be a
very robust detection scheme for low quantum efficiency.
\begin{table}[hbt]
\begin{center}
\begin{tabular}{|c|c|c|}
\hline \hline 
$O$ & $N[O]$ & $\delta O_\eta $ \\ \hline \hline 
$a^\dag a$ & $ \frac{1}{2}\left[\langle n^2\rangle+ 
\bar{n}\left(\frac{2}{\eta}-1\right)+\frac{1}{\eta^2}\right]$& $
\left(2+\frac{\eta\bar{n}}{2}+\frac{1}{2\eta\bar{n}}\right)^{1/2}$ \\ \hline
$X$ & $\frac{1}{2}\left[\bar{n}+\frac{1}{2\eta}\right]$
&$\left[2\left(1+\eta\bar{n}\right)\right]^{1/2}$\\ \hline
$a$ &$\frac{1}{2}\bar{n}$&$\left(
1+\eta\bar{n}\right)^{1/2}$\\ \hline
$\varphi $&$\frac{\pi}{12}-
\frac{1}{2\eta\bar{n}}$&$\pi\sqrt{\frac{\eta\bar{n}}{6}}$\\ 
\hline \hline
\end{tabular} 
\end{center}
\caption[fake]{Added noise $N[O] $ in tomographic measurement of $O$ 
and noise ratio $\delta O _\eta $for
coherent states. For the phase the results are valid in the asymptotic 
regime ${\bar{n} \gg 1}$ (From Ref. \cite{added}).  }\label{t:two}
\end{table}
\par
\begin{figure}[h]
\begin{center}
\epsfxsize=.4\hsize\leavevmode\epsffile{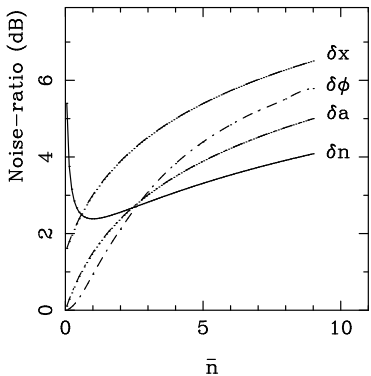}
\end{center}       
\caption{The coherent-state noise ratio (in dB) for all the quantities 
considered in this section (From Ref. \cite{added}). }\label{f:cmp}
\end{figure}

 In Fig. \ref{f:cmp} the coherent-state noise ratio (in dB) for
all the considered quantities are plotted for unit quantum efficiency
versus $\bar{n}$.
\par In conclusion, homodyne tomography adds larger noise for highly excited
states, however, it is not too noisy in the quantum regime of low
$\bar{n}$. It is then very useful in this regime, where currently available 
photodetectors suffer most limitations. Indeed, it has been adopted in
recent experiments of photodetection \cite{raymer95,sch}.

\section{Comparison between homodyne tomography and heterodyning}\label{compare}
We have seen that homodyne tomography allows to measure any field
observable $f\equiv f(a,a^{\dag})$ having {\em normal ordered}
expansion $f\equiv f^{(N)}(a,a^{\dag})=
\sum_{nm=0}^{\infty}f^{(N)}_{nm}a^{\dag}{}^n a^m$ and bounded integral
in Eq. (\ref{generalf}). On the other hand, as shown in
Sec. \ref{hetsec}, heterodyne detection allows to measure field
observables that admit {\em anti-normal ordered} expansion $f\equiv
f^{(A)}(a,a^{\dag})= \sum_{nm=0}^{\infty}f^{(A)}_{nm}a^ma^{\dag}{}^n$,
in which case the expectation value is obtained through the heterodyne
average
\begin{eqnarray}
\langle f\rangle =\int _{\mathbb C}{\frac {d^2\alpha}{\pi}}
f^{(A)}(\alpha,\alpha ^*)
\langle \alpha|\rho|\alpha\rangle \;.\label{QI}
\end{eqnarray}
As shown in Sec. \ref{hetsec}, for $\eta=1$ the heterodyne probability
is just the $Q$-function 
$Q(\alpha,\alpha ^*)={\frac 1\pi}\langle \alpha|\rho|\alpha\rangle $, 
whereas for $\eta <1$ it is Gaussian
convoluted with rms $(1 -\eta)/\eta $, thus giving the Wigner function 
$W_s(\alpha ,\alpha ^*)$, with $s=1-\frac 2\eta$. 

\par Indeed, the problem of measurability of the observable $f$
through heterodyne detection is not trivial, since one needs the
admissibility of anti-normal ordered expansion and the convergence of
the integral in Eq. (\ref{QI}). We refer the reader to
Refs. \cite{baltin,tokio} for more details and to Refs. \cite{qsm,dmr} for 
analysis of quantum state estimates based on heterodyne
detection.

\par The additional noise in homodyning the complex field $a$ has been
evaluated in Eq. (\ref{amp6}), where we found that homodyning is
always more noisy than heterodyning. On the other hand, for other
field observables it may happen that homodyne tomography is less noisy
than heterodyne detection. For example, the added noise in homodyning
the intensity $a^\dag a$ with respect to direct detection has been
evaluated in Eq. (\ref{pho6}). Analogously, one can easily evaluate
the added noise $N_{het}[n]$ when heterodyning the photon number
$n=a^{\dag}a$. According to Eq. (\ref{exhet}),  
the random variable corresponding to the photon number for heterodyne
detection with quantum efficiency $\eta $ is $\nu(\alpha )=|\alpha
|^2-\frac 1\eta $. From the relation 
\begin{eqnarray}
\overline{|\alpha |^4}=\langle a^2 a^{\dag 2} \rangle + 4 \frac
         {1-\eta }{\eta  }\langle a a^{\dag}\rangle +2
\left( \frac{1-\eta }{\eta }\right )^2
\; 
\end{eqnarray}
one obtains 
\begin{eqnarray}
\overline {\Delta\nu (\alpha )}^2=
\langle \Delta n ^2\rangle + \bar n \left(\frac 2 \eta -1 \right )
+\frac {1}{\eta ^2} 
\;. 
\end{eqnarray}
Upon comparing with Eq. (\ref{pho4}), one concludes that the added
noise  in heterodyning the photon number is given by 
\begin{eqnarray}
N_{het}[n]=\overline{\Delta \nu^2(z)} - 
\langle \Delta I^2 _\eta \rangle  =
\frac{1}{\eta ^2}(\eta \bar n - 1)
\;. 
\end{eqnarray}
With respect to the added noise in homodyning of Eq. (\ref{pho6}) one
has
\begin{eqnarray}
N_{het}[n]=N[n] -\frac 1 2\left( \langle n^2 \rangle -\bar n -\frac
{1}{\eta ^2}\right)\;.
\end{eqnarray} 
Since $\langle n^2 \rangle \geq \bar n ^2$, we can conclude that
homodyning the photon number is less noisy than heterodyning it for
sufficiently low mean photon number $\langle n \rangle < \frac 12
\left( 1+\sqrt{1+\frac {4}{\eta ^2}}\right)$.

%% file: cap5.tex
\chapter{Multimode homodyne tomography}
 The generalization of homodyne tomography from a
single-mode to a multimode field is quite obvious, the estimator of
simple operator tensors $ O=O_1\otimes O_2\otimes \ldots \otimes O_n$
being just the product of the estimators of each single-mode
operator $O_1,O_1,\ldots, O_n$. By linearity, one then obtains also
the estimator for arbitrary multimode operators. Such a simple
generalization, however, requires a separate homodyne detector for
each mode, which is unfeasible when the modes of the field are not
spatio-temporally separated. This is the case, for example of pulsed
fields, for which a general multimode tomographic method is especially
needed, also due to the problem of mode matching between the local
oscillator and the detected fields (determined by their relative
spatio-temporal overlap) \cite{matched-lo}, which produces a dramatic
reduction of the overall quantum efficiency.

\par In this Chapter we review the general method of
Ref. \cite{homtom} for homodyning observables of a multimode
electromagnetic field using a {\em single} local oscillator (LO),
providing the rule to evaluate the estimator of an arbitrary
multimode operator. The expectation value of the operator 
can then be obtained by averaging the estimator over the homodyne
outcomes that are collected using a single LO whose mode randomly
scans all possible linear combinations of incident modes.
We will then specifically consider some observables for a two-mode
field in a state corresponding to a twin-beam produced by parametric
downconversion, and prove the reliability of the method on the basis
of computer simulations. 

\par Finally, we report some experimental results \cite{kumarandme}
obtained in the Prem Kumar's lab at Northwestern University. Such
experiment actually represents the first measurement of the joint
photon-number probability distribution of the twin-beam state. 

\section{The general method}
 The Hilbert-Schmidt operator expansion in Eq. (\ref{glauber})  can be
generalized to any number of modes as follows
\begin{eqnarray}
O&=&\int _{\mathbb C}
\frac{d^2z_0}{\pi}\int _{\mathbb C}\frac{d^2z_1}{\pi}\ldots\int
_{\mathbb C}\frac{d^2z_M}{\pi}
\mbox{Tr}\left\{  O \exp\left[\sum_{l=0}^M\left(-z_l
a_l^{\dag}+z^*_l a_l\right)\right]\right\} \nonumber \\&\times & 
\exp\left[\sum_{l=0}^M\left(
z_l a_l^{\dag}-z^*_l a_l\right)\right] \;,\label{tomoidmany}
\end{eqnarray}
where $a_l$ and $a_l^{\dag}$, with $l=0,\ldots,M$ and
$[a_l,a_{l'}^{\dag}]=\delta_{ll'}$, are the annihilation and creation
operators of $M+1$ independent modes, and $O$ now denotes an operator
over all modes. Using the following hyper-spherical parameterization
for $z_l\in\mathbb C$
\begin{eqnarray}
z_0=\frac i2 k\, u_0({\vec\theta})e^{i\psi_0}&\doteq& \frac i2 k\,
e^{i\psi_0}\cos\theta_1\;, \\ z_1=\frac i2 k\,
u_1({\vec\theta})e^{i\psi_1}&\doteq& \frac i2 k\,
e^{i\psi_1}\sin\theta_1\cos\theta_2\;,\nonumber\\ z_2=\frac i2 k\,
u_2({\vec\theta})e^{i\psi_2}&\doteq& \frac i2 k\,
e^{i\psi_2}\sin\theta_1\sin\theta_2\cos\theta_3\;,\nonumber\\
&\ldots&\nonumber\\ z_{M-1}=\frac i2 k\,
u_{M-1}({\vec\theta})e^{i\psi_{M-1}}&\doteq& \frac i2 k\,
e^{i\psi_{M-1}}\sin\theta_1\sin\theta_2\ldots\sin\theta_{M-1}\cos\theta_M\;,
\nonumber\\
z_M=\frac i2 k\, u_M({\vec\theta})e^{i\psi_M}&\doteq& \frac i2 k\,
e^{i\psi_M}\sin\theta_1\sin\theta_2\ldots\sin\theta_{M-1}\sin\theta_M\;,\nonumber
\label{hyper}
\end{eqnarray}
where $k\in[0,\infty)$; $\psi_l\in[0,2\pi]$ for $l=0,1,\ldots ,M$; 
and $\theta_l\in[0,\pi/2]$ for $l=1,2,\ldots,M$, Eq.~(\ref{tomoidmany}) can be
rewritten as follows: 
\begin{eqnarray} 
O=\int d\mu[{\vec\psi}]\int d\mu[{\vec\theta}]\int_0^{+\infty}d k\,
\left(\frac{k}{2}\right)^{2M+1}\frac{1}{M!}
\mbox{Tr}[ O\,e^{-ik  X({\vec\theta},{\vec\psi})}]\,
e^{ik  X({\vec\theta},{\vec\psi})}\;.\label{thpsi1}
\end{eqnarray}
Here we have used the notation
\begin{eqnarray}
&&\int d\mu[{\vec\psi}]\doteq\prod_{l=0}^M\int_0^{2\pi}\frac{d\psi_l}
{2\pi}\;,\\
&&\int d\mu[{\vec\theta}] \doteq 2^M
\,M!\prod_{l=1}^M\int_0^{\pi/2}d\theta_l\,
\sin^{2(M-l)+1}\theta_l\cos\theta_l\;,\\
&&  X({\vec\theta},{\vec\psi})={\frac 1 2}\left[ 
A^{\dag}({\vec\theta},{\vec\psi})+ 
A({\vec\theta},{\vec\psi})\right]\;,\label{X}\\  
&&A({\vec\theta},{\vec\psi})= \sum_{l=0}^M
e^{-i\psi_l}u_l({\vec\theta})a_l\;.
\end{eqnarray}
From the parameterization in Eq.~(\ref{hyper}), one
has $\sum_{l=0}^M u^2_l({\vec\theta})=1$, and hence $[
A({\vec\theta},{\vec\psi}), A^{\dag}({\vec\theta},{\vec\psi})]=1$,
namely $A({\vec\theta},{\vec\psi})$ and $
A^{\dag}({\vec\theta},{\vec\psi})$ themselves are annihilation and
creation operators of a bosonic mode. By scanning all values of
$\theta_l\in[0,\pi/2]$ and $\psi_l\in[0,2\pi]$, all possible linear
combinations of modes $a_l$ are obtained.  \par For the quadrature
operator $ X({\vec\theta},{\vec\psi})$ in Eq.~(\ref{X}), one has the
following identity for the moments generating function
\begin{eqnarray}
\langle e^{ik  X({\vec\theta},{\vec\psi})
}\rangle=\exp\left(\frac{1-\eta}{8\eta}k^2\right)
\int_{-\infty}^{+\infty}d x\, e^{ikx}\,p_{\eta}(x;{\vec\theta},{\vec\psi})
\;,\label{momM}
\end{eqnarray}
where $p_{\eta}(x;{\vec\theta},{\vec\psi})$ denotes the homodyne
probability distribution of the quadrature $
X({\vec\theta},{\vec\psi})$ with quantum efficiency $\eta$.
Generally, $\eta $ can depend on the mode itself, i.e., it is a
function $\eta=\eta({\vec\theta},{\vec\psi})$ of the selected mode. In
the following, for simplicity, we assume $\eta$ to be mode
independent, however. By taking the ensemble average on each side of
Eq.~(\ref{thpsi1}) and using Eq.~(\ref{momM}) one has
\begin{eqnarray}
\langle  O\rangle=\int d\mu[{\vec\psi}]\int d\mu[{\vec\theta}]\,
\int _{-\infty}^{+\infty}d x \,p_\eta (x;{\vec\theta},{\vec\psi})
\,{\cal R}_{\eta}[  O](x;{\vec\theta},{\vec\psi})\;,\label{K}
\end{eqnarray}
where the estimator ${\cal R} _{\eta}[
O](x;{\vec\theta},{\vec\psi})$ has the following expression
\begin{eqnarray}
{\cal R}_{\eta}[ O](x;{\vec\theta},{\vec\psi})=\frac{\kappa^{M+1}}{M!}
\int_0^{+\infty} d t\, e^{-(1-\frac \kappa 2)t +2i\sqrt{\kappa
t}\,x}\, t^M\, \mbox{Tr}[ O\,e^{-2i\sqrt{\kappa t}
X({\vec\theta},{\vec\psi})}],\label{MAIN}
\end{eqnarray}
with $\kappa=2\eta /(2\eta-1)$. Eqs. (\ref{K}) and (\ref{MAIN}) allow
to obtain the expectation value $\langle O\rangle$ for any unknown
 state of the radiation field by averaging over the homodyne outcomes
of the quadrature $ X({\vec\theta},{\vec\psi})$ for
${\vec\theta}$ and ${\vec\psi}$ randomly distributed according to
$d\mu[{\vec\psi}]$ and $d\mu[{\vec\theta}]$. Such outcomes can be
obtained by using a single LO that is prepared in the multimode
coherent state $\otimes _{l=0}^M |\gamma _l \rangle $ with $\gamma _l
=e^{i\psi _l}u_l(\theta) K/2$ and $K \gg 1$.  In fact, in this case
the rescaled zero-frequency photocurrent at the output of a balanced
homodyne detector is given by
\begin{eqnarray}
I =\frac 1K \sum_{l=0}^M(\gamma ^*_l a_l+\gamma _l a_l^\dag )
\;,\label{dcc}
\end{eqnarray}
which corresponds to the operator $ X({\vec\theta},{\vec\psi})$. In
the limit of a strong LO ($K\rightarrow \infty$), all moments of the
current $I$ correspond to the moments of 
$X({\vec\theta},{\vec\psi})$, and the exact measurement of $
X({\vec\theta},{\vec\psi})$ is then realized.  Notice that for modes
$a_l$ with different frequencies, in the d.c.  photocurrent in
Eq. (\ref{dcc}) each LO with amplitude $\gamma _l$ selects the mode
$a_l$ at the same frequency (and polarization).  For less-than-unity
quantum efficiency, Eq.~(\ref{momM}) holds.  

\par Equation~(\ref{MAIN}) can be applied to some observables of
interest. In particular, one can estimate the matrix element
$\langle\{n_l\}| R|\{m_l\}\rangle$ of the multimode density operator
$R$. This will be obtained by averaging the estimator
\begin{eqnarray}
&&{\cal R}_{\eta}[|\{m_l\}\rangle\langle\{n_l\}|](x;{\vec\theta},{\vec\psi})=
e^{-i\sum_{l=0}^M(n_l-m_l)\psi_l}\,
\frac{\kappa^{M+1}}{M!} \nonumber \\& &\times  
\prod_{l=0}^M\left\{[-i\sqrt{\kappa} u_l({\vec\theta})]^{\mu_l-\nu_l}
\sqrt{\frac{\nu_l!}{\mu_l!}}\right\}\nonumber\\ &&\times 
\int_0^{+\infty}d t\,e^{-t+2i\sqrt{\kappa t}\,x}\,
t^{M+\sum_{l=0}^M(\mu_l-\nu_l)/2}\prod_{l=0}^ML_{\nu_l}^{\mu_l-\nu_l}
[\kappa u_l^2({\vec\theta})t]\;,\label{gasp}
\end{eqnarray}
where $\mu_l=\mbox{max}(m_l,n_l)$, $\nu_l=\mbox{min}(m_l,n_l)$, and
$L_n^{\alpha}(z)$ denotes the generalized Laguerre
polynomial. For diagonal matrix elements, Eq.~(\ref{gasp}) simplifies
to
\begin{eqnarray}
{\cal R}_{\eta}[|\{n_l\}\rangle\langle\{n_l\}|](x;{\vec\theta},{\vec\psi})=
\frac{\kappa^{M+1}}{M!}
\int_0^{+\infty}d t\,e^{-t+2i\sqrt{\kappa t}\,x}\,
t^M\prod_{l=0}^ML_{n_l}[\kappa u_l^2({\vec\theta})t]\;\label{minigasp}
\end{eqnarray}
with $L_n(z)$ denoting the customary Laguerre polynomial in $z$.
Using the following identity \cite{gradshtein}
\begin{eqnarray}
&&L_n^{\alpha_0+\alpha_1+\ldots+\alpha_M+M}(x_0+x_1+\ldots+x_M)
  \nonumber \\& & =
\sum_{i_0+i_1+\ldots+i_M=n}
L_{i_0}^{\alpha_0}(x_0)L_{i_1}^{\alpha_1}(x_1)\ldots
L_{i_M}^{\alpha_M}(x_M)\;,
\end{eqnarray}
from Eq.~(\ref{minigasp}) one can easily derive the estimator of the
probability distribution of the total number of photons $ 
N=\sum_{l=0}^Ma^{\dag}_la_l$ 
\begin{eqnarray}
{\cal R}_{\eta}[|n\rangle\langle n|](x;{\vec\theta},{\vec\psi})&=&
\frac{\kappa^{M+1}}{M!}
\int_0^{+\infty}d t\,e^{-t+2i\sqrt{\kappa t}\,x}\,
t^M L^{M}_n[\kappa t]\;,\label{N}
\end{eqnarray}
where $|n\rangle$ denotes the eigenvector of $ N$ with eigenvalue $n$.
Notice that the estimator in Eq.~(\ref{minigasp}) does not depend on
the phases $\psi_l$; only the knowledge of the angles $\theta_l$ is
needed. For the estimator in Eq.~(\ref{N}), even the angles $\theta_l$
can be unknown.  \par Now we specialize to the case of only two modes
$a$ and $b$ (i.e., M=1 and $\vec \theta$ is a scalar $\theta $).  The
joint photon-number probability distribution is obtained by averaging
\begin{eqnarray}
&&{\cal R}_{\eta}[|n,m\rangle\langle n,m|](x;\theta,\psi_0,\psi_1) =
\nonumber
\\&&
\kappa^2\int_0^{+\infty}d t\,e^{-t+2i\sqrt{\kappa t}\,x}\,
t \,L_n(\kappa t\cos^2\theta)L_m(\kappa t\sin^2\theta)\;.\label{twogasp}
\end{eqnarray}
The estimator (\ref{N}) of the probability distribution of the total
number of photons can be written as
\begin{eqnarray}
{\cal R}_{\eta}[|n\rangle\langle n|](x;\theta,\psi_0,\psi_1)=
\kappa^2\int_0^{+\infty}d t\,e^{-t+2i\sqrt{\kappa t}\,x}\,
t\,L^1_n[\kappa t]\;.\label{N1}
\end{eqnarray}
For the total number of photons one can also derive the
estimator of the moment generating function, using the generating
function for the Laguerre polynomials \cite{gradshtein}. One obtains
\begin{eqnarray}
{\cal R}_{\eta}[z^{a^{\dag}a+b^{\dag}b}](x;\theta,\psi_0,\psi_1)= 
\frac{1}{(z+\frac{1-z}{\kappa})^2}\Phi\left(2,{\frac 1 2};-
\frac{1-z}{z+\frac{1-z}{\kappa}}\,x^2\right)\;.\label{gen}
\end{eqnarray}
For the first two moments one obtains the simple expressions
\begin{eqnarray}
&&{\cal R}_{\eta}[a^{\dag}a+b^{\dag}b](x;\theta,\psi_0,\psi_1)= 
4x^2+{\frac{2}{\kappa}}-2\;,\\
&&{\cal R}_{\eta}[(a^{\dag}a+b^{\dag}b)^2](x;\theta,\psi_0,\psi_1)= 
8x^4+\left({\frac{24}{\gamma}}-20
\right)x^2+{\frac{6}{\gamma^2}}-{\frac {10}{\gamma}}+4 \nonumber \;.
\end{eqnarray}
It is worth noting that analogous estimators of the photon-number
difference between the two modes are singular and one needs a cutoff
procedure, similar to the one used in Ref.~\cite{our} for recovering
the correlation between the modes by means of the customary two-mode
tomography.  In fact, in order to extract information pertaining to a
single mode only one needs a delta-function at $\theta=0$ for mode
$a$, or $\theta=\pi/2$ for mode $b$, and, in this case, one could
better use the standard one-mode tomography by setting the LO to the
proper mode of interest.  \par Finally, we note that for two-mode
tomography the estimators can be averaged by the integral
\begin{eqnarray}
\langle  O\rangle &=&\int_0^{2\pi}\frac{d\psi_0}{2\pi}
\int_0^{2\pi}\frac{d\psi_1}{2\pi}\int_{-1}^1\frac{d(\cos 2\theta)}{2}\, 
\int_{-\infty }^{+\infty }d x\, p_{\eta }(x;\theta,\psi_0,\psi_1)\,
\nonumber \\&\times  & 
{\cal R}_{\eta}[  O](x;\theta,\psi_0,\psi_1)\;\label{K1}
\end{eqnarray}
over the random parameters $\cos(2\theta),\psi_0$, and $\psi_1$. For
example, in the case of two radiation modes having the same frequency
but orthogonal polarizations, $\theta $ represents a random rotation
of the polarizations, whereas $\psi_0$ and $\psi_1$ denote the
relative phases between the LO and the two modes, respectively.
\subsection{Numerical results for two-mode fields}
In this section we report some Monte-Carlo simulations from
Ref. \cite{homtom} to judge the experimental working
conditions for performing the single-LO tomography on two-mode
fields. We focus our attention on the twin-beam state, usually
generated by spontaneous parametric downconversion, namely 
\begin{eqnarray}
|\Psi\rangle=S(\chi )|0 \rangle _a|0 \rangle_b =
\sqrt{1-|\xi|^2}\sum_{n=0}^{\infty}\xi^n\,
|n\rangle_a|n\rangle_b\;,\label{Psi}
\end{eqnarray}
where $ S(\chi )=\exp(\chi a^\dag b^\dag -\chi ^*ab)$ and $\xi =
e^{i\arg \chi}\,\hbox{tanh}|\chi|$.  The parameter $\xi $ is related
to the average number of photons per beam $\bar n =|\xi |^2/(1-|\xi
|^2)$. For the simulations we need to derive the homodyne probability
distribution $p(x;\theta,\psi_0,\psi_1)$ which is given by
\begin{eqnarray}
p(x;\theta,\psi_0,\psi_1)&=&
\mbox{Tr}[ U^\dag \,|x\rangle_a {}_a\langle x|\otimes 1_b\,  U\,
|\Psi \rangle  \langle\Psi| ]   \label{a1} \\&=&
{}_a\langle 0|{}_b\langle 0|  \,S^{\dag }(\chi )
  \,U^\dag \,[|x \rangle {}_a {}_a\langle x|\otimes 1_b ]
  \,U \,  S(\chi)\,
|0\rangle_a|0\rangle_b   \nonumber\;,
\end{eqnarray}
where $|x\rangle_a$ is the eigenvector of the quadrature $
x={\frac 12}(a^{\dag}+a)$ with eigenvalue $x$ and $U$ is the unitary
operator achieving the mode transformation
\begin{eqnarray}
  U^{\dag}{a\choose b}  U= \left(
\begin{array}{ll}
e^{-i\psi_0}\cos\theta &e^{-i\psi_1}\sin\theta \\ 
-e^{i\psi_1}\sin\theta &e^{i\psi_0}\cos\theta 
\end{array}
\right){a\choose b}\;.\label{M}
\end{eqnarray}
In the case of two radiation modes having the same frequency but
orthogonal polarizations---the case of Type II phase-matched
parametric amplifier---Eq.~(\ref{a1}) gives the theoretical
probability of outcome $x$ for the homodyne measurement at a
polarization angle $\theta $ with respect to the polarization of the
$a$ mode, and with $\psi_0$ and $\psi_1$ denoting the relative phases
between the LO and the two modes, respectively.  By using the
Dirac-$\delta$ representation of the $ X$-quadrature projector
\begin{eqnarray}
|x \rangle \langle x|=\int_{-\infty}^{+\infty} 
\frac{d \lambda}{2\pi}\,\exp[i
\lambda (  X -x)] \;, 
\end{eqnarray}
Eq.~(\ref{a1}) can be rewritten as follows \cite{homtom}
\begin{eqnarray}
&&p(x;\theta,\psi_0,\psi_1)= \int_{-\infty}^{+\infty}\frac{d
\lambda}{2\pi}\, {}_a\langle 0|{}_b\langle 0| \,S^{\dag }(\chi)\,  
U^\dag \,e^{i \lambda (  X_a-x)}   \,U \,  S(\chi)\,
|0\rangle_a|0\rangle_b \nonumber \\& & =\int_{-\infty}^{+\infty}\frac{d
\lambda}{2\pi}\,e^{-i \lambda x}
{}_a\langle
0|{}_b\langle 0| \exp\left\{i \frac \lambda 2 \left[(e^{-i\psi_0}\mu
\cos\theta +e^{i\psi_1}\nu ^* \sin\theta )a  \right.\right.
\nonumber \\& & 
+ \left. \left.(e^{i\psi_0}\nu ^*\cos\theta +
e^{-i\psi_1}\mu\sin\theta )b + \hbox{H.c.}\right]\frac{}{}\!\!\!
\right \}
|0\rangle_a|0\rangle_b \;,\label{a2} 
\end{eqnarray}
where we have used Eq.~(\ref{M}) and the transformation
\begin{eqnarray}
  S^{\dag}(\chi ){a\choose \,b^\dag } S(\chi )= \left(
\begin{array}{ll}
\mu &\nu
\\ \nu ^* &\mu 
\end{array}
\right){a\choose \,b^\dag }\;\label{M2}
\end{eqnarray}
with $\mu=\hbox{cosh}|\chi |$ and $\nu =e^{i\arg\chi }\,\hbox{sinh}|\chi |$. 
Upon defining 
\begin{eqnarray}
&&K C=e^{-i\psi_0}\mu\cos\theta
+e^{i\psi_1}\nu ^*\sin\theta \;,\nonumber \\
&&KD=e^{i\psi_0}\nu ^*\cos\theta +
e^{-i\psi_1}\mu\sin\theta 
\;, 
\end{eqnarray}
where $K\in \mathbb R$ and $C,D\in\mathbb C$, with $|C|^2+|D|^2=1$ one
has 
\begin{eqnarray}
K^2=\mu ^2+|\nu |^2 +2\mu|\nu |\sin 2\theta\cos(\psi_0+\psi_1-\arg
\nu)
\;. 
\end{eqnarray}
Now, since the unitary transformation 
\begin{eqnarray}
\left(\begin{array}{ll}
C &D \\  D^* & C^* \end{array}
\right){a\choose b}\longrightarrow 
{a\choose b}\;
\end{eqnarray}
has no effect on the vacuum state, 
Eq.~(\ref{a2}) leads to the following Gaussian distribution
\begin{eqnarray}
&&p(x;\theta,\psi_0,\psi_1) \nonumber \\& & = 
\int_{-\infty}^{+\infty}\frac{d
\lambda}{2\pi}\,e^{-i \lambda x} {}_a\langle
0|{}_b\langle 0| \exp\left\{i K\frac \lambda 2 
\left[(C\, a+ D\,b)+ \hbox{H.c.}\right]\right\}
|0\rangle_a|0\rangle_b \nonumber \\&&=
\int_{-\infty}^{+\infty}\frac{d
\lambda}{2\pi}\,e^{-i \lambda x} {}_a\langle
0|\exp\left[i K\frac \lambda 2 \left(a+a^\dag\right)\right]
|0\rangle_a =\frac 1K \left|{}_a\langle 0 |x/K \rangle _{a}\right |^2 
\nonumber \\&&=
{\frac {1} {\sqrt{2\pi\Delta^2(\theta,\psi_0,\psi_1)}}}
\exp\left(-\frac{x^2}{2\Delta^2(\theta,\psi_0,\psi_1)}\right)
\;,\label{quasifinal}
\end{eqnarray}
where the variance  $\Delta^2(\theta,\psi_0,\psi_1)$ is given by
\begin{eqnarray}
\Delta^2(\theta,\psi_0,\psi_1)=\frac {K^2}{4}=
\frac{1+|\xi|^2+2|\xi|
\sin 2\theta\cos(\psi_0+\psi_1-\arg\xi)}{4(1-|\xi|^2)}\;.
\end{eqnarray}
Taking into account the Gaussian convolution that results from
less-than-unity quantum efficiency, the variance just increases as
\begin{eqnarray}
\Delta^2(\theta,\psi_0,\psi_1)\to\Delta_{\eta}^2(\theta,\psi_0,\psi_1)=
\Delta^2(\theta,\psi_0,\psi_1)+\frac{1-\eta}{4\eta}
\;.
\end{eqnarray}
Notice that the
probability distribution in Eq.~(\ref{quasifinal}) corresponds to a
squeezed vacuum for $\theta={\frac \pi 4}$ and $\psi_0+\psi_1-\arg\xi=0$
or $\pi$.
\begin{figure}[hbt]\begin{center}
\epsfxsize=.45\hsize\leavevmode\epsffile{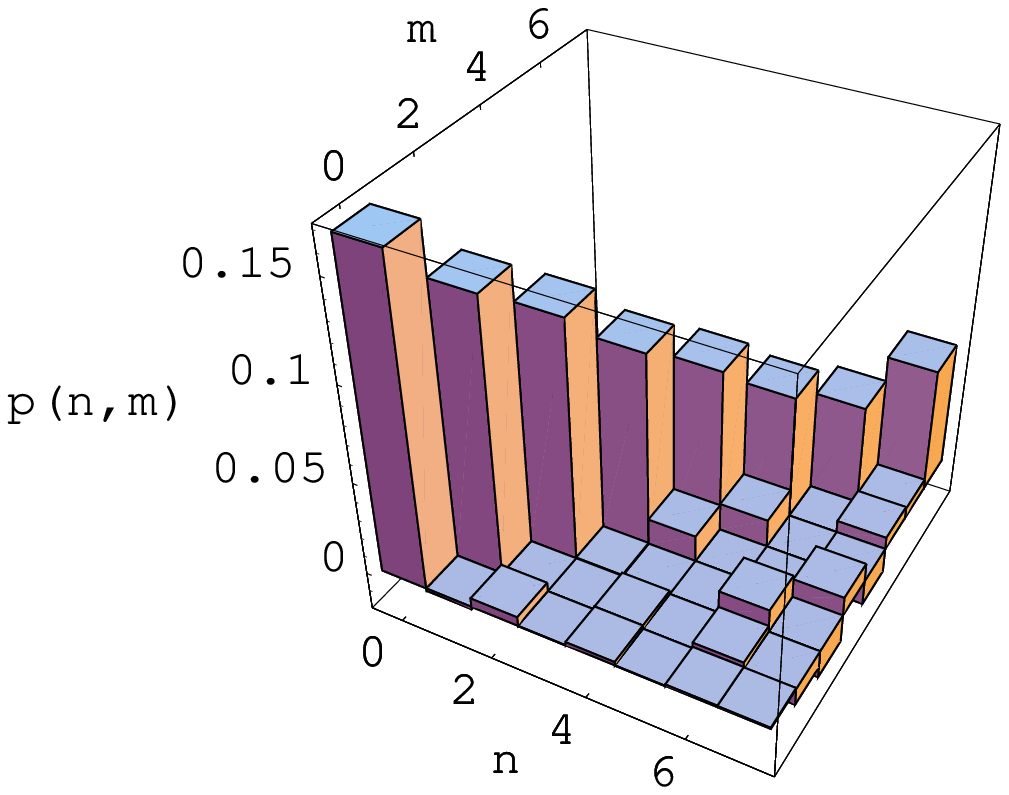}\hspace{12pt}
\epsfxsize=.45\hsize\leavevmode\epsffile{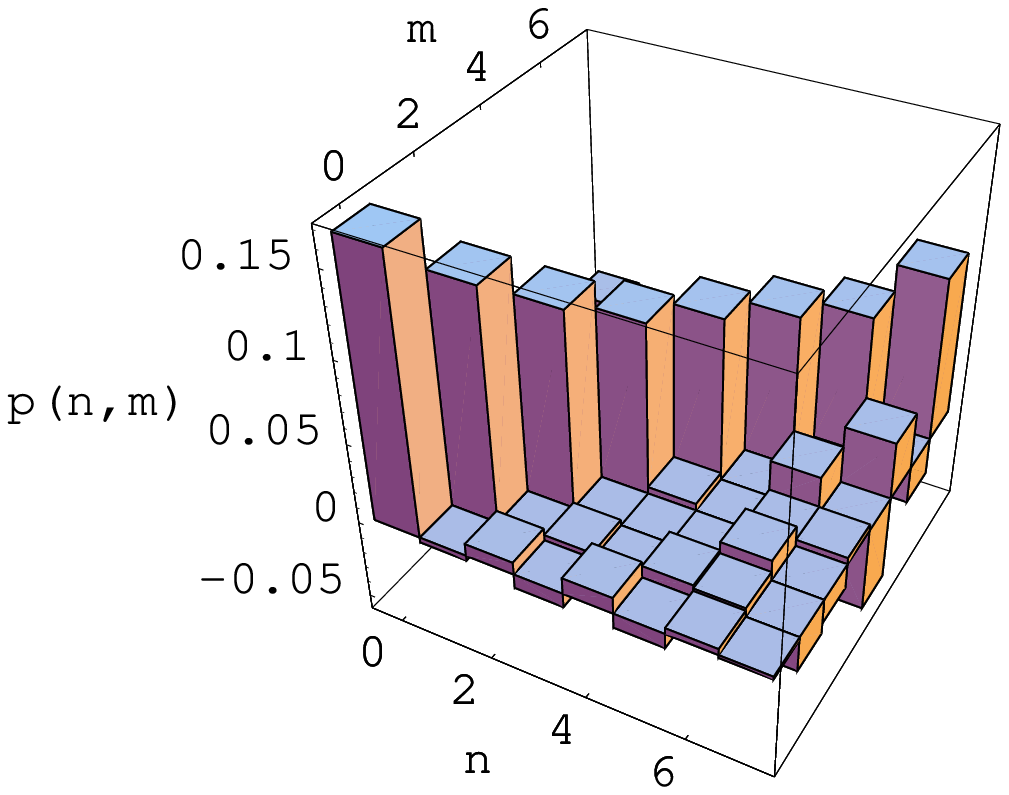}\end{center}
\caption{Two-mode photon-number probability $p(n,m)$ of the twin-beam
state in Eq.~(\ref{Psi}) for average number
of photons per beam $\overline{n}=5$ obtained by a
Monte-Carlo simulation with the estimator in Eq.~(\ref{twogasp}) and
random parameters $\cos 2\theta$, $\psi_0$, and $\psi_1$. On the
left: quantum efficiency $\eta=1$ and $10^6$ data samples were
used in the reconstruction.  On the right: 
$\eta=0.9$, and $5\times 10^6$ data samples (From
Ref. \cite{homtom}). }
\label{f:matrix}\end{figure}
\par We study the tomographic measurement of the joint photon-number
probability distribution and the probability distribution for the
total number of photons with use of the estimators in Eqs.~(\ref{twogasp})
and (\ref{N1}), respectively. Moreover, using the estimator in
Eq.~(\ref{gasp}) we reconstruct the matrix elements  
\begin{eqnarray}
C_{n,m}\equiv {}_a\langle m|{}_b \langle m|\Psi \rangle \langle 
\Psi |n \rangle _a |n \rangle _b,\label{cnm}
\end{eqnarray}
which reveal the coherence of the twin-beam state. Theoretically one should have 
\begin{eqnarray}
C_{n,m}=(1-|\xi |^2)\xi^m\,\xi^{*n}
\;.\label{cnmxi}
\end{eqnarray} 
The estimators have been numerically
evaluated by applying the Gauss method for calculating the integral in
Eq.~(\ref{gasp}), which results in a fast and sufficiently precise
algorithm with use of just 150 evaluation points.
\begin{figure}[hbt]
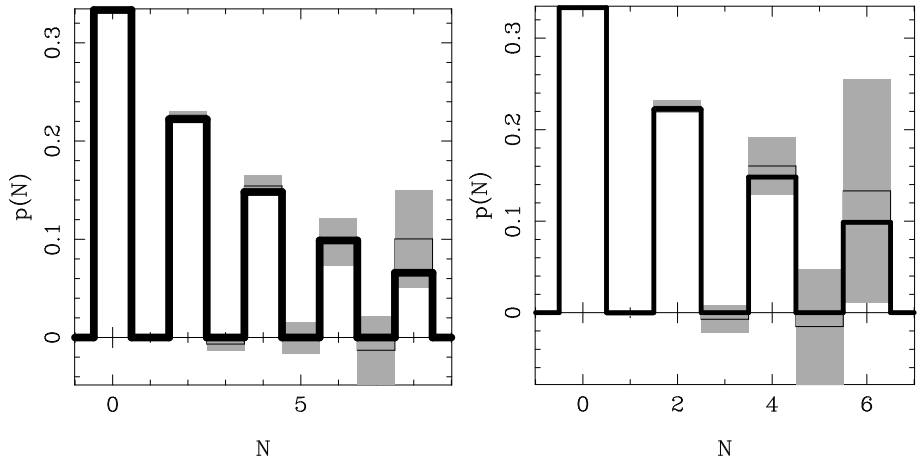

\begin{tabular}{cc}
\psfig{file=cap5fig3.ps,width=6cm}
\psfig{file=cap5fig4.ps,width=6cm}
\end{tabular}
\caption{Probability distribution for the total number of photons of
the twin beams in Eq.~(\ref{Psi}) for average number of photons per
beam $\overline{n}=2$ obtained using the estimator in Eq.~(\ref{N1}).
The oscillation of the total photon-number probability due to the
perfect correlation of the twin beams has been reconstructed by
simulating $10^7$ data samples with quantum efficiency $\eta=0.9$ (on
the left), and $2\times 10^7$ data samples $\eta=0.8$ (on the
right). The theoretical probability (thick solid line) is superimposed
onto the result of the Monte-Carlo experiment; the latter is shown by
the thin solid line. Notice the dramatic increase of errors (in
gray shade) versus N and for smaller $\eta$ (From
Ref. \cite{homtom}). }
\label{f:sum2}  
\end{figure}
\par In Fig. \ref{f:matrix} a Monte-Carlo simulation of the joint
photon-number probability distribution is  reported. The simulated
values compare very well with the theoretical ones. In
Ref. \cite{homtom} a careful analysis of the statistical errors has
been done for various twin-beam states by constructing histograms of
deviations of the results from 
different simulated experiments from the theoretical ones. In
comparison to the customary two-LO tomography of Ref. \cite{our}, where for
$\eta=1$ the statistical errors saturate for increasingly large $n$
and $m$, here we have statistical errors that are slowly increasing
versus $n$ and $m$. This is due to the fact that the range of the
estimators in Eq.~(\ref{twogasp}) increases versus $n$ and
$m$. Overall we find that for any given quantum efficiency the
statistical errors are generally slightly larger than those obtained
with the two-LO method. The convenience of using a single LO then
comes with its own price tag.
\par By using the estimator in Eq.~(\ref{N1})  the
probability distribution for the total number of photons $N$ of the
twin beams has been also constructed (Fig. \ref{f:sum2}). Notice the
dramatic increase of error bars versus N and for smaller $\eta$.
\begin{figure}[hbt]\begin{center}
\epsfxsize=.45\hsize\leavevmode\epsffile{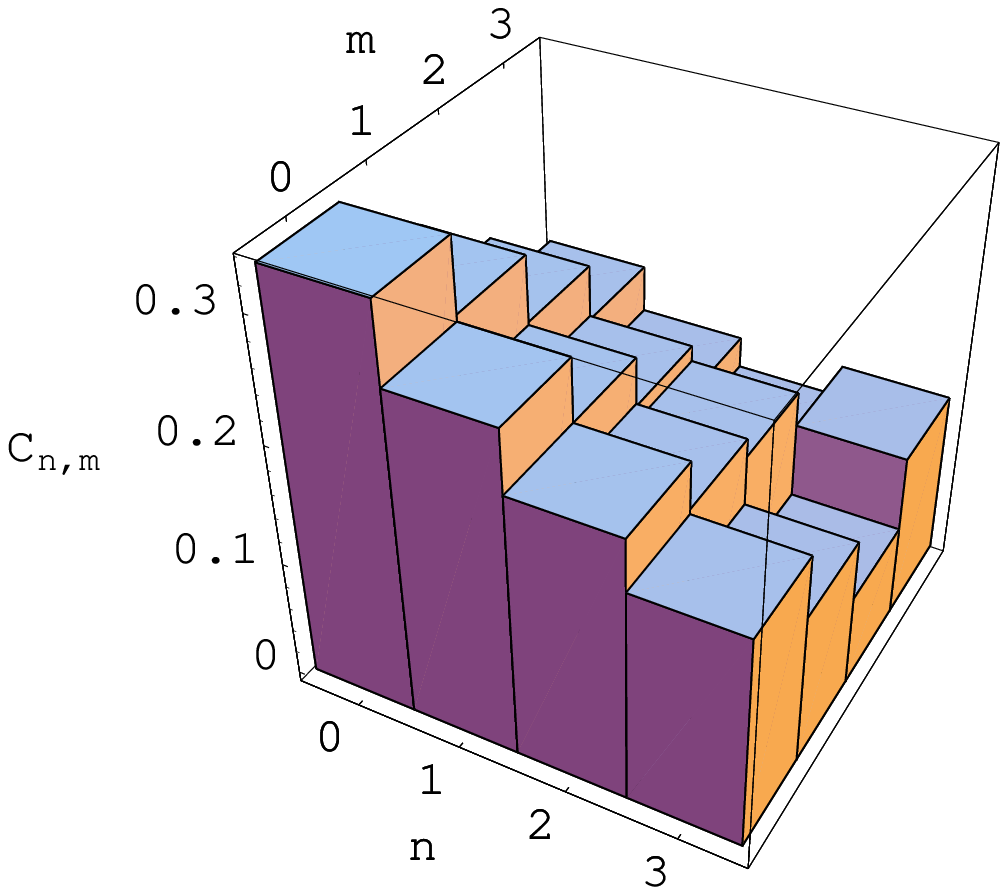}
\hspace{12pt}
\epsfxsize=.45\hsize\leavevmode\epsffile{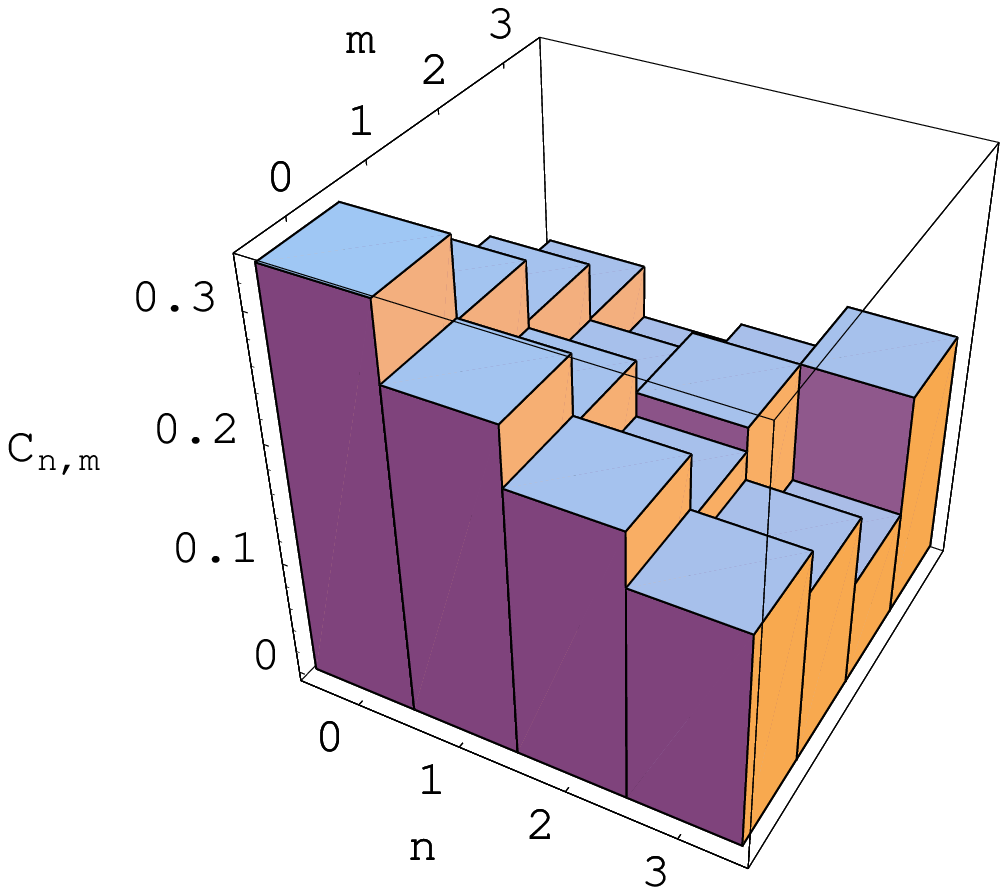}
\end{center}
\caption{Tomographic reconstruction of the matrix elements
$C_{n,m}\equiv {}_a\langle m|{}_b \langle m|\Psi \rangle \langle \Psi
|n \rangle _a |n \rangle _b $ of the twin beams in Eq.~(\ref{Psi}) for
average number of photons per beam $\overline{n}=2$, obtained using
the estimator in Eq.~(\ref{gasp}). On the left we used $10^6$
simulated data samples and quantum efficiency $\eta=0.9$; on the right
$3\times 10^6$ data samples and $\eta=0.8$. The coherence of the
twin-beam state is easily recognized as $C_{n,m}$ varies little for
$n+m=\mbox{constant}$ [$\xi $ in Eq.~(\ref{Psi}) has been chosen
real]. For a typical comparison between theoretical and experimental
matrix elements and their relative statistical errors, see results in
Fig. \ref{f:sum2} (From Ref. \cite{homtom}).}
\label{f:matrix2}\end{figure}
\par Finally, in Fig. \ref{f:matrix2} we report the results of the
tomographic measurement of $C_{n,m}$ defined in
Eq.~(\ref{cnm}). Because the reconstructed $C_{n,m}$ is close to the
theoretically expected value in Eq.~(\ref{cnmxi}), these reveal the
purity of the twin beams, which cannot be inferred from the thermal
diagonal distribution of Fig. \ref{f:matrix}.

\par
\begin{figure}[hbt]\begin{center}
\epsfxsize=.85\hsize\leavevmode\epsffile{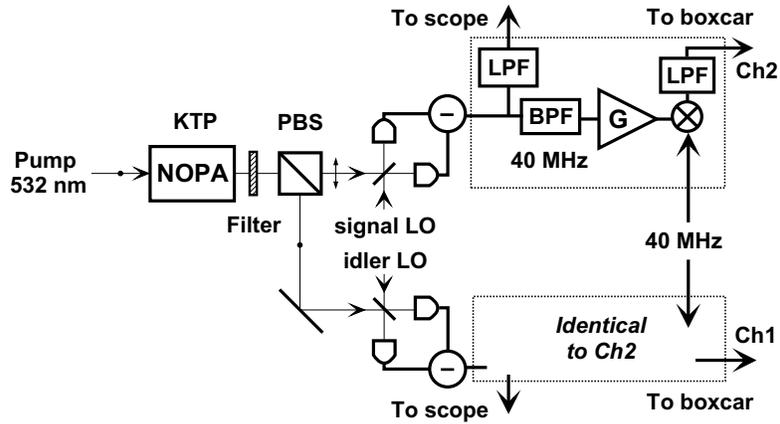}
\end{center}
\caption{A schematic of the experimental setup. NOPA:
non-degenerate optical parametric amplifier; LOs: local oscillators;
PBS: polarizing beam splitter; LPFs: low-pass filters; BPF: band-pass
filter; G: electronic amplifier. Electronics in the two channels are
identical (From Ref. \cite{kumarandme}). \label{f:exper1}}\end{figure}

\begin{figure}[hbt]\begin{center}
\begin{tabular}{cc}
\psfig{file=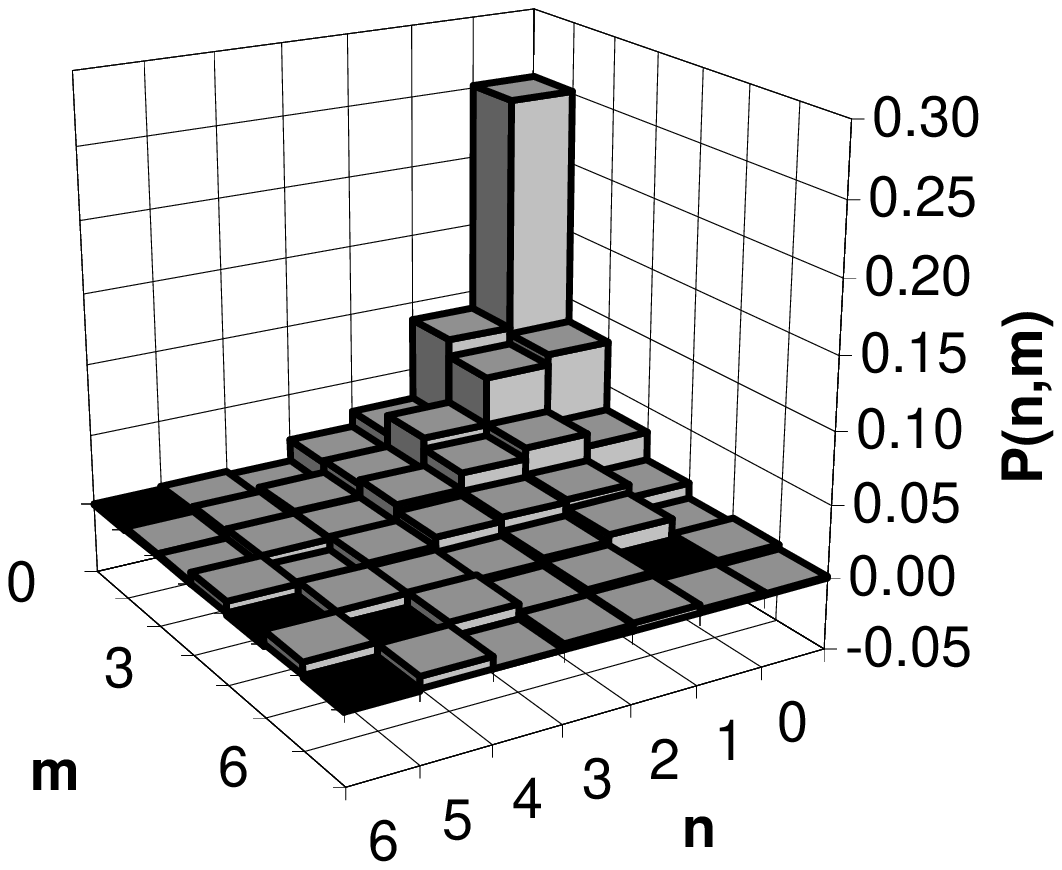,width=6cm}
\psfig{file=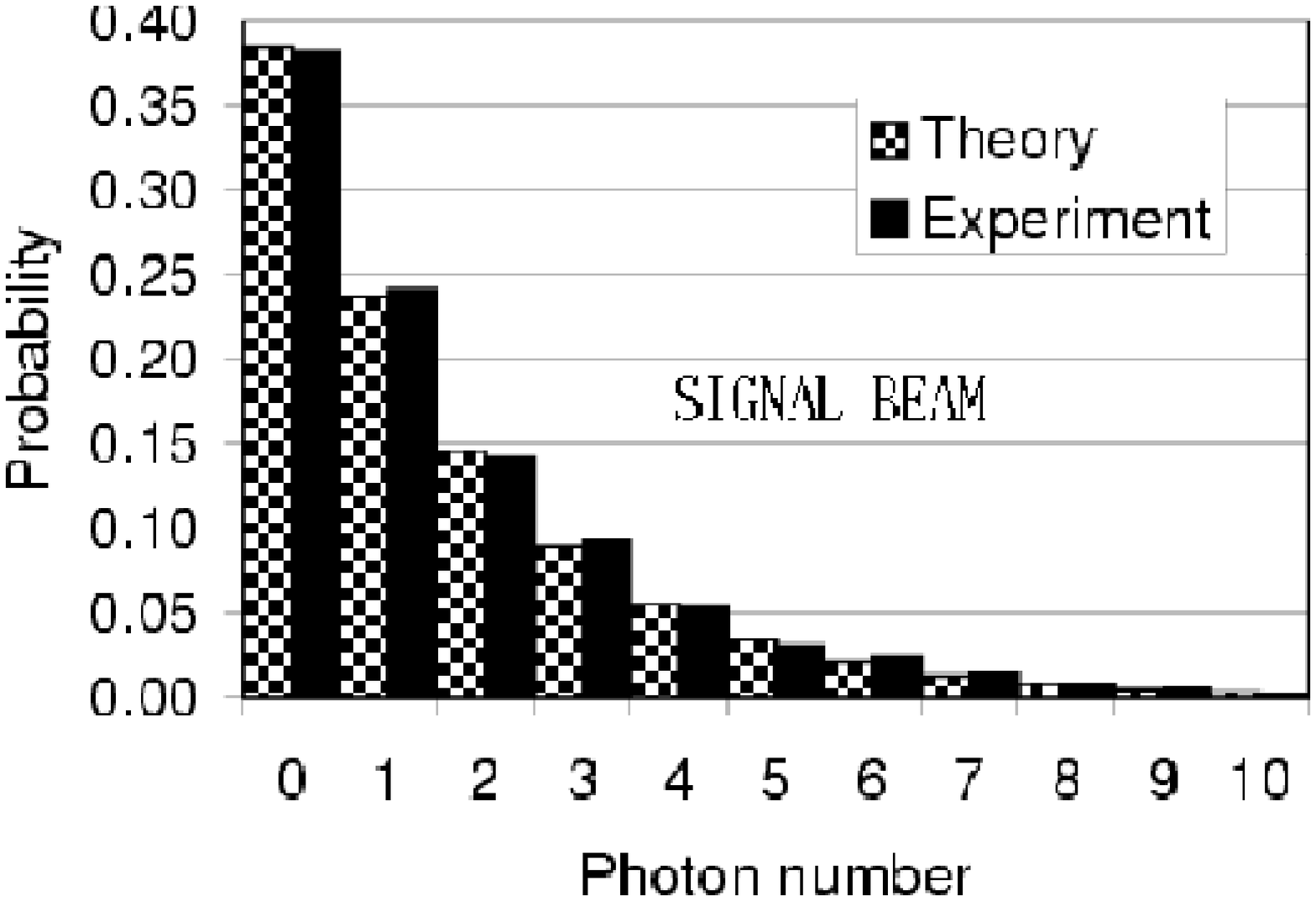,width=6cm}
\end{tabular}
\caption[fake]{Left: Measured joint photon-number probability
distribution for the twin-beam state with average number of photons
per beam $\bar n=1.5$ and $400000$ samples. 
Right: marginal distribution for the signal beam for the same
data. The theoretical distribution is also shown. Very similar results
are obtained for the idler beam (From Ref. \cite{kumarandme}).
\label{f:exper2}}\end{center}
\end{figure}

 The first experimental results of a measurement of the joint
photon-number probability distribution for a two-mode quantum state
created by a nondegenerate optical parametric amplifier has been
presented in Ref. \cite{kumarandme}. 
In this experiment, however, the twin beams are detected separately by
two balanced-homodyne detectors. A schematic of the experimental setup
is reported in Fig. \ref{f:exper1}, and some experimental results are
reported in Fig. \ref{f:exper2}. As expected for parametric
fluorescence, the experiment has shown a measured joint photon-number
probability distribution that exhibited up to 1.9 dB of quantum
correlation between the two modes, with thermal marginal
distributions.

%% file: cap6.tex
\chapter{Applications to quantum measurements}

In this chapter we review a number of applications of quantum
tomography related to some fundamental tests in quantum mechanics. 

\par First, we report the proposal of Ref. \cite{noncl} for testing
the nonclassicality of quantum states by means of an operational
criterion based on a set of quantities that can be measured
experimentally with some given level of confidence, even in the
presence of loss, noise, and less-than-unity quantum efficiency.

\par Second, we report the experiment proposed in Ref. \cite{sr} for
testing quantum state reduction.   The state-reduction rule is tested
using optical homodyne tomography by directly measuring the fidelity
between the theoretically-expected reduced state and the experimental
state. 

\par Finally, we review some experimental results obtained at the
Quantum Optics Lab of the University of Naples \cite{TomoNa} about the
reconstruction of coherent signals, together with application to the
estimation of the losses introduced by simple optical components.

\section{Measuring the nonclassicality of a quantum state} 
The concept of nonclassical states of light has drawn much attention
in quantum optics
\cite{lee1,mand,mand2,hill,spe,tomb,agar,kly1,iss,jpa}.  The
customary definition of nonclassicality is given in terms of the $P$-function
presented in Sec. \ref{wigsec}: a nonclassical state does not admit a
regular positive $P$-function representation, namely, it cannot be
written as a statistical mixture of coherent states. Such states
produce effects that have no classical analogue.  These kinds of
states are of fundamental relevance not only for the demonstration of
the inadequacy of classical description, but also for applications,
e.g., in the realms of information transmission and interferometric
measurements \cite{spe,tomb,iss}.

\par We are interested in testing the nonclassicality of a quantum
state by means of a set of quantities that can be measured
experimentally with some given level of confidence, even in the
presence of loss, noise, and less-than-unity quantum efficiency. The
positivity of the $P$-function itself cannot be adopted as a test,
since there is no  viable method to measure it. As proved in
Sec. \ref{s:ob} only the generalized Wigner functions of order
$s<1-\eta ^{-1}$ can be measured, $\eta $ being the quantum efficiency
of homodyne detection. Hence, through this technique, all functions
from $s=1$ to $s=0$ cannot be recovered, i.e., we cannot obtain the
$P$-function and all its smoothed convolutions up to the customary
Wigner function.  For the same reason, the nonclassicality parameter
proposed by Lee \cite{lee1}, namely, the maximum $s$-parameter that
provides a positive distribution, cannot be experimentally measured.
\par Among the many manifestations of nonclassical effects, one finds
squeezing, antibunching, even-odd oscillations in the photon-number
probability, and negativity of the Wigner
function~\cite{mand2,hill,spe,iss,schwh,yue,yam,igo}.  Any of these
features alone, however, does not represent the univocal criterion we
are looking for.  Neither squeezing nor antibunching provides a
necessary condition for nonclassicality~\cite{agar}. The negativity of
the Wigner function, which is well exhibited by the Fock states and
the Schr\"odinger-cat-like states, is absent for the squeezed states.
As for the oscillations in the photon-number probability, some
even-odd oscillations can be simply obtained by using a statistical
mixture of coherent states.  \par Many authors~\cite{agar,kly1,jpa}
have adopted the non-positivity of the phase-averaged $P$-function
$F(I)={\frac{1}{2\pi}}\int _0^{2\pi} d\phi\,P(I^{1/2}e^{i\phi})$ as the
definition for a nonclassical state, since $F(I)< 0$ invalidates
Mandel's semiclassical formula~\cite{mand} of photon counting, i.e.,
it does not allow a classical description in terms of a stochastic
intensity. Of course, some states can exhibit a ``weak''
nonclassicality~\cite{jpa}, namely, a positive $F(I)$, but with a
non-positive $P$-function (a relevant example being a coherent state
undergoing Kerr-type self-phase modulation). However, from the point
of view of the detection theory, such ``weak'' nonclassical states
still admit a classical description in terms of positive intensity
probability $F(I)>0$. For this reason, we adopt non-positivity of
$F(I)$ as the definition of nonclassicality.
\subsection{Single-mode nonclassicality}
\par The authors of 
Refs.~\cite{agar,kly1,jpa} have pointed out some relations between $F(I)$
and generalized moments of the photon distribution, which, in turn,
can be used to test the nonclassicality. The problem is reduced to an
infinite set of inequalities that provide both necessary and
sufficient conditions for nonclassicality~\cite{kly1}.  In terms of
the photon-number probability $p(n)=\langle n| \rho|n\rangle$ of
the state with density matrix $ \rho$, the simplest sufficient
condition involves the following three-point relation \cite{kly1,jpa}
\begin{eqnarray}
B(n)\equiv(n+2)p(n)p(n+2) -(n+1)[p(n+1)]^2<0 \;.\label{p2p}
\end{eqnarray}
Higher-order sufficient conditions involve five-, seven-, \dots, 
$(2k+1)$-point relations, always for adjacent values of $n$.  It is
sufficient that just one of 
these inequalities is satisfied in order to assure the negativity of
$F(I)$. Notice that for a coherent state $B(n)=0$ identically for all
$n$.  \par In the following we show that quantum tomography can be
used as a powerful tool for performing the nonclassicality test in
Eq.~(\ref{p2p}).  For less-than-unity quantum efficiency ($\eta <1$),
we rely on the concept of a ``noisy state'' $ \rho_{\eta}$,
wherein the effect of quantum efficiency is ascribed to the
quantum state itself rather than to the detector. In this model, the
effect of quantum efficiency is treated in a Schr\"odinger-like
picture, with the state evolving from $ \rho$ to
$ \rho_{\eta}$, and with $\eta$ playing the
role of a time parameter.  Such lossy evolution is described by the
master equation \cite{gard}
\begin{eqnarray}
\partial_t \rho (t)=
{\frac \Gamma 2}\left[2
  a \rho (t)  a^{\dag}-
  a^{\dag}  a  \rho(t)-   \rho (t)  a^{\dag}  a
\right]\;,\label{mm1}
\end{eqnarray}
wherein $ \rho (t)\equiv \rho _{\eta }$ 
with $t=-\ln\eta/\Gamma $. 
\par For the nonclassicality test, reconstruction in terms of the
noisy state has many advantages. In fact, for non-unit quantum
efficiency $\eta <1$ the tomographic method introduces errors for
$p(n)$ which are increasingly large versus $n$, with the additional
limitation that quantum efficiency must be greater than the minimum
value $\eta =0.5$. On the other hand, the reconstruction of the
noisy-state probabilities $p_{\eta}(n)=\langle n|
\rho_{\eta}|n\rangle$ does not suffer such limitations, and even
though all quantum features are certainly diminished in the
noisy-state description, nevertheless the effect of non-unity quantum
efficiency does not change the sign of the $P$-function, but only
rescales it as follows:
\begin{eqnarray}
P(z)\rightarrow P_{\eta }(z)={\frac 1 \eta}P(z/\eta ^{1/2})
\;.\label{peta}
\end{eqnarray} 
Hence, the inequality (\ref{p2p}) still represents a sufficient
condition for nonclassicality when the probabilities $p(n)=\langle n|
\rho|n\rangle$ are replaced with $p_{\eta }(n)=\langle n|
\rho_{\eta}|n\rangle$, the latter being given by a Bernoulli
convolution, as shown in Eq. (\ref{conv_n}). When referred to
the noisy-state probabilities $p_{\eta }(n)$, the inequality in
Eq.~(\ref{p2p}) keeps its form and simply rewrites as follows
\begin{eqnarray}
B_{\eta }(n)\equiv (n+2)p_{\eta }(n)p_{\eta }(n+2)
-(n+1)[p_{\eta }(n+1)]^2 < 0 \;.\label{p2pm}
\end{eqnarray}
The quantities $B(n)$ and $B_{\eta }(n)$ are nonlinear in the density
matrix. Then, they cannot be measured by averaging a suitable
estimator over the homodyne data.  Hence, in the evaluation of $B(n)$ 
one has to reconstruct the photon-number probabilities $p(n)$, using
the estimator ${\cal R}_\eta [|n \rangle \langle n|](x,\varphi)$
in Eq. (\ref{estimat}).  The noisy-state probabilities $p_{\eta}(n)$
are obtained by using the same estimator for $\eta=1$, namely
without recovering the convolution effect of non-unit quantum
efficiency. Notice that the estimator does not depend on the
phase of the quadrature. Hence, the knowledge of the phase of the
local oscillator in the homodyne detector is not needed for the
tomographic reconstruction, and it can be left fluctuating in a real
experiment. 
\par Regarding the estimation of statistical errors, they are
generally obtained by dividing the set of homodyne data into blocks,
as shown in Sec. \ref{s:stat}. However, in the present case, the
nonlinear dependence on the photon number probability introduces a
systematic error that is vanishingly small for increasingly larger
sets of data. Therefore, the estimated value of $B(n)$ is obtained
from the full set of data, instead of averaging the mean value of the
different statistical blocks.  

\begin{figure}[htb]
\begin{center}
\epsfxsize=0.4\hsize\leavevmode\epsffile{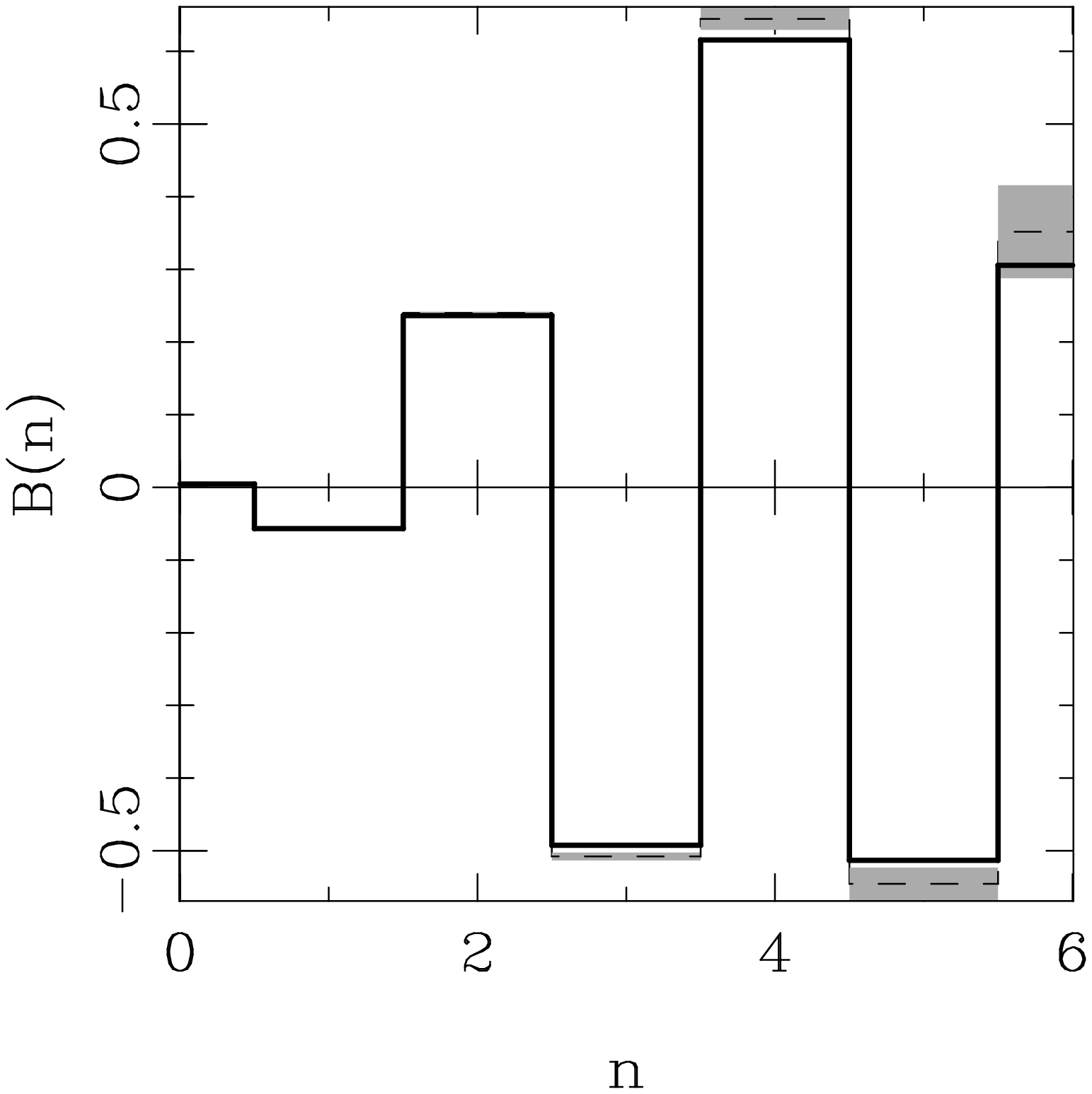}\hfill
\epsfxsize=0.4\hsize\leavevmode\epsffile{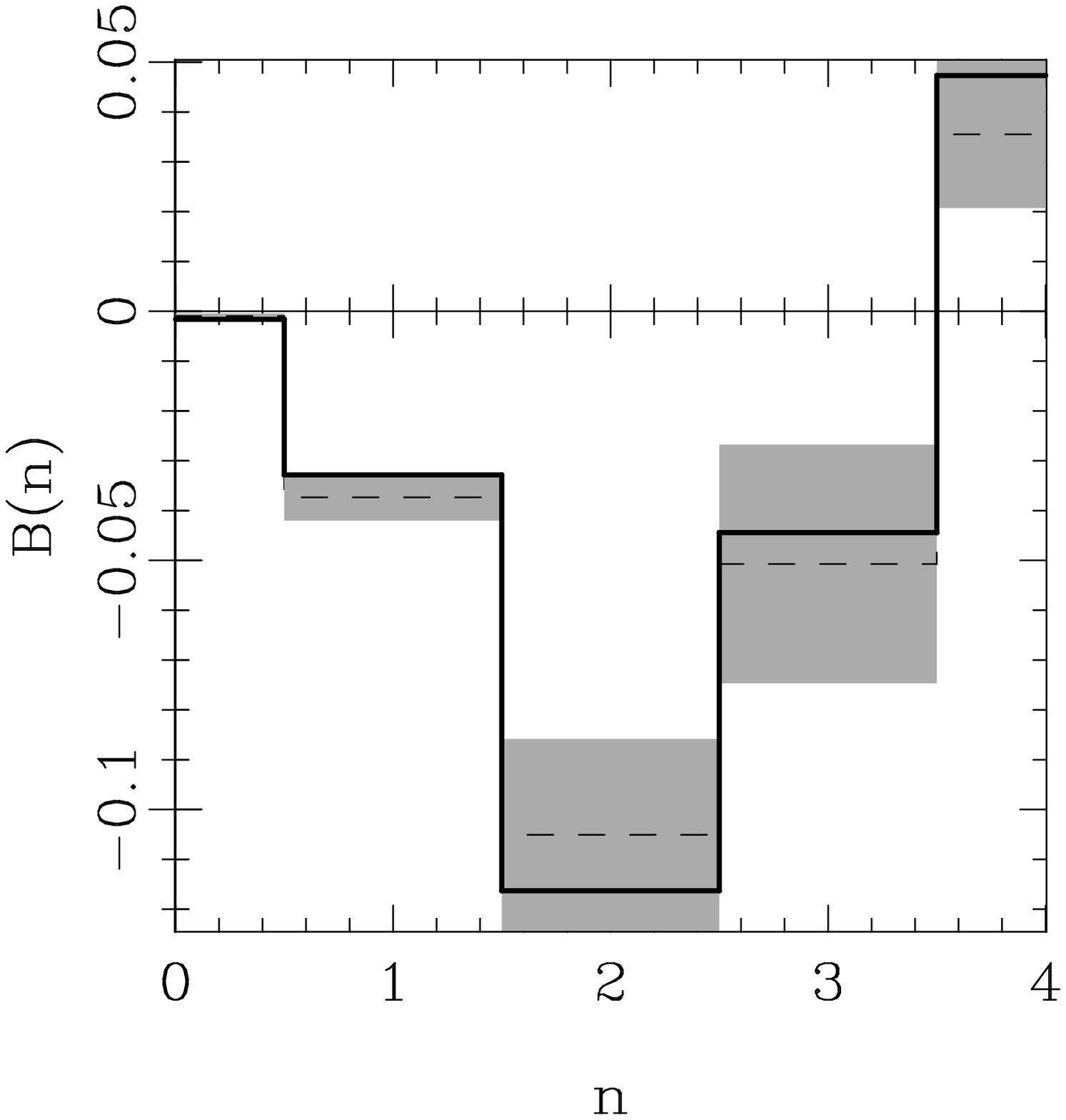}
\end{center}
\caption{Tomographic measurement of $B(n)$ (dashed trace) 
with the respective error bars (superimposed in gray-shade) 
along with the theoretical values (solid trace) 
for a Schr\"odinger cat state with average photon number 
$\bar n=5$ (left); for a phase-squeezed state with $\bar n=5$
and $\bar n_{\rm sq}= \sinh^2 r =3$ squeezing photons (right). 
In both cases the quantum efficiency is $\eta =0.8$ and the number of 
simulated experimental data is $10^7$ (From Ref. \cite{noncl}).}
\label{f:fig1}
\end{figure}
\par
\begin{figure}[htb]
\begin{center}
\epsfxsize=0.4\hsize\leavevmode\epsffile{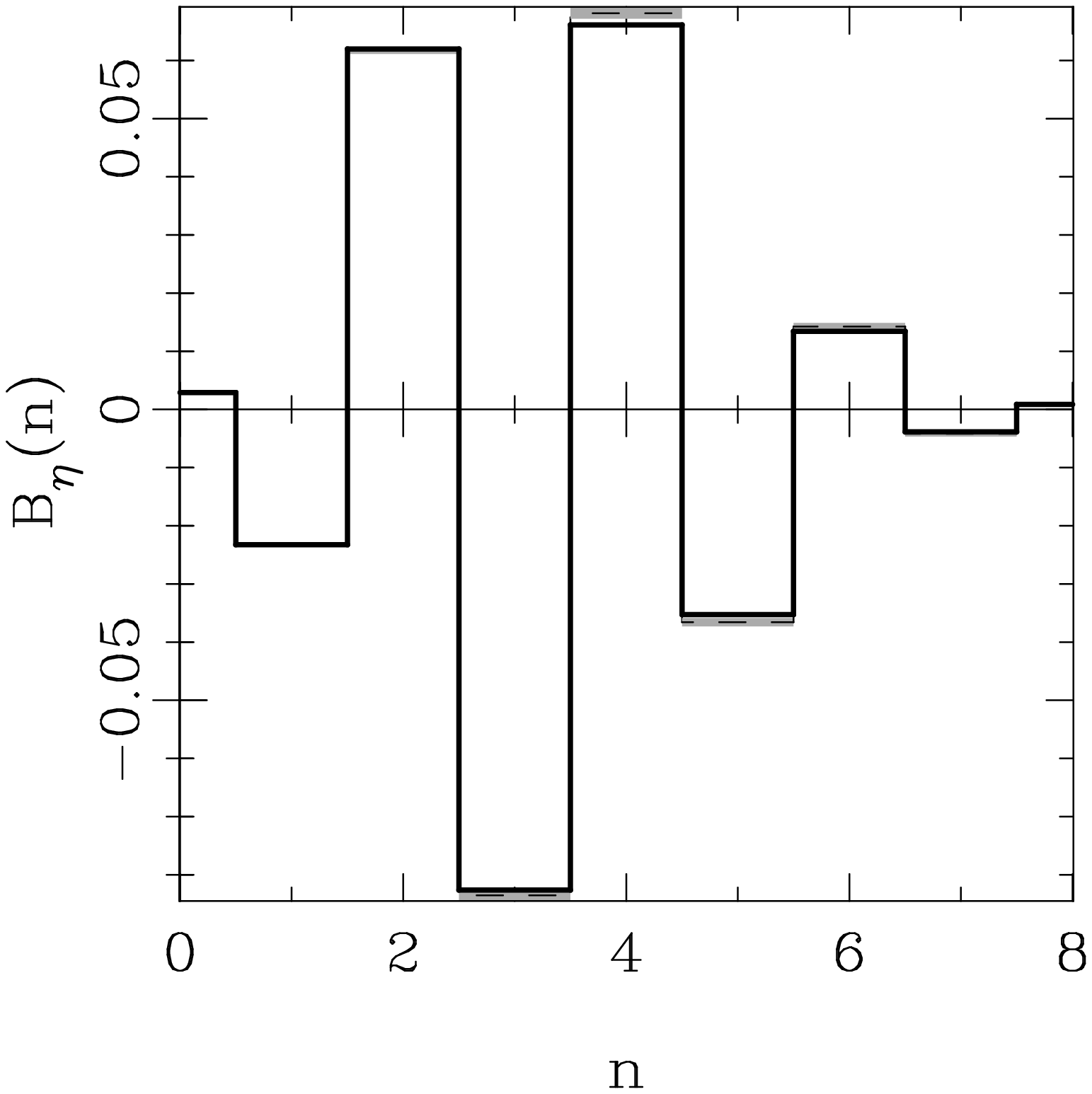}\hfill
\epsfxsize=0.4\hsize\leavevmode\epsffile{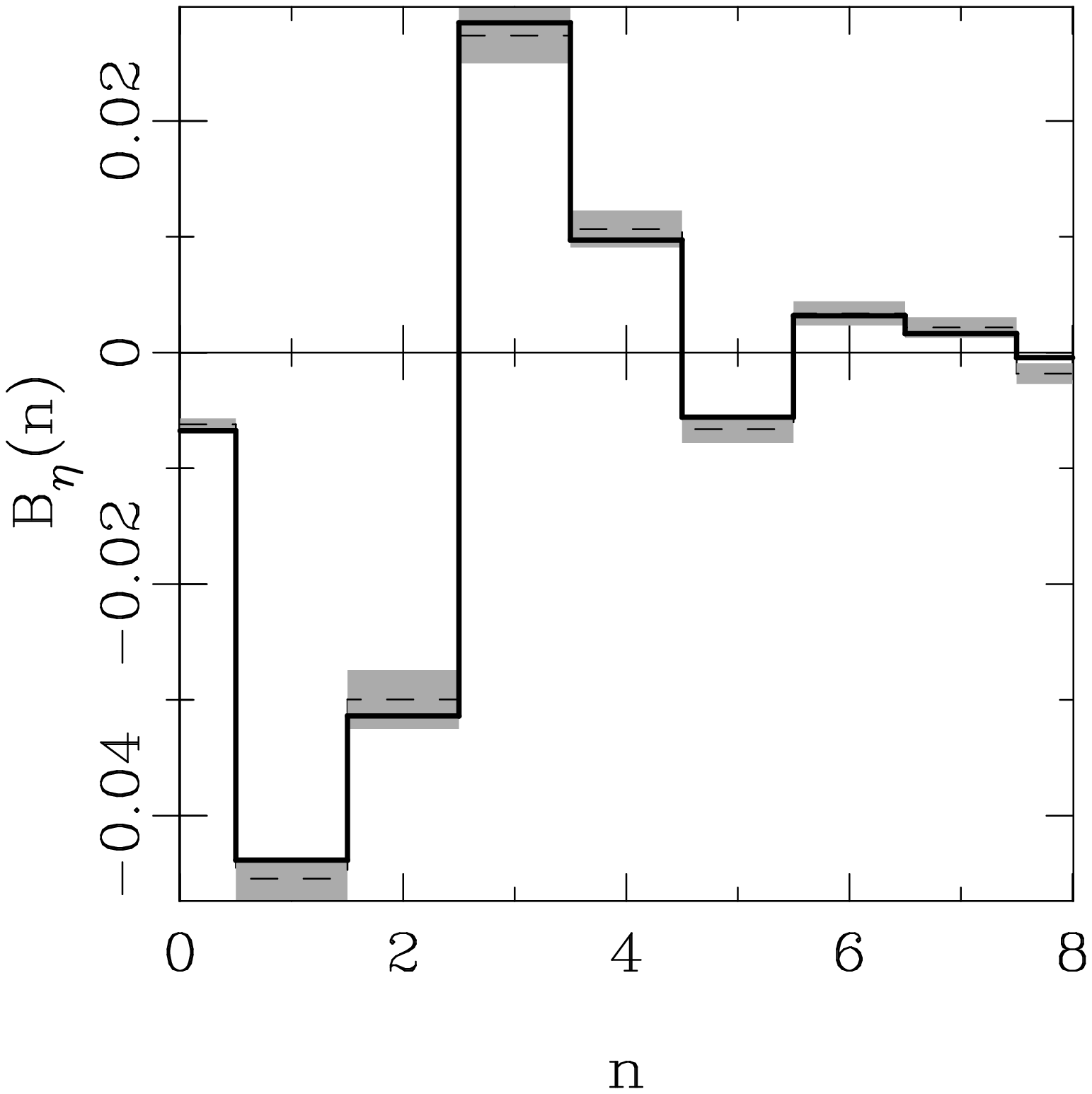}
\end{center}
\caption{Same as Fig. \ref{f:fig1}, but here for  $B_{\eta}(n)$ (From Ref. \cite{noncl}).}
\label{f:fig2}
\end{figure}

 In Figs.~\ref{f:fig1}--\ref{f:fig2} some numerical results from
Ref. \cite{noncl} are reported, which
are obtained by a Monte-Carlo simulation of a quantum 
tomography experiment. The nonclassicality criterion is tested either
on a Schr\"odinger-cat state $|\psi (\alpha )\rangle \propto
(|\alpha\rangle +|-\alpha\rangle )$ or on a squeezed state
$|\alpha,r\rangle\equiv D(\alpha )S(r)|0\rangle $, wherein $|\alpha
\rangle$, $D(\alpha )$, and $S(r)$ denote a coherent state with
amplitude $\alpha $, the displacement operator $D(\alpha )=e^{\alpha
a^{\dag }-\alpha ^* a}$, and the squeezing operator $S(r)=e^{r(
a^{\dag 2}- a^2)/2}$, respectively. Fig.~\ref{f:fig1} shows
tomographically-obtained values of $B(n)$, with the respective error
bars superimposed, along with the theoretical values for a
Schr\"odinger-cat state and for a phase-squeezed state ($r>0$). For
the same set of states the results for $B_{\eta}(n)$ obtained by
tomographic reconstruction of the noisy state are reported in
Fig.~\ref{f:fig2}. Let us compare the statistical errors that affect
the of $B(n)$ and $B_{\eta }(n)$ on the original and the noisy states,
respectively. In the first case the error increases with $n$, whereas
in the second it remains nearly constant, albeit with less marked
oscillations in $B_{\eta }(n)$ than those in $B(n)$.
\par The nonclassicality of the states here analyzed is experimentally
verifiable, as $B_{\eta }(0)< 0$ by more than five standard
deviations. In contrast, for coherent states one obtains small
statistical fluctuations around zero for all $n$.  Finally, we remark
that the simpler test of checking for antibunching or oscillations in
the photon-number probability in the case of the phase-squeezed state
(left of Figs.~\ref{f:fig1} and \ref{f:fig2}) would
not reveal the nonclassical features of such a state.
\subsection{Two-mode nonclassicality}
\par In Ref. \cite{noncl} it is also shown how quantum homodyne
tomography can also be employed to test the nonclassicality of two-mode states. For a
two-mode state nonclassicality is defined in terms of non-positivity of
the following phase-averaged two-mode $P$-function~\cite{jpa}:
\begin{eqnarray}
F(I_1,I_2,\phi )=\frac{1} {2\pi}\int_0^{2\pi}d\phi_1\,
P(I_1^{1/2}e^{i\phi _1},I_2^{1/2}e^{i(\phi _1+\phi )})
\;.\label{non2} 
\end{eqnarray}
In Ref.~\cite{jpa} it is also proved that a sufficient condition for 
nonclassicality is
\begin{eqnarray}
C=\langle (n_1 -  n_2)^2\rangle-
(\langle   n_1 -  n_2\rangle )^2-
\langle   n_1 +  n_2\rangle <0
\;,\label{n22} 
\end{eqnarray}
where $ n_1$ and $ n_2$ are the photon-number operators of the two
modes.  

\par A tomographic test of the inequality in Eq.~(\ref{n22}) can be
performed by averaging the estimators for the involved operators
using Table \ref{td1}.  Again, the value $\eta=1$ can be used to
reconstruct the ensemble averages of the noisy state $ \rho_{\eta}$.
As an example, we consider the twin-beam state of Eq. (\ref{Psi}).
The theoretical value of $C$ is given by $C=-2|\xi |^2/(1-|\xi
|^2)<0$.  With regard to the effect of quantum efficiency $\eta <1$,
the same argument still holds as for the single-mode case: one can
evaluate $C_{\eta}$ for the twin beams degraded by the effect of loss,
and use $\eta =1$ in the estimators.  In this case, the
theoretical value of $C_\eta$ is simply rescaled, namely
\begin{eqnarray}
C_\eta =-2\eta ^2 |\xi |^2/(1-|\xi |^2)\;.
\label{n2}
\end{eqnarray}
\par
\begin{figure}[htb]
\begin{center}
\epsfxsize=0.4\hsize\leavevmode\epsffile{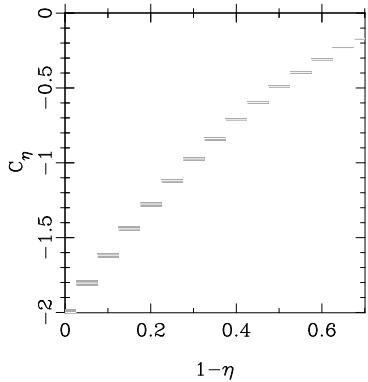}
\end{center}
\caption{Tomographic measurement of the nonclassical parameter
  $C_{\eta}$ for twin beams in Eq.~(\ref{Psi}) with $|\xi |^2=0.5$.
  The results are shown for different values of the quantum efficiency
  $\eta $ (in steps of 0.05), and for each value the number of
  simulated data is $4\times 10^5$. Statistical errors are shown in
  the gray shade (From Ref. \cite{noncl}).}
\label{f:fig9}
\end{figure}

 In Fig.~\ref{f:fig9} we report $C_{\eta }$ {\em vs.}  $1-\eta $,
 with $\eta $ ranging from 1 to 0.3 in steps of 0.05, for the
 twin beam in Eq.~(\ref{Psi}) with $|\xi |^2=0.5$, corresponding
 to a total average photon number $\langle n_1 +n_2 \rangle =2$. The
 values of $C_{\eta }$ result from a Monte-Carlo simulation of a
 homodyne tomography experiment with a sample of $4\times 10^5$ data.
 The nonclassicality test in terms of the noisy state gives values of
 $C_{\eta }$ that are increasingly near the classically positive
 region for decreasing quantum efficiency $\eta $. However, the
 statistical error remains constant and is sufficiently small to allow
 recognition of the nonclassicality of the twin beams up to $\eta
 =0.3$.

\par We conclude that quantum homodyne tomography allows one to 
perform nonclassicality tests for single- and two-mode radiation 
states, even when the quantum efficiency of homodyne detection is rather low. 
The method involves reconstruction of the photon-number probability or of 
some suitable function of the number operators pertaining to the 
noisy state, namely, the state degraded by the less-than-unity
quantum efficiency. The noisy-state reconstruction is 
affected by the statistical errors; however, they are sufficiently small
that the nonclassicality of the state can be tested even for low values of 
$\eta $. For the cases considered here, we have shown that the 
nonclassicality of the states can be proved (deviation from classicality 
by many error bars) with $10^5$--$10^7$ 
homodyne data. Moreover, since the knowledge of the phase of the local 
oscillator in the homodyne detector is not needed for the tomographic 
reconstruction, it can be left fluctuating in a real experiment. 
\section{Test of state reduction}
In quantum mechanics the state reduction (SR) is still a very
discussed rule. The so--called ``projection postulate'' was introduced
by von Neumann \cite{vonNeumann} to explain the results from the
Compton-Simons experiment, and it was generalized by L\"uders
\cite{Luders} for measurements of observables with degenerate
spectrum. The consistency of the derivation of the SR rule and its
validity for generic measurements have been analyzed with some
criticism \cite{bibsr}. In a very general context, the SR rule was
derived in a physically consistent way from the Schr\"odinger equation
for the composite system of object and measuring apparatus
\cite{Ozawa97}.  An experiment for testing quantum SR is therefore a
very interesting matter. Such a test in general is {\em not}
equivalent to a test of the repeatability hypothesis since the latter
holds only for measurements of observables that are described by
self-adjoint operators.  For example, joint measurements like the
Arthurs-Kelly \cite{Arthurs} are not repeatable, as the reduced states
are coherent states, which are not orthogonal.

Quantum optics offers a possibility of testing the SR, because several
observables can be chosen to perform different measurements on a fixed
system. For instance, one can decide to perform either homodyne or
heterodyne, or photon-number detection.  This is a unique opportunity;
in contrast, in particle physics the measurements are mostly
quasi-classical and restricted to only a few observables.  In
addition, optical homodyne tomography allows a precise determination
of the quantum system after the SR.

A scheme for testing the SR could be based on tomographic measurements
of the radiation density matrix after nondemolition
measurements. However, such a scheme would reduce the number of
observables that are available for the test.  Instead, one can take
advantage of the correlations between the twin beams of
Eq. (\ref{Psi}) produced by a non-degenerate optical parametric
amplifier (NOPA), in which case one can test the SR even for
demolitive-type measurements. Indeed, if a measurement is performed on
one of the twin beams, the SR can be tested by homodyne tomography on
the other beam. This is precisely the scheme for an
experimental test of SR proposed in Ref. \cite{sr}, which is reviewed
in the following.  

\begin{figure}[hbt]
\begin{center}
\epsfxsize=.6 \hsize\leavevmode\epsffile{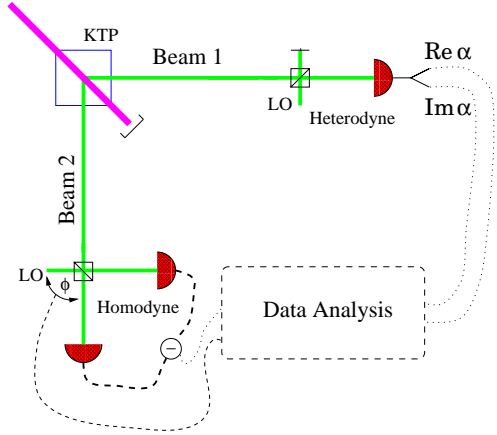}
\end{center}
\caption{Schematic of the proposed scheme for testing the SR for
heterodyne detection. A NOPA generates a pair of twin beams (1 and
2). After heterodyning beam 1, the reduced state of beam 2 is analyzed
by homodyne tomography, which is conditioned by the heterodyne
outcome. In place of the heterodyne detector one can put any other
kind of detector for testing the SR on different observables. We also
consider the case of direct photodetection (From Ref.\cite{sr}). }
\label{f:expsr}
\end{figure}

The scheme for the SR test is given in Fig.~\ref{f:expsr}.  Different
kinds of measurements can be performed on beam 1, even
though here the SR only for heterodyne detection and photon-number
detection will be considered.

\par For a system described by a density operator $ \rho$, the
probability $p(\lambda) d\lambda$ that the outcome of a quantum
measurement of an observable is in the interval
$[\lambda,\lambda+d\lambda)$ is given by Born's rule
$p(\lambda)d\lambda=\mbox{Tr}[ \rho \, \Pi_{\lambda}d\lambda]$, where
$ \Pi _{\lambda}$ is the POVM pertaining to the measurement that
satisfies $ \Pi _{\lambda}\geq 0$ and $\int d\lambda\, \Pi _{\lambda}=
I$.  For an exact measurement of an observable, which is described by
a self-adjoint operator, $ \Pi _{\lambda}$ is just the projector over
the eigenvector corresponding to the outcome $\lambda$.  In the case
of the photon number $ a^{\dagger} a$ the spectrum is discrete and the
POVM is $ \Pi _{m}=|m\rangle \langle m|$ for integer eigenvalue $m$.
For the Arthurs-Kelly joint measurement of the position and momentum
(corresponding to a joint measurement of two conjugated quadratures of
the field) we have the coherent-state POVM $ \Pi
_{\alpha}=\pi^{-1}|\alpha\rangle \langle\alpha|$.

\par When on beam 1 we perform a measurement described by $ \Pi
_{\lambda}$, the reduced normalized state of beam 2 is
\begin{equation}
 \rho(\lambda)=
\frac{\mbox{Tr}_1[|\xi\rangle \langle\xi|( \Pi _{\lambda}\otimes  1 )]}
{\mbox{Tr}_{1,2}[|\xi\rangle \langle\xi|
( \Pi _{\lambda}\otimes  1 )]} =
\frac{ \Xi  \Pi ^\tau _{\lambda}  \Xi^{\dagger}}{p(\lambda)} \,,
\label{sr1}
\end{equation}
where $O^\tau $ denotes the transposed operator (on a fixed basis), $
\Xi=(1-|\xi|^2)^{1/2}\xi^{ a^{\dagger} a}$, and
$p(\lambda)=\mbox{Tr}_{1,2}[ \Xi \Pi ^\tau _{\lambda} \Xi^{\dagger}]$
is the probability density of the measurement outcome $\lambda$.  In
the limit of infinite gain $|\xi |\to 1$ one has $ \rho(\lambda)
\propto \Pi _{\lambda}^\tau $. For example, for heterodyne detection
with outcome $\alpha$, we have $ \rho(\alpha)= |\alpha ^*\rangle
\langle\alpha ^*|$.

If the readout detector on beam 1 has quantum efficiency $\eta_r$,
Eq. (\ref{sr1}) is replaced with
\begin{equation}
 \rho^{\eta_r}(\lambda)=
\frac{ \Xi ( \Pi _{\lambda}^{\eta_r})^\tau  \Xi^{\dagger}}
{p^{\eta_r}(\lambda)} \,,
\label{sr}
\end{equation}
where $p^{\eta_r}(\lambda)=\mbox{Tr}_{1,2}[ \Xi (
\Pi _{\lambda}^{\eta_r})^\tau  \Xi^{\dagger}]$, and $ \Pi _{\lambda}^{\eta_r}$
is the POVM for measurement with quantum efficiency $\eta _r$. As
shown in Sec. \ref{hetsec}, for heterodyne detection one has the Gaussian
convolution 
\begin{equation}
 \Pi _{\alpha}^{\eta_r}=\frac{1}{\pi} \int _{\mathbb C}\frac{d^2 z}
{\pi \Delta_{\eta_r}^2}\,e^{-\frac{|z-\alpha|^2}{\Delta_r^2}}
|z\rangle \langle z| \,,
\label{het}
\end{equation}
with $\Delta_{\eta_r}^2=(1-\eta_r)/\eta_r$. For direct photodetection
$ \Pi _{m}=|m\rangle \langle m|$ is replaced with the Bernoulli
convolution 
\begin{equation}
 \Pi _m^{\eta_r}=\sum_{j=m}^\infty \left(\begin{array}{c} j \\ m
\end{array}\right) \eta_r^m (1-\eta_r)^{j-m} |j\rangle \langle j|\;.
\end{equation}

The experimental test proposed here consists of performing
conditional homodyne tomography on beam 2, given the outcome $\lambda$
of the measurement on beam 1. We can directly measure the ``fidelity
of the test''
\begin{equation}
F(\lambda)=\mbox{Tr}[ \rho^{\eta_r}(\lambda) \,
 \rho_{\rm meas}(\lambda)]\,,
\label{fid}
\end{equation}
where $ \rho^{\eta_r}(\lambda)$ is the theoretical state in
Eq.~(\ref{sr}), and $ \rho_{\rm meas}(\lambda)$ is the experimentally
measured state on beam 2.  Notice that we use the term ``fidelity''
even if $F(\lambda)$ is a proper fidelity when at least one of the two
states is pure, which occurs in the limit of unit quantum efficiency
$\eta_r$.  In the following we evaluate the theoretical value of
$F(\lambda)$ and compare it with the tomographic measured value.

The fidelity (\ref{fid}) can be directly measured by homodyne
tomography using the estimator for the operator
$\rho^{\eta_r}(\lambda)$, namely 
\begin{equation}
F(\lambda)=\int^{\pi}_0\frac{\mbox{d}\varphi}{\pi} 
\int^{+\infty}_{-\infty} dx\, p_{\eta_h}(x,\varphi;\lambda)
{\cal R}_{\eta_h}[ \rho^{\eta_r}(\lambda)](x,\varphi)\;,
\label{condfidel}
\end{equation}
where $p_{\eta_h}(x,\varphi;\lambda)$ is the conditional homodyne probability
distribution for outcome $\lambda$ at the readout detector.

For heterodyne detection on beam 1 with outcome $\alpha\in\mathbb C$,
the reduced state on beam 2 is given by the displaced thermal state
\begin{equation}
 \rho^{\eta_r}(\alpha)=\eta_{\xi}
 D(\gamma)
(1-\eta_{\xi})^{ a^{\dagger} a}
 D^{\dagger}(\gamma) \,,
\label{srhet}
\end{equation}
where
\begin{equation}
\eta_{\xi}=1+(\eta_r-1)|\xi|^2\,,\qquad
\gamma = \frac{\xi\eta_r}{\eta_{\xi}}\alpha^*\,.\label{etaxigamma}
\end{equation}
The estimator in Eq. (\ref{condfidel}) is given by 
\begin{eqnarray}
{\cal R}_{\eta_h}[ \rho^{\eta_r}(\alpha)](x,\varphi)=
\frac{2\eta_h\eta_{\xi}}{2\eta_h-\eta_{\xi}}\Phi\left(1,{\frac 1 2};
-\frac{2\eta_h\eta_{\xi}}{2\eta_h-\eta_{\xi}}(x-\gamma_{\varphi})^2\right)\;,
\label{trhet}
\end{eqnarray}
where ${\gamma}_{\varphi}=\mbox{Re}({\gamma}e^{-i\varphi})$, and
$\Phi(a,b;z)$ denotes the customary confluent hypergeometric
function. The estimator in Eq.~(\ref{trhet}) is bounded for $\eta_h >
{\frac 1 2}\eta_{\xi}$, then one needs to have
\begin{equation}
\eta_h > \frac{1}{2}\left[1-{|\xi|^2}(1-\eta_r)\right] \, .
\label{vallim}
\end{equation}
As one can see from Eq.~(\ref{vallim}), for $\eta_h>0.5$ the fidelity
can be measured for any value of $\eta_r$ and any gain parameter $\xi$
of the NOPA. We recall that the condition $\eta_h>0.5$ is required for
the measurement of the density matrix. However, in this direct
measurement of the fidelity, the reconstruction of the density matrix is
bypassed, and we see from Eq.~(\ref{vallim}) that the bound
$\eta_h=0.5$ can be lowered.

\par The measured fidelity $F(\alpha)$ in Eq.~(\ref{condfidel}) with $
\rho^{\eta_r}(\alpha)$ as given in Eq.~(\ref{srhet}) must be compared
with the theoretical value
\begin{equation}
F_{\rm th}=\eta_{\xi}/(2-\eta_{\xi})\;,
\label{fidth}
\end{equation}
that is independent of $\alpha$.  

\par For direct photodetection on beam 1 with outcome $n$, the reduced
state on beam 2 is given by
\begin{eqnarray}
 \rho^{\eta_r}(n)=\eta_{\xi}\left(\frac{\eta_{\xi}}{1-\eta_{\xi}}
\right)^n\left(\begin{array}{c} a^{\dag} a \\ n \end{array}\right)
(1-\eta_{\xi})^{ a^{\dag} a} \,.
\label{srnum}
\end{eqnarray}
The estimator for the fidelity measurement is
\begin{eqnarray}
{\cal R}_{\eta_h}[ \rho^{\eta_r}(n)](x,\varphi)=
\left.\frac{(\eta_{\xi}\partial_z)^n}{n!}\right|_{z=0}
\frac{2\eta_h\eta_{\xi}}{2\eta_h-\eta_{\xi}+z}
\Phi\left(1,{\frac 1 2};
-\frac{2\eta_h(\eta_{\xi}-z)}{2\eta_h-\eta_{\xi}+z}x^2\right)
\,. \label{trnum1}
\end{eqnarray}
We see that the same bound of Eq.~(\ref{vallim}) holds. In this case
the measured fidelity $F(n)$ must be compared with the theoretical
value
\begin{equation}
F_{\rm th}(n)=\eta_{\xi}^{2+2n}
F\left(2n+1,2n+1;1;(1-\eta_{\xi})^2\right)\,,
\label{fidthn}
\end{equation}
where $ F(a,b;c;z)$ denotes the customary hypergeometric function.

\par Several simulations have been reported in
Ref. \cite{sr} for both heterodyne and photodetection on beam 1. In
the former case the quadrature probability distribution has been
simulated, pertaining to the reduced state (\ref{srhet}) on beam 2,
and averaged the estimators in Eq. (\ref{trhet}).  In the latter case
the reduced state (\ref{srnum}) and the estimators in
Eq. (\ref{trnum1}) have been used.

Numerical results for the fidelity was thus obtained for different
values of the quantum 
efficiencies $\eta_r$ and $\eta_h$, and of the NOPA gain parameter
$\xi $. A decisive test can be performed with samples of just a few
thousand measurements. The statistical error in the measurement was
found rather insensitive to both quantum efficiencies and NOPA gain.

\section{Tomography of coherent signals and applications}
 Quantum homodyne tomography has been proved useful in various
experimental situations, such as for measuring the photon statistics
of a semiconductor laser \cite{raymer95}, for determining the density
matrix of a squeezed vacuum \cite{sch} and the joint photon-number
probability distribution of a twin beam created by a non-degenerate
optical parametric amplifier \cite{kumarandme}, and, finally, for
reconstructing the quantum states of spatial modes with an array
detector \cite{beckArray}. In this section we review some experimental
results about homodyne tomography with coherent states, with
application to the estimation of the loss introduced by simple optical
components \cite{TomoNa}.  \par
\begin{figure}[htb]
\begin{center}
\epsfxsize=.8\hsize\leavevmode\epsffile{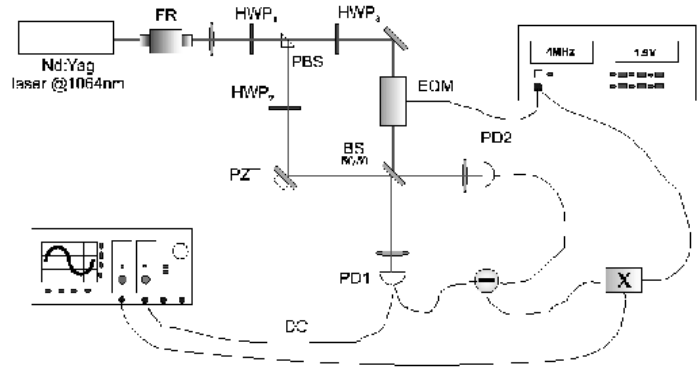}
\end{center}
\caption{Schematic of the experimental set-up. A Nd:YAG laser beam is
divided into two beams, the first acting as a strong local oscillator,
the second representing the signal beam. The signal is modulated at
frequency $\Omega$ with a defined modulation depth to control the
average photon number in the generated coherent state. The tomographic
data are collected by a homodyne detector whose difference photocurrent
is demodulated and then acquired by a digital oscilloscope 
(From Ref. \cite{TomoNa}). }\label{f:exp} 
\end{figure}

 The experiment has been performed in
the Quantum Optics Lab of the University of Naples, and its the
schematic is presented in Fig. \ref{f:exp}. The principal radiation
source is provided by a monolithic Nd:YAG laser ($ \approx $ 50 mW at
1064 nm; Lightwave, model 142). The laser has a linewidth of less than
10 kHz/ms with a frequency jitter of less than 300 kHz/s, while its
intensity spectrum is shot--noise limited above 2.5 MHz.

\par The laser emits a linearly polarized beam in a TEM00 mode,
which is split in two parts by a beam splitter. One part provides the
strong local oscillator for the homodyne detector.  The other part,
typically less than 200\thinspace $\mu $W, is the homodyne signal. The
optical paths traveled by the local oscillator and the signal beams
are carefully adjusted to obtain a visibility typically above $75\%$
measured at one of the homodyne output port. The signal beam is
modulated, by means of a phase electro--optic modulator (EOM, Linos
Photonics PM0202), at 4MHz, and a halfwave plate (HWP$_{2}$,
HWP$_{3}$) is mounted in each path to carefully match the polarization
state at the homodyne input.  \par The detector is composed by a
50$\div $50 beam splitter (BS), two amplified photodiodes (PD1, PD2),
and a power combiner. The difference photocurrent is demodulated at
4MHz by means of an electrical mixer. In this way the detection occurs
outside any technical noise and, more important, in a spectral region
where the laser does not carry excess noise.  

\par The phase modulation added to the signal beam moves a certain
number of photons, proportional to the square of the modulation depth,
from the carrier optical frequency $\omega $ to the side bands at
$\omega \pm \Omega $ so generating two weak coherent states with
engineered average photon number at frequencies $\omega \pm \Omega
$. The sum sideband modes is then detected as a controlled
perturbation attached to the signal beam. The demodulated current is
acquired by a digital oscilloscope (Tektronix TDS 520D) with 8 bit
resolution and record length of 250000 points per run. The acquisition
is triggered by a triangular shaped waveform applied to the PZT
mounted on the local oscillator path. The piezo ramp is adjusted to
obtain a $2\pi $ phase variation between the local oscillator and the
signal beam in an acquisition window.  \par The homodyne data to be
used for tomographic reconstruction of the state have been calibrated
according to the noise of the vacuum state. This is obtained by
acquiring a set of data leaving the signal beam undisturbed while
scanning the local oscillator phase.  It is important to note that in
case of the vacuum state no role is played by the visibility at the
homodyne beam--splitter.  \par The tomographic samples consist of $N$
homodyne data $\{ x_{j},\varphi _{j}\}_{j=1,...,N}$ with phases
$\varphi _{j}$ equally spaced with respect to the local
oscillator. Since the piezo ramp is active during the whole
acquisition time, we have a single value $x_{j}$ for any phase
$\varphi _{j}$. From calibrated data we first reconstruct the quantum
state of the homodyne signal. According to the experimental setup, we
expect a coherent signal with nominal amplitude that can be adjusted
by varying the modulation depth of the optical mixer. However, since
we do not compensate for the quantum efficiency of photodiodes in the
homodyne detector ($\eta \simeq 90\% $) we expect to reveal coherent
signals with reduced amplitude.  In addition, the amplitude is further
reduced by the non-maximum visibility (ranging from $75\%$ to $85\%$)
at the homodyne beam--splitter.  \par
\begin{figure}[htb]
\begin{tabular}{lcr}
\psfig{file=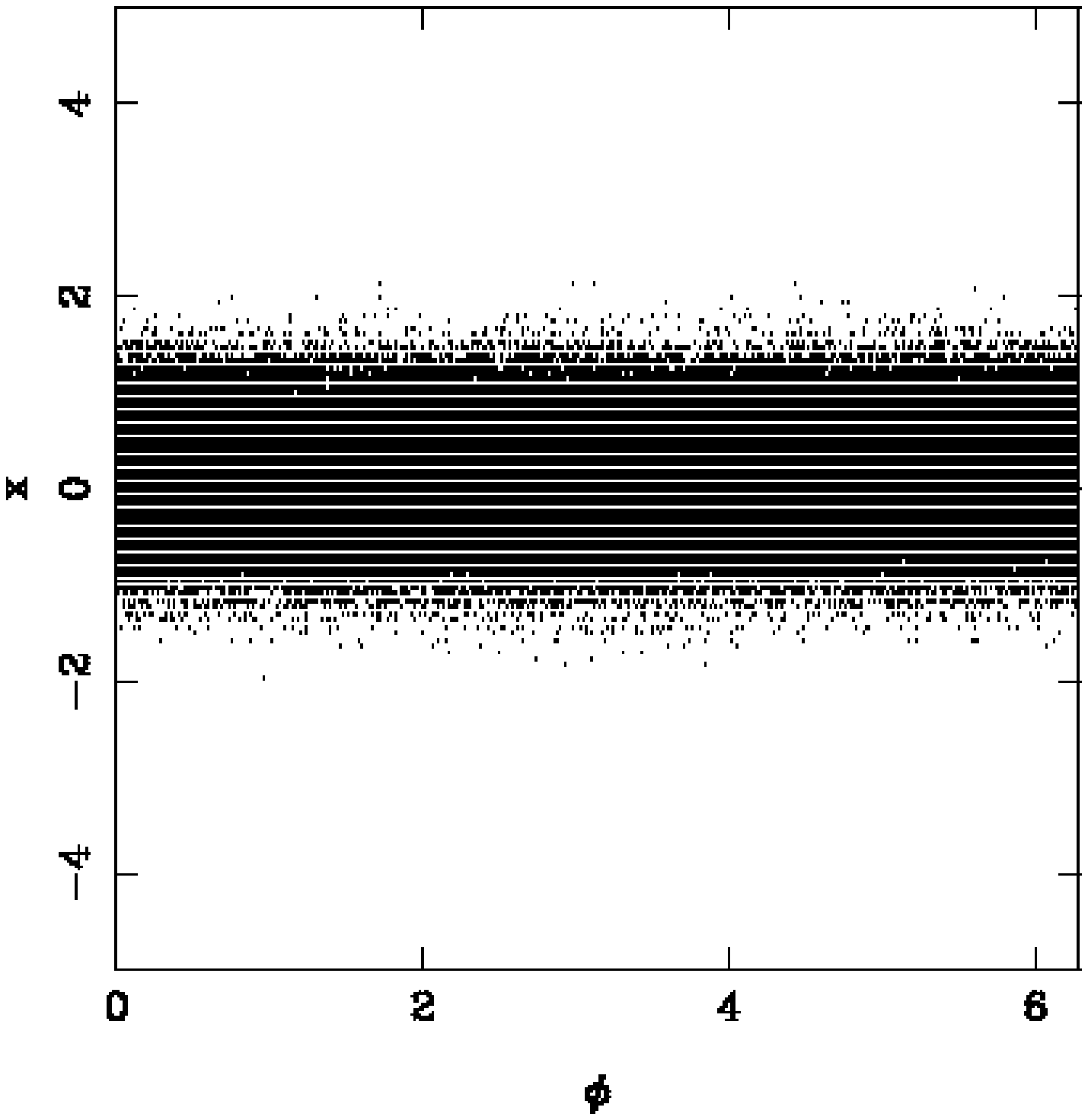,width=4cm} &
\psfig{file=cap6fig9.ps,width=4cm} &
\psfig{file=cap6fig10.ps,width=4cm} \\
\psfig{file=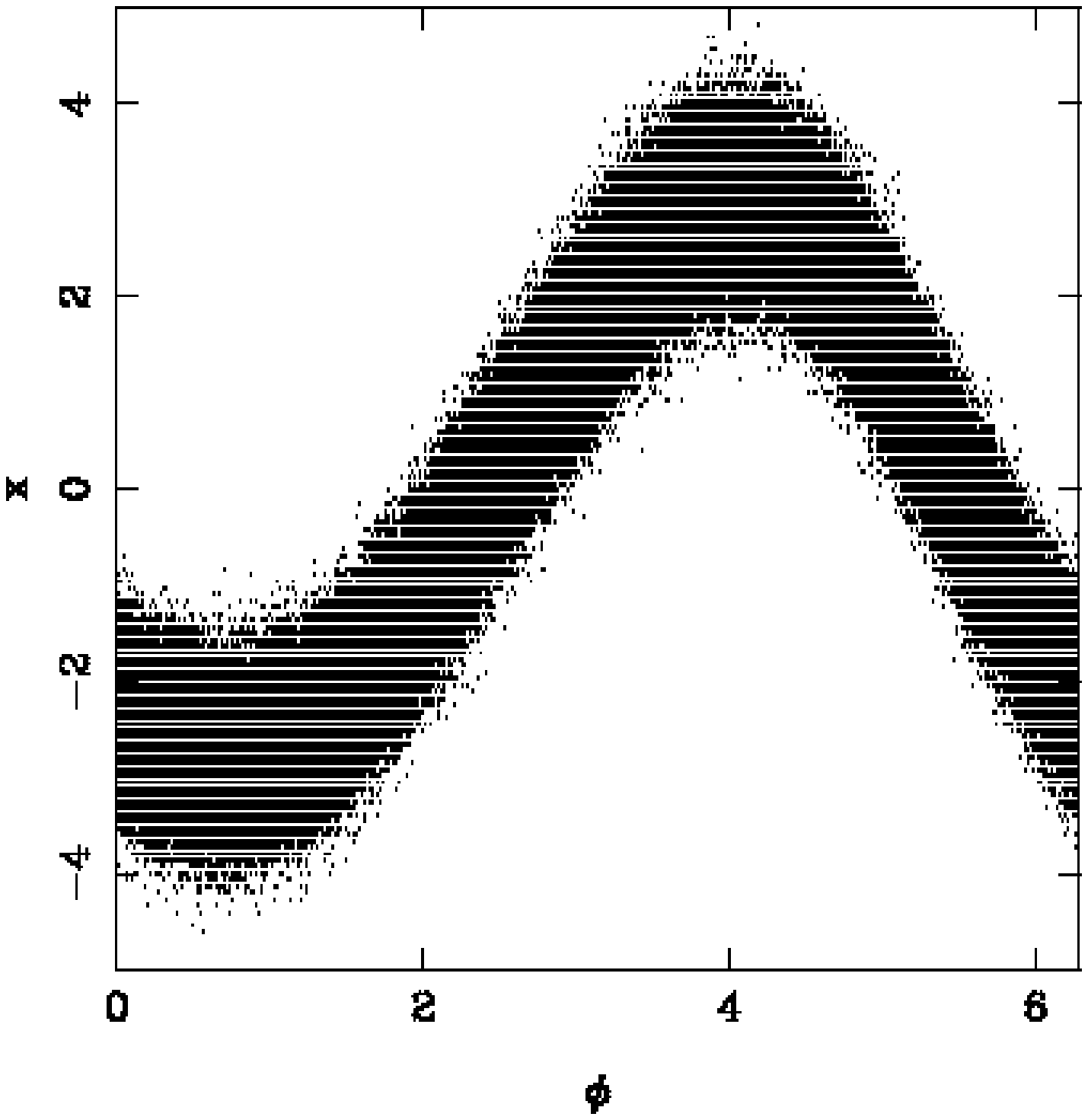,width=4cm} &
\psfig{file=cap6fig12.ps,width=4cm} &
\psfig{file=cap6fig13.ps,width=4cm}
\end{tabular}
\caption{Reconstruction of the quantum state of the signal, and of the
vacuum state used for calibration. For both states, from left to
right, we report the raw data, a histogram of the photon number
distribution, and a contour plot of the Wigner function. The
reconstruction has been performed by a sample of $N=242250$ homodyne
data. The coherent signal has an estimated average photon number equal
to $\langle a^\dag a\rangle=8.4$. The solid line denotes the
theoretical photon distribution of a coherent state with such number
of photons. Statistical errors on matrix elements are about $2\%$
\label{f:rec}. The slight phase asymmetry in the Wigner distribution
corresponds to a value of about $2\%$ of the maximum (From Ref. \cite{TomoNa}).}
\end{figure}

 In Fig. \ref{f:rec} we
 report a typical reconstruction, together with the
reconstruction of the vacuum 
state used for calibration. For both states, we report the raw data,
the photon number distribution $\rho_{nn}$, and a contour plot of
the Wigner function. The matrix elements are obtained by sampling the
corresponding estimators in Eq. (\ref{estimat}), whereas the
confidence intervals for diagonal elements are given by
$\delta\rho_{nn}=\Delta\rho/\sqrt{N}$, $\Delta\rho$ being the
rms deviation of the estimator over data. For off-diagonal elements the
confidence intervals are evaluated for the real and imaginary part
separately.

\par In order to see the quantum state as a whole, we also report the
reconstruction of the Wigner function of the field, which can be
expressed in terms of the matrix elements as the discrete Fourier
transform 
\begin{eqnarray}
W(\alpha , \alpha ^* )=
\mbox{Re} \sum_{d=0}^{\infty} e^{id\varphi} \sum_{n=0}^{\infty} \Lambda
(n,d;|\alpha |) \rho_{n,n+d} \;  \label{w2}
\end{eqnarray}
where $\varphi =\arg \alpha $, and 
\begin{eqnarray}
\Lambda (n,d;|\alpha |) = (-)^n 2 (2-\delta_{d0}) |2 \alpha  |^d 
\sqrt{\frac{n!}{(n+d)!}}
e^{-2|\alpha |^2} L_n^d (|2 \alpha  |^2) \;,  \label{w3}
\end{eqnarray}
$L_n^d (x)$ denoting the Laguerre polynomials. Of course, the series in
Eq. (\ref{w2}) should be truncated at some point, and therefore the Wigner
function can be reconstructed only at some finite resolution. 
\par Once the coherence of the signal has been established we may use
homodyne tomography to estimate the loss imposed by a passive optical
component like an optical filter. The procedure may be outlined as
follows. We first estimate the initial mean photon number
$\bar{n}_{0}=|\alpha _{0}|^{2}$ of the signal beam, and then the same
quantity inserting an optical neutral density filter in the signal
path. If $\Gamma $ is the loss parameter, then the coherent amplitude
is reduced to $\alpha _{\Gamma }=\alpha _{0}e^{-\Gamma }$, and the
intensity to $\bar{n}_{\Gamma }=\bar{n}_{0}e^{-2\Gamma }$.  \par The
estimation of the mean photon number can be performed adaptively on
data, using the general method presented in Sec. \ref{s:adapt}.  One
takes the average of the estimator
\begin{eqnarray}
{\cal R}[a^{\dag} a](x,\varphi )=2 x^2 - \frac{1}{2} + 
\mu e^{i2\varphi} + \mu ^* e^{-i2\varphi} \;,
\end{eqnarray}
where $\mu$ is a parameter to be determined in order to minimize
fluctuations. As proved in Ref. \cite{adapt} one has $\mu=-1/2 \langle
a^{\dag 2}\rangle$, which itself can be obtained from homodyne
data. In practice, one uses the data sample twice: first to evaluate
$\mu$, then to obtain the estimate for the mean photon number.
\begin{figure}[htb]
\begin{center}
\psfig{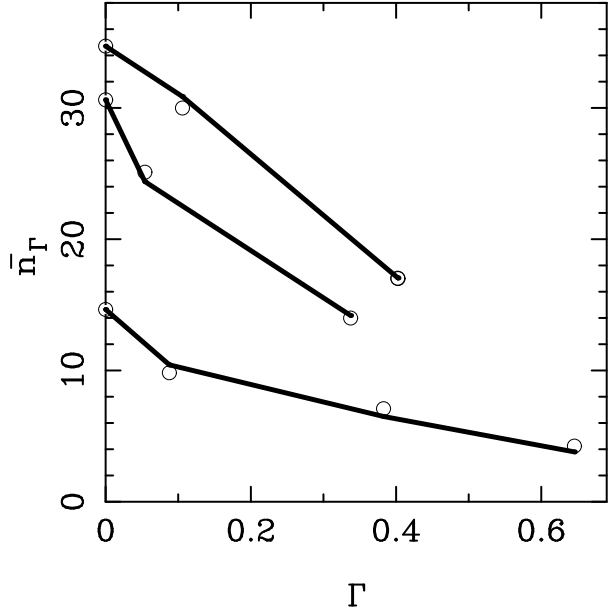}
\end{center}
\caption{Estimation of the mean photon number of a coherent signal as a
function of the loss imposed by an optical filter. Three set of experiments,
corresponding to three different initial amplitudes are reported. Open circles 
are the tomographic determinations, whereas solid line
denotes the expected values, as follow from nominal values of loss and
visibility at homodyne detector. Statistical errors are within
the circles (From Ref. \cite{TomoNa}).}\label{f:loss} 
\end{figure}

\par In Fig. \ref{f:loss} the tomographic determinations of
$\bar{n}_\Gamma$ are compared with the expected values for three set
of experiments, corresponding to three different initial
amplitudes. The expected values are given by $\bar{n}_{\Gamma
}=\bar{n}_{0}e^{-2\Gamma }{\cal V}$, where $\Gamma $ is the value
obtained by comparing the signal dc currents $I_{0}$ and $I_{\Gamma }$
at the homodyne photodiodes and ${\cal V=V}_{\Gamma }/{\cal V}_{0}$ is
the relative visibility. The solid line in Fig. \ref{f:loss} denotes
these values. The line is not continuous due to variations of
visibility. It is apparent from the plot that the estimation is
reliable in the whole range of values we could explore.  It is worth
noting that the estimation is absolute, {\em i.e.} it does not
 depend on the knowledge of the initial amplitude, and it is robust, since it
can be performed independently on the quantum efficiency of the homodyne
detector.  \par One may notice that the estimation of loss can be
pursued also by measuring an appropriate observable, typically the
intensity of the light beam with and without the filter. However, this
is a concrete possibility only for high amplitude signals, whereas
losses on weak coherent states cannot be properly characterized
neither by direct photocounting using photodiodes (due to the low
quantum efficiency and large fluctuations), nor by avalanche
photodetectors (due to the impossibility of discriminating among the
number of photons).  On the contrary, homodyne tomography provides the
mean intensity (actually the whole photon distribution) independently
 on the signal level, thus allowing a precise
characterization also in the quantum regime. Indeed, in
Ref. \cite{adapt} adaptive tomographic determination of the mean
photon number has been extensively applied to (numerically simulated)
homodyne data for coherent states of various amplitudes. The analysis
has shown that the determination is reliable also for small samples
and that precision is not much affected by the intensity of the signal.

%% file: cap7.tex
\chapter{Tomography of a quantum device}
 If we want to determine experimentally the operation of 
a quantum device, we need, by definition, quantum tomography. In fact,
the characterization of the device operation could be done by 
running a {\em basis} of possible known inputs, and determining the
corresponding outputs by quantum tomography. In quantum mechanics the
inputs are density operators, and the role of the transfer matrix is
played by the so-called {\em quantum operation} of the device, here
denoted by $\cal E$. Thus the output state $\rho_{out}$ (a part from
a possible normalization) is given by the quantum operation applied to
the input state as follows 
\begin{equation}
\rho_{out}={\cal{E}}(\rho_{in}).
\end{equation}
Since the set of states $\rho$ actually belongs to a {\em space of
operators}, this means that if we want to characterize $\cal {E}$
completely, we need to run a complete orthogonal basis of quantum
states $|n\rangle$ ($n=0,1,2,\ldots$), along with their linear combinations
$\frac{1}{\sqrt2}(|n'\rangle+i^{k} |n''\rangle)$, with $k=0,1,2,3$ and $i$
denoting the imaginary unit. However, the availability of such a set
of states in the lab is, by itself, a very hard technological problem. For example,
for an optical device, the states $|n\rangle$ are those with a precise
number $n$ of photons, and, a part from very small $n$---say at most 
n=2---they have never been achieved in the lab, whereas preparing their 
superpositions remains a dream for experimentalists, especially
if $n\gg 1$ (a kind of {\em Schr\H{o}dinger kitten} states).
\par The idea of achieving the quantum operation of a device by
scanning the inputs and making tomography of the corresponding output
is the basis of the early methods proposed in Refs. \cite{nielsen,macca}. 
Due to the mentioned problems in the availability of input states,
both methods have limited application. The method of
Ref. \cite{nielsen} has been designed for NMR quantum processing,
whereas the method of Ref. \cite{macca} was conceived for determining
the Liouvillian of a phase-insensitive amplifier, namely for a case in
which the quantum operation has no off-diagonal matrix elements,
to evaluate which one needs the superpositions
$\frac{1}{\sqrt2}(|n'\rangle+i^{k} |n''\rangle)$ with $k=0,1,2,3$
mentioned above. The problem of availability of input spates and their
superpositions was partially solved by the method of
Ref. \cite{miolik}, where it was suggested to use randomly drawn 
coherent states to estimate the quantum operation of an optical device
via a maximum likelihood approach. This method, however, cannot be
used for quantum systems different from the em radiation---such as
finite dimensional systems, i.e. qubits---due to the peculiarity of
coherent states. The solution to the problem came with the recent
method of Ref. \cite{cptomo}, where the problem of the availability of
input states was solved by using a single bipartite entangled input,
which is equivalent to run all possible input states in a kind of
``quantum parallel'' fashion (bipartite entangled states are 
nowadays easily available in most quantum systems of interest).
The method is also very simple and effective, and its experimental
feasibility (for single-photon polarization-encoded qubits) has
been already demonstrated in a recent experiment performed in the
Francesco De Martini laboratory in Roma La Sapienza 
\cite{tomo_chan_exp}. In the next sections we will review the general
method and report some computer simulated results from Ref. \cite{cptomo}. 
\section{The method}
As already mentioned, the most description of a general
state-transformation in quantum mechanics is given in terms of the
so-called  {\em quantum operation}. The state transformation due to
the quantum operation ${\cal E}$ is given as follows
\begin{equation}
\rho \rightarrow 
\frac{{\cal E}(\rho)}{\mbox{Tr}\bigl({\cal E}(\rho)\bigr)}.\label{map}
\end{equation}
The transformation occurs with probability given by $p=\hbox{Tr}[{\cal
  E}(\rho)]\le 1$.  The quantum operation ${\cal E}$ is a linear,
trace-decreasing completely positive (CP) map.  We remind that a map
is completely positive if it preserves positivity generally when
applied locally to an entangled state. In other words, upon denoting
by ${\cal I}$ the identical map on the Hilbert space ${\cal K}$ of a
second quantum system, the extended map ${\cal E}\otimes{\cal I}$ on
${\cal H}\otimes{\cal K}$ is positive for any extension ${\cal K}$.  
Typically, the CP map is written using a Kraus decomposition
\cite{Kraus83a} as follows
\begin{eqnarray}
{\cal E}(\rho) = \sum_n K_n \rho K_n^{\dagger}\;,\label{kraus-mix} 
\end{eqnarray}
where the operators $K_n$ satisfy 
\begin{eqnarray}
\sum_n K_n^\dagger K_n\le I\;.\label{sum}
\end{eqnarray}
The transformation (\ref{kraus-mix}) occurs with generally non-unit
probability $\hbox{Tr}[{\cal E}(\rho)]\le 1$, and the probability is
unit independently on $\rho$ when ${\cal E}$ is trace-preserving,
i.e. when we have the equal sign in Eq. (\ref{sum}). The particular 
case of unitary transformations corresponds to having just one term $K_1 = U$
in the sum (\ref{kraus-mix}), with $U$ unitary. However, one can consider also
non-unitary operations with one term only, namely
\begin{eqnarray}
{\cal E}(\rho) = A \rho A^{\dagger}\;,\label{kraus-pure} 
\end{eqnarray}
where $A$ is a {\em contraction}, i. e. $||A||\le 1$.  Such operations
leave pure states as pure, and describes, for example, the state
reduction from a measurement apparatus for a particular fixed outcome
that occurs with probability $\mbox{Tr}[\rho A^{\dagger} A]\le 1$.
 
\par In the following we will use the notation for bipartite pure
states introduced in Eq. (\ref{iso}), and we will denote by $O^\tau 
$ and $O^*$ the transposed and the conjugate operator of $O$ with
respect to some pre-chosen orthonormal basis.

\par
\begin{figure}[hbt]
\begin{center}
\setlength{\unitlength}{800sp}%
\begingroup\makeatletter\ifx\SetFigFont\undefined%
\gdef\SetFigFont#1#2#3#4#5{%
  \reset@font\fontsize{#1}{#2pt}%
  \fontfamily{#3}\fontseries{#4}\fontshape{#5}%
  \selectfont}%
\fi\endgroup%
\begin{picture}(19506,5446)(1549,-5109)
\thicklines
\put(1801,-3961){\line( 1, 0){8400}}
\put(12601,-961){\vector( 2,-1){2160}}
\put(12601,-3961){\vector( 2, 1){2160}}
\put(1801,-961){\line( 1, 0){2700}}
\put(6901,-961){\line( 1, 0){3300}}
\put(14701,-3961){\framebox(6300,3000){}}
\put(15200,-2761){COMPUTER}
\put(1026,-2761){\Large $|\psi \rangle\! \rangle 
$}
\put(5701,-961){\oval(2372,2372)}\put(5400,-1261){\large ${\cal E}$}
\put(11401,-886){\oval(2372,2372)}\put(11000,-1000){$X_\lambda$}
\put(11401,-3886){\oval(2372,2372)}\put(11000,-4050){$X_\lambda'$}
\end{picture}
\end{center}
\caption{General scheme of the method for the tomographic estimation
of a quantum operation. Two identical quantum systems are prepared in
a bipartite state $|\psi\rangle \!\rangle $, with invertible $\psi
$. One of the two systems undergoes the quantum operation $\cal E$,
whereas the other is left untouched. At the output one performs a
quantum tomographic estimation, by measuring jointly two observables
$X_\lambda$ and $X'_\lambda$ from two quorums $\{X_\lambda\}$ and
$\{X'_\lambda\}$ for the two Hilbert spaces, such as two different
quadratures of the two field modes in a two-mode homodyne tomography 
(From Ref. \cite{cptomo}).}
\label{f:qoscheme}\end{figure}
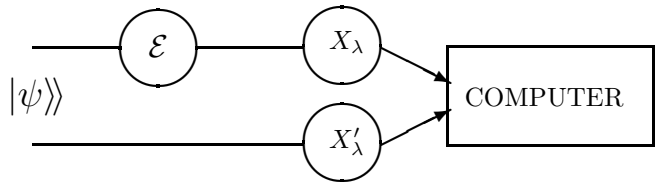

 The basic idea of the method in Ref. \cite{cptomo} is
the following. An unknown quantum operation $\cal E$ can be determined 
experimentally through quantum tomography, by exploiting the following 
one-to-one correspondence ${\cal E}\leftrightarrow R_{{\cal E}}$
between quantum operations $\cal E$ and positive operators $R_{\cal
E}$ on two copies of the Hilbert space ${\cal H}\otimes{\cal H}$
\begin{equation}
R_{{\cal E}}={\cal E}\otimes
{\cal I} (|I{\rangle\!\rangle}{\langle\!\langle} I|),\qquad
{\cal E}(\rho)=\hbox{Tr}_2[I\otimes{\rho ^\tau } R_{\cal E}]\;. 
\end{equation}
Notice that the vector $|I {\rangle\!\rangle}$ represents a
(unnormalized) maximally entangled state.  If we consider a bipartite
input state $|\psi{\rangle\!\rangle}$ and operate with $\cal E$ only
on one Hilbert space as in Fig. \ref{f:qoscheme}, the output state is
given by
\begin{equation}
R(\psi)\equiv {\cal E}\otimes {\cal I}
(|\psi{\rangle\!\rangle}{\langle\!\langle}\psi|). 
\end{equation}
For invertible $\psi $ the two matrices $R(I)\equiv R_{\cal E}$ and
$R(\psi)$ are related  as follows
\begin{equation}
R(I)=(I\otimes {\psi^{-1}}^\tau R(\psi)(I\otimes\psi^{-1*})\;.
\end{equation}
Hence, the (four-index) quantum operation  matrix $R_{\cal E}$ can be
obtained by estimating via quantum tomography the
following ensemble averages
\begin{equation}
{\langle\!\langle} i,j|R(I)|l,k{\rangle\!\rangle}
=\hbox{Tr}\left [R(\psi ) \left (
|l\rangle\langle i|\otimes
\psi^{-1*}|k \rangle\langle j|\psi^{-1*} \right )\right ]\;.\label{ensav}
\end{equation}
Then one simply has to perform a quantum tomographic estimation, by
measuring jointly two observables $X_\lambda$ and $X'_\lambda$ from
two quorums $\{X_\lambda\}$ and $\{X'_\lambda\}$ for
 the two entangled quantum systems.
\section{An example in the optical domain}

In Ref. \cite{cptomo} it is shown that the proposed method for quantum
tomography of a device can be actually performed using joint homodyne
tomography on a twin-beam from downconversion of vacuum, with an
experimental setup similar to that used in the experiment in
Ref. \cite{kumarandme}. The feasibility analysis considers, as an
example, the experimental determination of the quantum operation
corresponding to the unitary displacement operator 
$D(z)=e^{za^{\dag}- z^*a}$. The pertaining matrix $R(I)$ 
is given by 
\begin{eqnarray}
R(I)= |D(z) {\rangle\!\rangle} {\langle\!\langle}D(z)|\;, \label{}
\end{eqnarray}
which is the (unnormalizable) eigenstate of the operator $a-b^\dag $
with eigenvalue $z$, as shown in Sec. \ref{hetsec}. As an input
bipartite state, one uses the twin-beam from parametric
downconversion of Eq. (\ref{Psi}), which is clearly invertible, since
\begin{eqnarray}
\psi =\sqrt{1-|\xi |^2} \,\xi ^{a^\dag a}\;, \qquad 
\psi ^{-1}=\frac {1}{\sqrt{1-|\xi |^2)}}\, \xi ^{-a^\dag a}\;.
\end{eqnarray}
\par The experimental apparatus is the same as in the experiment of
 Ref. \cite{kumarandme}, where the twin-beam is provided by a
non-degenerate optical parametric amplifier (a KTP crystal) pumped by
the second harmonic of a Q-switched mode-locked Nd:YAG laser, which
produces a 100-MHz train of 120-ps duration pulses at 1064 nm. The
orthogonally polarized twin beams emitted by the KTP crystal (one of
which is displaced of $D(z)$ by a nearly transparent beam splitter
with a strong local oscillator) are separately detected by two
balanced homodyne detectors that use two independent local oscillators
derived from the same laser. This provides the joint tomography of
quadratures $X_{\phi'}\otimes X_{\phi''}$ needed for the
reconstruction. The only experimental problem which still need to be
addressed (even though is practically solvable) with respect to the
original experiment of Ref. \cite{kumarandme} is the control of the
quadrature phases $\phi'$ and $\phi''$ with respect to the LO, which
in the original experiment were random. 

\par
\begin{figure}[hbt]\begin{center}
\epsfxsize=.4\hsize\leavevmode\epsffile{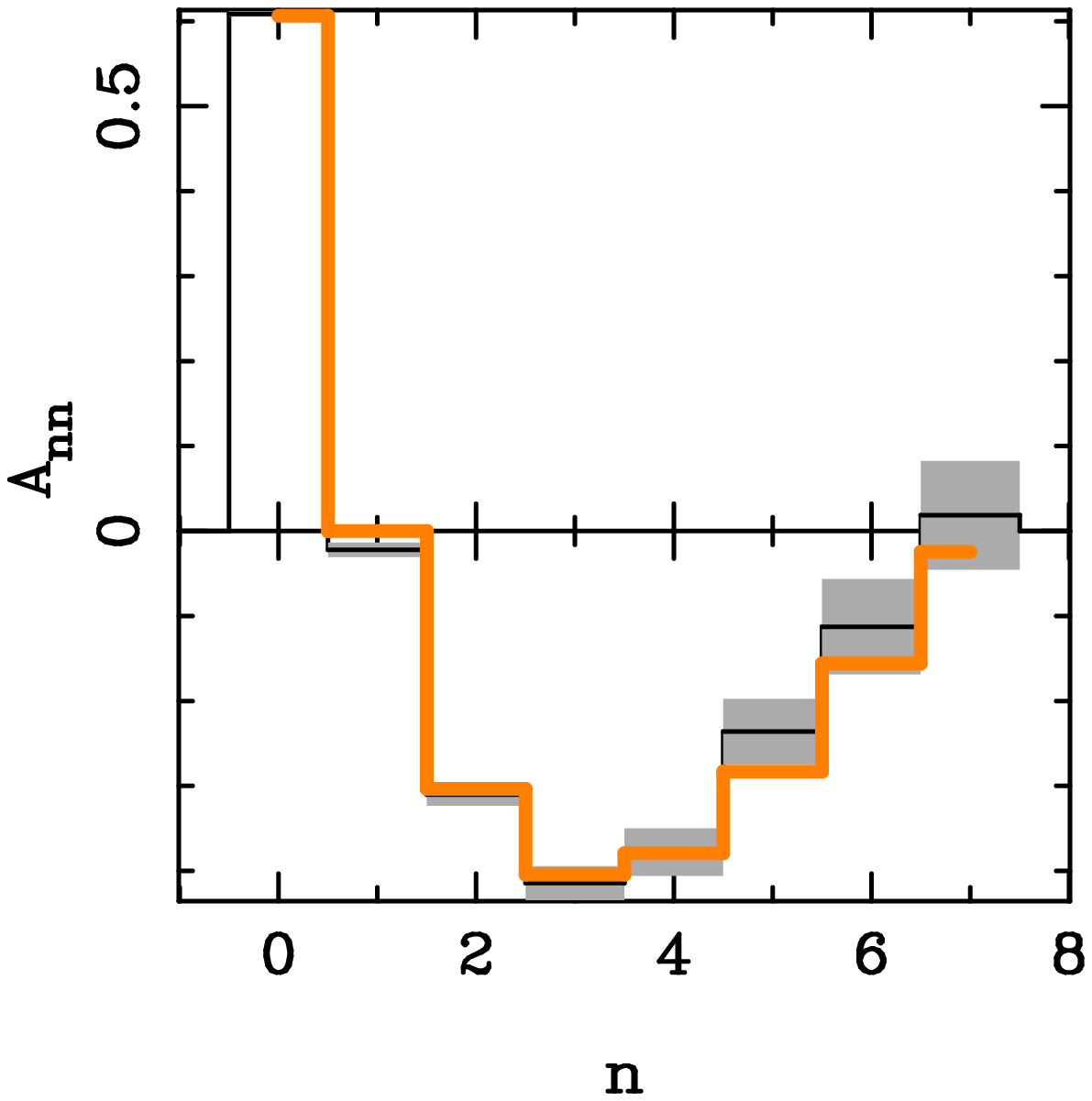}\hfil
\epsfxsize=.4\hsize\leavevmode\epsffile{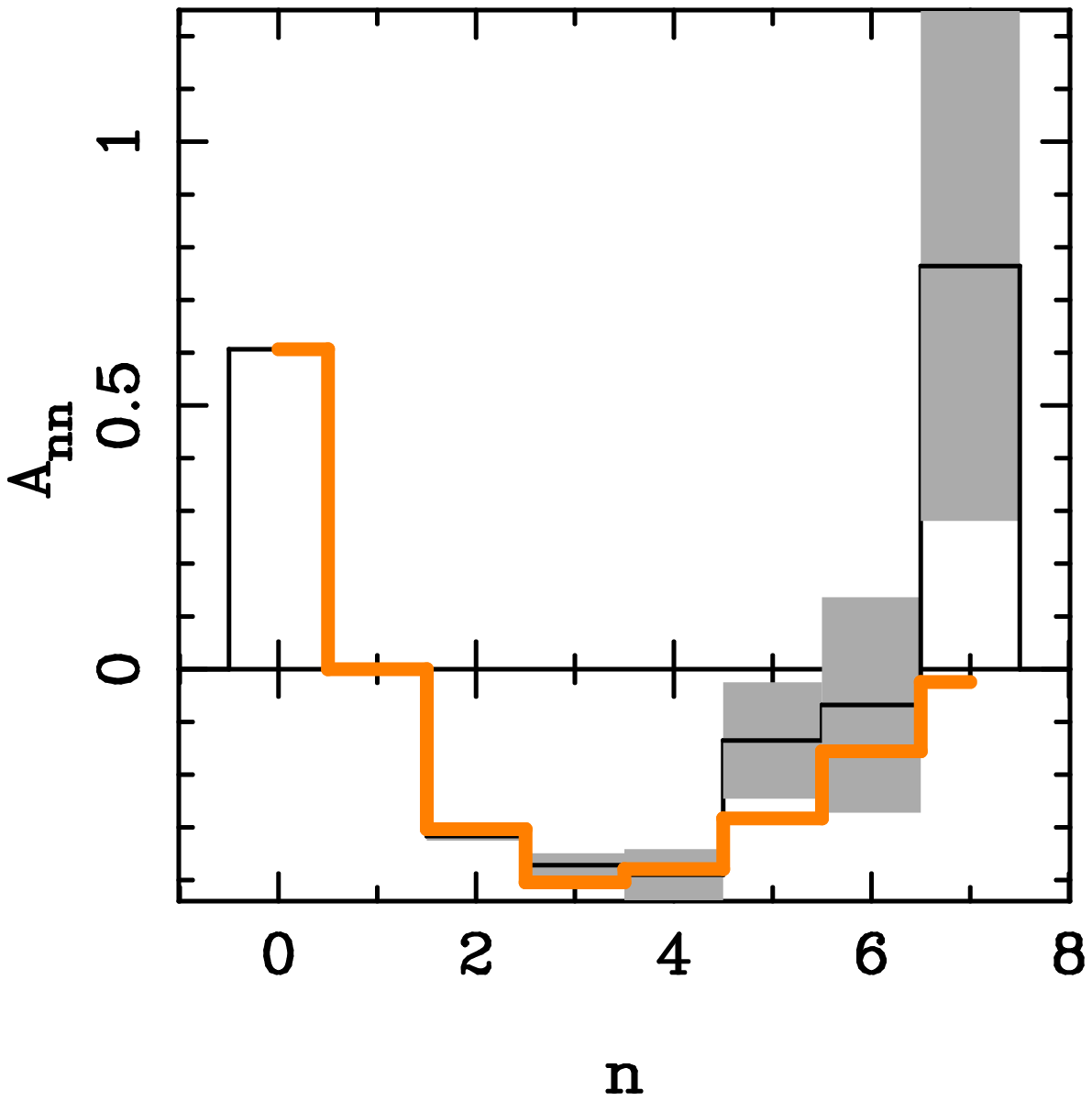}
\caption{Homodyne tomography of the quantum operation corresponding to
the unitary displacement operator $D(z)$, with $z=1$.  The
reconstructed diagonal elements $A_{nn}= \langle n|D(z)| n \rangle $
are shown (thin solid line on an extended abscissa range, with their
respective error bars in gray shade), compared to the theoretical
value (thick solid line). Similar results are obtained for
off-diagonal terms.  The reconstruction has been achieved using at the
input the twin beam state of Eq. (\ref{Psi}), with total average
photon number $\bar n$ and quantum efficiency at homodyne detectors
$\eta $. Left: $\bar n=5$, $\eta=0.9$, and $150$ blocks of $10^4$ data
have been used. Right: $\bar n=3$, $\eta=0.7$, and $300$ blocks of
$2\cdot 10^5$ data have been used (From  Ref. \cite{cptomo}).
\label{simul}}
\end{center}
\end{figure}

 In Fig. \ref{simul} the results of a simulated experiment are
reported, for displacement parameter $z=1$, and for some typical
values of the quantum efficiency $\eta$ at homodyne detectors and of
the total average photon number $\bar n$ of the twin beam.  The
diagonal elements $A_{nn}= \langle n|D(z)| n \rangle = [\langle 
n|\langle n |R_{D(z)}| n \rangle n \rangle]^{1/2}$ are
plotted for the displacement operator with $z=1$. 
The reconstructed values are shown by thin solid
line on an extended abscissa range, with their respective error bars
in gray shade, and compared to the theoretical probability (thick
solid line).  A good reconstruction of the matrix can be achieved in
the given range with $\bar n\sim 1$, quantum efficiency as low as
$\eta=0.7$, and $10^6\div10^7$ data. The number of data can be
decreased by a factor $100-1000$ using the tomographic max-likelihood
techniques of Ref. \cite{maxlik}, however at the expense of the
complexity of the algorithm.  Improving quantum efficiency and
increasing the amplifier gain (toward a maximally entangled state)
have the effect of making statistical errors smaller and more uniform
versus the photon labels $n$ and $m$ of the matrix $A_{nm}$.

\par It is worth emphasizing that the quantum tomographic method of
Ref. \cite{cptomo} for measuring the matrix of a quantum operation can
be much improved 
by means of a max-likelihood strategy aimed at the estimation of some
unknown parameters of the quantum operation. In this case, instead of
obtaining the matrix elements of $R(I)$ from the ensemble averages in
Eq. (\ref{ensav}), one parametrizes $R(I)$ in terms of unknown
quantities to be experimentally determined, and the likelihood is
maximized for the set of experimental data at various randomly
selected (tensor) quorum elements, keeping the same fixed bipartite
input state. This method is especially useful for a very precise
experimental comparison between the characteristics of a given device
(e.g. the gain and loss of an active fiber) with those of a {\em
quantum standard} reference.

%% file: cap8.tex
\chapter{Maximum-likelihood method in quantum estimation}
Quantum estimation of states, observables and parameters is, from very
basic principles, matter of statistical inference from a population
sampling, and the most comprehensive quantum estimation procedure is 
quantum tomography. As we have shown in Chapter 3, the expectation
value of an operator is obtained by averaging an estimator  over
the experimental data of a ``quorum'' of observables. The method is
very general and efficient, however, in the averaging procedure, we
have fluctuations which result in relatively large statistical errors.

\par Another relevant strategy, the maximum-likelihood (ML) method,
can be used for measuring unknown parameters of transformation on a
given state \cite{parlik}, or for measuring the matrix elements of the
density operator itself \cite{maxlik}. The ML strategy
\cite{MaxLik,BanaszekPRA98} is an entirely
different approach to quantum state measurement compared to the
standard quantum-tomographic techniques.
The ML procedure consists in finding the quantum state, or the value of
the parameters, that are most likely to generate the observed data.
This idea can be quantified and implemented using the concept of the
likelihood functional.

\par As regards state estimation, the ML method estimates the quantum
state as a whole. Such a procedure incorporates {\em a priori}
knowledge about relations between elements of the density matrix. This
guarantees positivity and normalization of matrix, with the result of
a substantial reduction of statistical errors. Regarding the
estimation of specific parameters, we notice that in many cases the
resulting estimators are efficient, unbiased and consistent, thus
providing a statistically reliable determination.

\par As we will show, by using the ML method only small samples of
data are required for a precise determination. However, we want to
emphasize that such method is not always the optimal solution of the
tomographic problem, since it suffers from some major
limitations. Besides being biased due to the Hilbert space
truncation---even though the bias can be very small if, from other
methods, we know where to truncate---it cannot be generalized
to the estimation of any ensemble average, but just of a set of
parameters from which the density matrix depends. In addition, for
increasing number of parameters the method has exponential complexity.

\par In the following we will review the ML methods
proposed in Refs. \cite{maxlik} and \cite{parlik}, by deriving the
likelihood functional, and applying the ML method to the quantum state
reconstruction, with examples for both radiation and spin
systems, and, finally, considering the ML estimation for the
relevant class of Gaussian states in quantum optics. 

\section{Maximum likelihood principle} 
Here we briefly review the theory of the maximum-likelihood (ML)
estimation of a single parameter. The generalization to several
parameters, as for example the elements of the density matrix, is
straightforward. The only point that should be carefully analyzed is
the parameterization of the multidimensional quantity to be
estimated. In the next section the specific case of the density matrix
will be discussed.

\par Let $p(x | \lambda)$ the probability density of a random variable
$x$, conditioned to the value of the parameter $\lambda$.  The form of
$p$ is known, but the true value of $\lambda$ is unknown, and will be
estimated from the result of a measurement of $x$.  Let $x_1, x_2,
..., x_N$ be a random sample of size $N$. The joint probability
density of the independent random variable $x_1, x_2, ..., x_N$ (the
global probability of the sample) is given by
\begin{eqnarray}
{\cal L}(x_1, x_2, ..., x_N| \lambda)= \Pi_{k=1}^N  p(x_k |\lambda)
\label{likdef}\;,
\end{eqnarray}
and is called the likelihood function of the given data sample
(hereafter we will suppress the dependence of ${\cal L}$ on the
data). The maximum-likelihood estimator (MLE) of the parameter
$\lambda$ is defined as the quantity $\lambda_{ml} \equiv
\lambda_{ ml} (\{x_k\})$ that maximizes ${\cal L} (\lambda)$ for
variations of $\lambda$, namely $\lambda_{ml}$ is given by the
solution of the equations
\begin{eqnarray}
\frac{\partial {\cal L} (\lambda) } {\partial \lambda} = 0 \; ; \quad
\frac{\partial^2 {\cal L} (\lambda)} {\partial \lambda^2} < 0
\label{maxlikdef}\;.
\end{eqnarray}
The first equation is equivalent to $\partial L/ \partial \lambda = 0
$ where
\begin{eqnarray}
L(\lambda) = \log {\cal L} (\lambda) = \sum_{k=1}^N \log p(x_k |
\lambda)
\label{loglikfun}\;
\end{eqnarray}
is the so-called log-likelihood function. 

\par In order to obtain a measure for the confidence interval in the
determination of $\lambda_{ml}$ we consider the variance
\begin{eqnarray}
\sigma^2_\lambda = \int \left[\prod_k dx_k \, p(x_k|\lambda)\right] 
\left[\lambda_{ml} 
(\{x_k\})- \lambda \right]^2  \label{varMLEdef}\;.
\end{eqnarray}
In terms of the Fisher information
\begin{eqnarray}
F= \int dx \left[ \frac{\partial p(x |\lambda)}{\partial \lambda}\right]^2
\frac1{p(x | \lambda)}
\label{FisherDef}\;,
\end{eqnarray}
it is easy to prove that 
\begin{eqnarray}
\sigma^2_\lambda \geq \frac{1}{N F} \;,\label{manca}
\end{eqnarray}
where $N$ is the number of measurements. The inequality in
Eq. (\ref{manca}) is known as the Cram\'er-Rao bound \cite{Cramer} on
the precision of the ML estimation. Notice that this bound holds for
any functional form of the probability distribution $p(x|\lambda)$,
provided that the Fisher information exists $\forall \lambda$ and
$\partial_\lambda p(x|\lambda)$ exists $\forall x$. When an experiment
has "good statistics" (i.e. for a large enough data sample) the
Cram\'er-Rao bound is saturated.
\section{ML quantum state estimation}
 In this section we review the method of the maximum likelihood
estimation of the quantum state of Ref. \cite{maxlik}, focusing
attention to the cases of homodyne and spin tomography.

We consider an experiment consisting of $N$ measurements performed on
identically prepared copies of a given quantum system. Each
measurement is described by a positive operator-valued measure
(POVM). The outcome of the $i$th measurement corresponds to the
realization of a specific element of the POVM used in the
corresponding run, and we denote this element by $ \Pi_i$.  The
likelihood is here a functional of the density matrix ${\cal L}( {\rho
})$ and is given by the product
\begin{equation} {\cal
L}( {\rho }) = \prod_{i=1}^{N} \hbox{Tr}( {\rho }
 \Pi_i)\;, 
\end{equation} 
which represents the probability of the observed data.  The unknown
element of the above expression, which we want to infer from data, is
the density matrix describing the measured ensemble. The estimation
strategy of the ML technique is to maximize the likelihood functional
over the set of the density matrices.  Several properties of the
likelihood functional are easily found, if we restrict ourselves to
finite dimensional Hilbert spaces.  In this case, it can be easily
proved that ${\cal L}( {\rho })$ is a concave function defined on a
convex and closed set of density matrices.  Therefore, its maximum is
achieved either on a single isolated point, or on a convex subset of
density matrices. In the latter case, the experimental data are
insufficient to provide a unique estimate for the density matrix using
the ML strategy.  On the other hand, the existence of a single maximum
allows us to assign unambiguously the ML estimate for the density
matrix.  

The ML estimation of the quantum state, despite its elegant general
formulation, results in a highly nontrivial constrained optimization
problem, even if we resort to purely numerical means. The main
difficulty lies in the appropriate parameterization of the set of all
density matrices. The parameter space should be of the minimum dimension
in order to preserve the maximum of the likelihood function as a
single isolated point. Additionally, the expression of quantum
expectation values in terms of this parameterization should enable
fast evaluation of the likelihood function, as this step is performed
many times in the course of numerical maximization.

For such purpose one introduces \cite{maxlik} a parameterization of
the set of density matrices which provides an efficient algorithm for
maximization of the likelihood function.  We represent the density
matrix in the form
\begin{equation}
\label{Eq:rhoTT}
 {\rho } =  {T}^{\dagger}  {T}\;,
\end{equation}
which automatically guarantees that $ {\rho }$ is positive and
Hermitian. The remaining condition of unit trace $\hbox{Tr} {\rho } =
1$ will be taken into account using the method of Lagrange
multipliers. In order to achieve the minimal parameterization, we
assume that $ {T}$ is a complex lower triangular matrix, with real
elements on the diagonal. This form of $ {T}$ is motivated by the
Cholesky decomposition known in numerical analysis \cite{Cholesky} for
arbitrary non negative Hermitian matrix.  For an $M$-dimensional
Hilbert space, the number of real parameters in the matrix $ {T}$ is
$M+2M(M-1)/2=M^2$, which equals the number of independent real
parameters for a Hermitian matrix. This confirms that such
parameterization is minimal, up to the unit trace condition.

In numerical calculations, it is convenient to replace the likelihood
functional by its natural logarithm, which of course does not change
the location of the maximum. Thus the log-likelihood function
subjected to numerical maximization is given by
\begin{equation}
\label{eq:lt}
L( {T}) = \sum_{i=1}^{N} \ln \hbox{Tr}( {T}^\dagger  {T}
 \Pi _i) - \lambda \hbox{Tr}( {T}^\dagger  {T})\;,
\label{loglik}
\end{equation}
where $\lambda$ is a Lagrange multiplier accounting for normalization
of $ \rho $. Writing $\rho $ in terms of its eigenvectors $| \psi_\mu
\rangle $ as $\rho = \sum_{\mu } y_{\mu}^{2} | \psi_\mu \rangle
\langle \psi_\mu | $, with real $y_{\mu}$, the maximum likelihood
condition $\partial L/\partial y_{\nu} = 0$ reads 
\begin{eqnarray}
\lambda y_{\nu} =
\sum_{i=1}^{N} [y_\nu \langle \psi_\nu | \Pi _i | \psi_\nu
\rangle/ \hbox{Tr}(\rho \Pi _i)]
\;,
\end{eqnarray}
which, after
multiplication by $y_{\nu}$ and summation over $\nu $, yields $\lambda
= N$. The Lagrange multiplier then equals the total number of
measurements $N$.

\par This formulation of the maximization problem allows one to apply
standard numerical procedures for searching the maximum over the $M^2$
real parameters of the matrix $ {T}$. The examples presented below use
the downhill simplex method \cite{Ameba}.

\par The first example is the ML estimation of a single-mode radiation
field. The experimental apparatus used in this technique is the
homodyne detector.  According to Sec. \ref{hetsec} the homodyne
measurement is described by the positive operator-valued measure \par
\begin{equation} 
\label{Eq:Hxphi}
 {\cal H}(x;
\varphi) = \sqrt{\frac{2\eta }{\pi(1-\eta)}}
 \exp \left[ - \frac{2\eta }{1-\eta }(X_{\varphi}-x)^2\right]\;,
\end{equation}
where $\eta $ is the detector efficiency, and $ X_\varphi =(a^\dag
\,e^{i\varphi}+a\,e^{-i\varphi})/2$ is the quadrature operator at  phase
$\varphi $.

\begin{figure}[h] \begin{center}
\epsfxsize=.5\hsize\leavevmode\epsffile{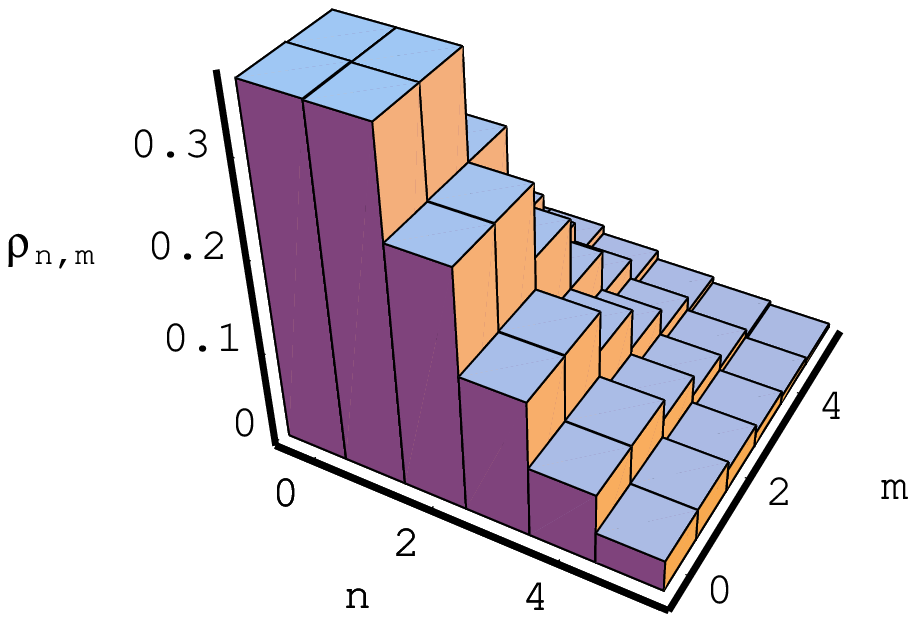}\hfill
\epsfxsize=.5\hsize\leavevmode\epsffile{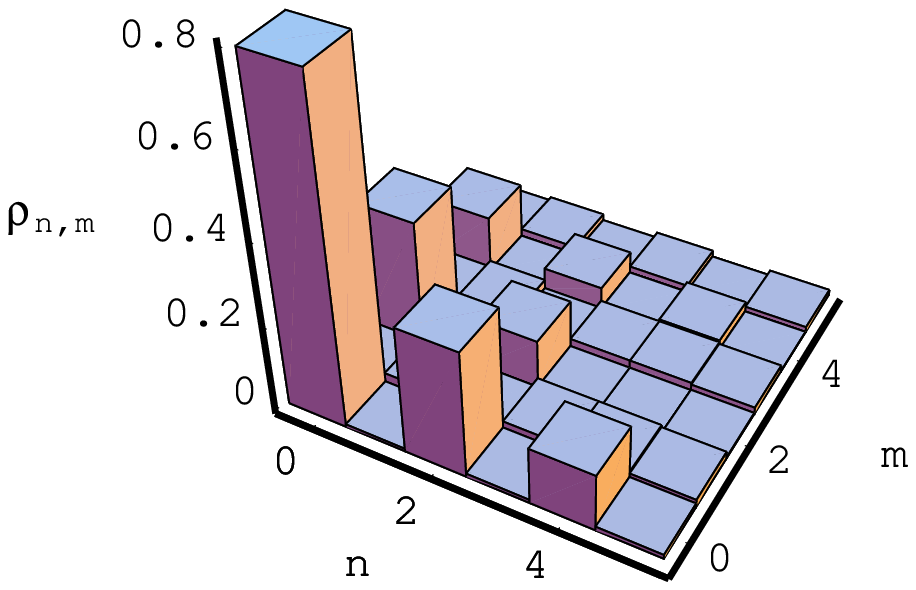}
\end{center} 
\caption{Reconstruction of the density matrix of a single-mode
radiation field by the ML method. The plot shows the matrix elements
of a coherent state (left) with $\langle  a^{\dag } a \rangle
=1$ photon, and for a squeezed vacuum (right) with $\langle  
a^{\dag } a \rangle =0.5$ photon. A sample of 50000 simulated
homodyne data for quantum efficiency $\eta=80\%$ has been
used (From Ref. \cite{maxlik}).}\label{Fig:QHT}
\end{figure}

\par After $N$ measurements, we obtain a set of pairs $(x_i; \varphi_i)$,
where $i=1,\ldots, N$. The log-likelihood functional is given by
Eq. (\ref{loglik}) with $ \Pi_i\equiv {\cal H}(x_i; \varphi_i)$.  Of
course, for a light mode it is necessary to truncate the Hilbert space
to a finite dimensional basis. We shall assume that the highest Fock
state has $M-1$ photons, i.e.\ that the dimension of the truncated
Hilbert space is $M$.  For the expectation $\mbox{Tr}[ {T}^{ \dagger}
{T} {\cal H}(x;\varphi)]$ it is necessary to use an expression which
is explicitly positive, in order to protect the algorithm against
occurrence of small negative numerical arguments of the logarithm
function. A simple derivation yields
\begin{eqnarray}
\mbox{Tr}[ {T}^{\dagger} {T} {\cal H}(x; \varphi)] =
\sqrt{\eta }\sum_{k=0}^{M-1}\sum_{j=0}^{k} \left|\sum_{n=0}^{k-j} \langle k |
 {T} | n+j \rangle B_{n+j,n} \langle n | \sqrt{\eta } x\rangle
e^{in\varphi}\right|^2 \;,
\end{eqnarray}
where 
\begin{eqnarray}
B_{n+j,n} = \left[{{n+j} \choose n} \eta^{n}
(1-\eta)^{j}\right]^{1/2} \;,
\end{eqnarray}
and 
\begin{eqnarray}
\langle n | x \rangle = \left(\frac 2 \pi 
\right )^{1/4}\frac{1}{\sqrt{2^n n! }} H_n(\sqrt 2 \,x)\,
\exp(-x^2) 
\;
\end{eqnarray} 
are the eigenstates of the
harmonic oscillator in the position representation---$H_n (x)$ being
the $n$th Hermite polynomial.

\par The ML technique can be applied to reconstruct the density
matrix in the Fock basis from Monte Carlo simulated homodyne
statistics.  Fig.~\ref{Fig:QHT} depicts the matrix elements of the
density operator as obtained for a coherent state and a squeezed
vacuum, respectively. Remarkably, only 50000 homodyne data have been
used for quantum efficiency $\eta=80\%$. We recall that in quantum
homodyne tomography the statistical errors are known to grow rapidly
with decreasing efficiency $\eta$ of the detector
\cite{added,nico}. In contrast, the elements of the density matrix
reconstructed using the ML approach remain bounded, as the whole
matrix must satisfy positivity and normalization constraints. This
results in much smaller statistical errors. As a comparison one could
see that the same precision of the reconstructions in
Fig.~\ref{Fig:QHT} could be achieved using $10^7$--$10^8$ data samples
with conventional quantum tomography. On the other hand, in order to
find numerically the ML estimate we need to set {\em a priori} the
cut-off parameter for the photon number, and its value is limited by
increasing computation time.

\begin{figure}[hbt]
\begin{center}
\epsfxsize=.5\hsize\leavevmode\epsffile{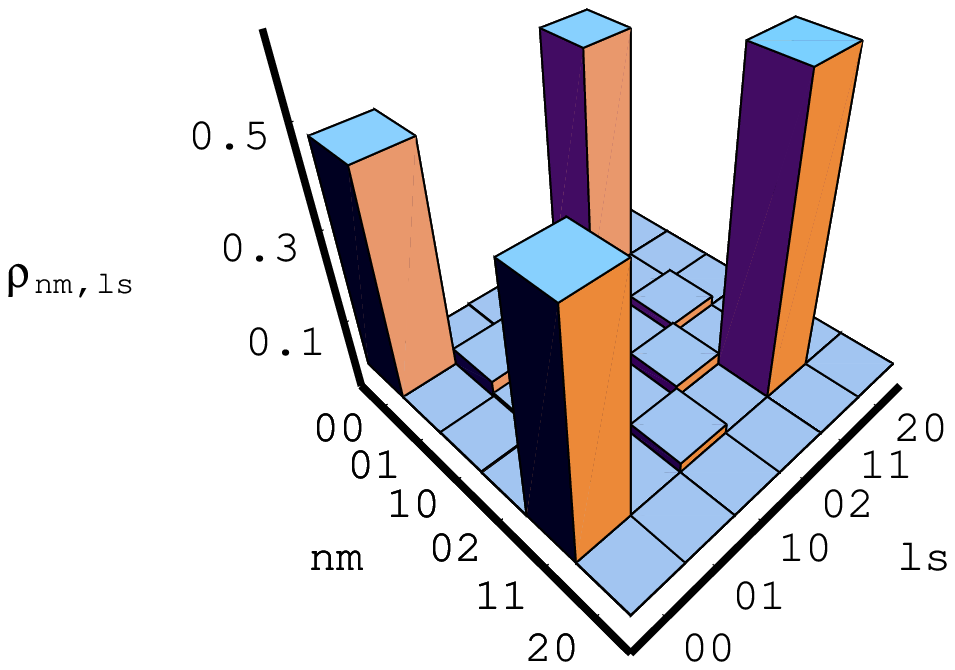}\hfill
\epsfxsize=.5\hsize\leavevmode\epsffile{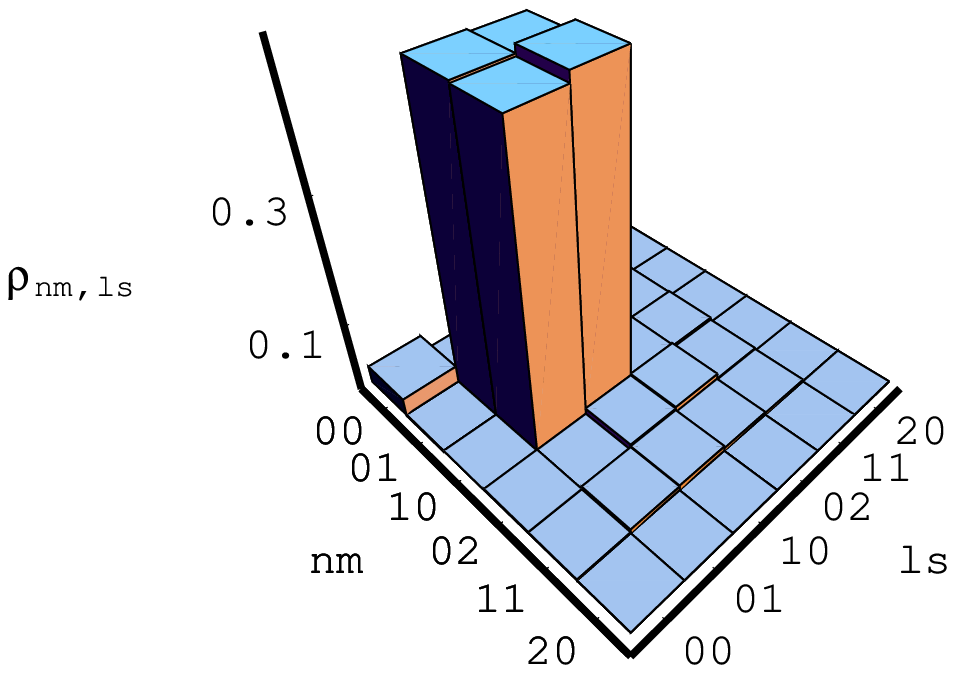}
\caption{ML reconstruction of the density matrix of a two-mode
radiation field. On the left the matrix elements obtained for the
state $|\Psi _1\rangle =(|00 \rangle + |11 \rangle )/\sqrt 2$; on the
right for $|\Psi _2\rangle =(|01 \rangle + |10 \rangle )/\sqrt 2$. 
For $|\Psi_1 \rangle $ we used
100000 simulated homodyne data and $\eta =80\%$; for $|\Psi_2 \rangle
$ we used 20000 data and $\eta =90\%$
 (From Ref. \cite{maxlik}).}\label{Fig:TwoMode}
\end{center}
\end{figure}

\par Another relevant example is the reconstruction of the quantum
state of two-mode field using single-LO homodyning of Chapter 5.
Here, the full joint density matrix can be measured by scanning the
quadratures of all possible linear combinations of modes. For two
modes the measured quadrature operator is given by 
\begin{eqnarray}
X(\theta,\psi_0,\psi_1) = \frac 12 ( {a} e^{-i\psi_0}\cos\theta + {b}
e^{-i\psi_1} \sin\theta + \mbox{h.c.} )
\;,
\end{eqnarray}
where $(\theta,\psi_0,\psi_1)\in S^2 \times [0,2\pi]$, $S^2$ being the
Poincar\'e sphere and one phase ranging between $0$ and $2\pi$. In
each run these parameters are chosen randomly. The POVM describing the
measurement is given by the right-hand side of Eq.~(\ref{Eq:Hxphi}),
with $ X_\varphi$ replaced by $X(\theta,\psi_0,\psi_1)$. 
 An experiment for the two orthogonal states $|\Psi 
_1\rangle =(|00 \rangle + |11 \rangle)/\sqrt 2$ and $|\Psi _2\rangle
=(|01 \rangle + |10 \rangle )/\sqrt 2$ has been simulated, in order to
reconstruct the density matrix in the two-mode Fock basis using the ML
technique.  The results are reported in Fig.~\ref{Fig:TwoMode}.

\par
\begin{figure}[hbt]
\begin{center}
\epsfxsize=.5\hsize\leavevmode\epsffile{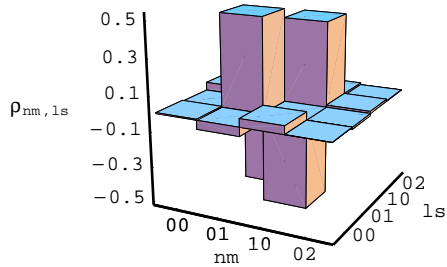}
\caption{ML reconstruction of the density matrix of a pair of spin-1/2
particles in the singlet state. 
The particles are shared by two parties. In each run, the
parties select randomly and independently from each other a direction
along which they perform spin measurement. 
The matrix elements has been obtained by a sample of 500 simulated
data (From Ref. \cite{maxlik}).}
\label{Fig:singlet} 
\end{center}
\end{figure}

 The ML procedure can also be applied for reconstructing the
density matrix of spin systems. For example, let us consider $N$
repeated preparations of a pair of spin-1/2 particles. The particles
are shared by two parties. In each run, the parties select randomly
and independently from each other a direction along which they perform
a spin measurement.  The obtained result is described by the joint
projection operator (spin coherent states \cite{arecchi}) $ {\cal F}_i
= |\Omega^A_{i}, \Omega^B_{i} \rangle \langle \Omega^A_{i},
\Omega^B_{i}|$, where $\Omega^A_{i}$ and $\Omega^B_{i}$ are the
vectors on the Bloch sphere corresponding to the outcomes of the $i$th
run, and the indices $A$ and $B$ refer to the two particles. As in the
previous examples, it is convenient to use an expression for the
quantum expectation value $\mbox{Tr}( {T}^{ \dagger} {T} {\cal F}_i$)
which is explicitly positive.  The suitable form is
\begin{equation}
\mbox{Tr}( {T}^{\dagger}  {T} {\cal F}_i)
= \sum_\mu |\langle \mu |  {T} | \Omega^A_{i}, \Omega^B_{i} \rangle
|^2\;,
\end{equation}
where $|\mu\rangle$ is an orthonormal basis in the Hilbert space of
the two particles. The result of a simulated experiment with only 500
data for the reconstruction of the density matrix of the singlet state
is shown in Fig. \ref{Fig:singlet}.

\par Summarizing, the ML technique can be used to estimate the density
matrix of a quantum system.  With respect to conventional quantum
tomography this method has the great advantage of needing much smaller
experimental samples, making experiments with low data rates feasible,
however with a truncation of the Hilbert space dimension. We have
shown that the method is general and the algorithm has solid
methodological background, its reliability being confirmed in a number
of Monte Carlo simulations. However, for increasing dimension of
Hilbert spaces the method has exponential complexity.

\section{Gaussian-state estimation}\label{s:gauss} 
 In this Section we review the ML determination method of
Ref. \cite{parlik} for the parameters of Gaussian states. Such states
represent the wide class of coherent, squeezed and thermal states, all
of them being characterized by a Gaussian Wigner function.  Apart from
an irrelevant phase, we consider Wigner functions of the form
\begin{eqnarray}
W(x,y)=\frac{2\Delta ^2}{\pi}\exp\left\{-2\Delta ^2\left[
e^{-2r}(x-\hbox{Re}\mu)^2
+e^{2r}(y-\hbox{Im}\mu)^2\right]\right\}\;,\label{wxy}
\end{eqnarray}
and the ML technique with homodyne detection is applied
to estimate the four real parameters 
$\Delta , r, \hbox{Re}\mu$ and $\hbox{Im}\mu$. 
The four parameters provide the number of thermal, squeezing and 
coherent-signal photons in the quantum state as follows
\begin{eqnarray}
& & n_{th}=\frac 12\left(\frac {1}{\Delta ^2}-1\right)\;,
\nonumber \\
&&n_{sq}=\sinh^2 r\;, \nonumber \\& & n_{coh}=|\mu|^2\;.
\end{eqnarray}
The density matrix $\rho $ corresponding to the Wigner function in
Eq. (\ref{wxy}) writes
\begin{eqnarray}
\rho  =D(\mu)\,S(r)\,\frac
{1}{n_{th}+1}\left(\frac{n_{th}}{n_{th}+1}\right)^{a^\dag a}
\,S^\dag (r)\,D^\dag (\mu)\;,
\end{eqnarray}
where $S(r)=\exp[r(a^2-a^{\dag 2})/2]$ and $D(\mu)=\exp(\mu a^\dag -\mu
^*a)$ denote the squeezing and displacement operators, respectively. 
\par The theoretical homodyne probability distribution at phase $\varphi
$ with respect to the local oscillator can be evaluated using
Eq. (\ref{margw}), and is given by the Gaussian 
\begin{eqnarray}
p(x,\varphi)&=&\sqrt{\frac{2 \Delta ^2}{\pi(e^{2r}\cos^2\varphi
+e^{-2r}\sin^2\varphi)}}\nonumber \\&\times &  
\exp\left\{-\frac{2 \Delta ^2}{e^{2r}\cos^2\varphi
+e^{-2r}\sin^2\varphi}\left[x-\hbox{Re}(\mu\,e^{-i\varphi})\right]^2\right\}
\;.\label{pxfi}
\end{eqnarray}
The log-likelihood function (\ref{loglikfun}) for a set of $N$
homodyne outcomes $x_i$ at random phase $\varphi _i$ then writes as follows
\begin{eqnarray}
L&=&\sum_{i=1}^N \frac 12 \log \frac{2\Delta^2}{\pi 
(e^{2r}\cos^2 \varphi_i
+e^{-2r}\sin^2\varphi_i)} \nonumber \\&- & 
\frac{2 \Delta ^2}{e^{2r}\cos^2\varphi_i
+e^{-2r}\sin^2\varphi_i}\left[x_i-\hbox{Re}(\mu\,e^{-i\varphi_i})\right]^2
\;.\label{lgau}
\end{eqnarray}
The ML estimators $\Delta _{ml}, r_{ml}, \hbox{Re}\mu _{ml}$ and
$\hbox{Im}\mu_{ml}$ are found upon maximizing Eq. (\ref{lgau}) versus
$\Delta, r, \hbox{Re}\mu$ and $\hbox{Im}\mu$.  \par
\begin{figure}[h]
\begin{center}
\epsfxsize=.45\hsize\leavevmode\epsffile{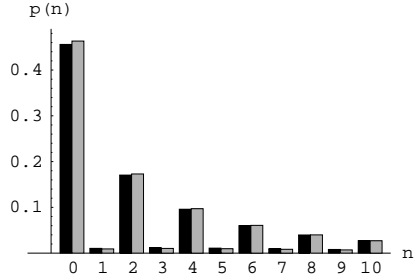}
\end{center} 
\caption{Photon-number probability of a squeezed-thermal state
(thermal photons $n_{th}=0.1$, squeezing photons $n_{sq}=3$). Compare
the reconstructed probabilities by means of the maximum likelihood
method and homodyne detection (gray histogram) with the theoretical
values (black histogram). Number of data samples $N=50000$, quantum
efficiency $\eta =80\%$. The statistical error affects the third
decimal digit, and it is not visible on the scale of the plot (From
Ref. \cite{parlik}).}
\label{f:sqth}
\end{figure}

 In order to
evaluate globally the state reconstruction, one considers 
the normalized overlap $\cal O$ between the theoretical and
the estimated state
\begin{eqnarray}
{\cal O}=\frac{\hbox{Tr}[\rho  \,\rho  _{ml}]}{\sqrt
{\hbox{Tr}[\rho  ^2]\,\hbox{Tr}[\rho  _{ml} ^2]}}\;.
\end{eqnarray}
Notice that ${\cal O}=1$ iff $\rho =\rho _{ml}$. Through Monte-Carlo
simulations, one always finds a value around unity, typically with
statistical fluctuations over the third digit, for number of data
samples $N=50000$, quantum efficiency at homodyne detectors
$\eta=80\%$, and state parameters with the following ranges:
$n_{th}<3$, $n_{coh}<5$, and $n_{sq}<3$. Also with such a small number
of data samples, the quality of the state reconstruction is so good
that other physical quantities that are theoretically evaluated from
the experimental values of $\Delta _{ml}, r_{ml}, \hbox{Re}\mu _{ml}$
and $\hbox{Im}\mu_{ml}$ are inferred very precisely.  For example, in
Ref. \cite{parlik} the photon number probability of a squeezed thermal
state has been evaluated, which is given by the integral
\begin{eqnarray}
\langle n|\rho  |n\rangle =\int _{0}^{2\pi}
\frac {d\phi}{2\pi} \frac{[C(\phi,n_{th},r)-1]^n}
{C(\phi,n_{th},r)^{n+1}}\;,
\end{eqnarray}
with $C(\phi,n_{th},r)=(n_{th}+\frac
12)(e^{-2r}\sin^2\phi+e^{2r}\cos^2\phi)+\frac 12$.  The comparison
of the theoretical and the experimental results for a state with
$n_{th}=0.1$ and $n_{sq}=3$ is reported in Fig. \ref{f:sqth}. The
statistical error of the reconstructed number probability affects the
third decimal digit, and is not visible on the scale of the plot.
\par The estimation of parameters of  Gaussian Wigner functions through
the ML method allows one to estimate the parameters in quadratic
Hamiltonians of the generic form
\begin{eqnarray}
H=\alpha a+\alpha ^* a^\dag + \varphi a^\dag a +\frac 12 \xi a^2+\frac
12 \xi^*a^{\dag 2}\;.\label{ham}
\end{eqnarray}
In fact, the unitary evolution operator $U=e^{-iHt}$ preserves the
Gaussian form of an input state with Gaussian Wigner function. In
other words, one can use a known Gaussian state to probe and characterize an
optical device described by a Hamiltonian as in Eq. (\ref{ham}).  
Assuming $t=1$ without loss of generality, 
the Heisenberg evolution of the radiation mode $a$ is given by
\begin{eqnarray}
U^\dag \,a\,U=\gamma a+\delta a^\dag +\mu\;,
\end{eqnarray}
with 
\begin{eqnarray}
&&\gamma =\cos (\sqrt{\varphi ^2-|\xi|^2})-i\frac {\varphi}
{\sqrt{\varphi ^2-|\xi|^2}}\sin (\sqrt{\varphi ^2-|\xi|^2})\;,
\label{3eq} \\& & \delta =-i \frac{\xi ^*}{\sqrt{\varphi ^2-|\xi|^2}}
\sin(\sqrt{\varphi ^2-|\xi|^2})\;, \nonumber \\& & 
\mu=\frac{\varphi\alpha ^*-\xi^*\alpha }{\varphi ^2-|\xi|^2}
(\cos (\sqrt{\varphi ^2-|\xi|^2})-1)-i\frac{\alpha ^*}
{\sqrt{\varphi ^2-|\xi|^2}}\sin(\sqrt{\varphi ^2-|\xi|^2})\;.\nonumber
\end{eqnarray}
For an input state $\rho  $ 
with known Wigner function $W_\rho  (\beta \,,\beta ^*)$, the
corresponding output Wigner function writes
\begin{eqnarray}
W_{U\rho  U^\dag }(\beta \,,\beta ^*)=
W_\rho  [(\beta -\mu)\gamma ^*-(\beta ^*-\mu ^*)\delta \,,
(\beta^* -\mu^*)\gamma -(\beta -\mu )\delta ^*]\;.\label{wout}
\end{eqnarray}
Hence, by estimating the parameters $\gamma ,\delta ,\mu $ and
inverting Eqs. (\ref{3eq}), one obtains the ML values for $\alpha
,\varphi $, and $\xi $ of the Hamiltonian in Eq. (\ref{ham}). 
The present example can be used in practical applications for the
estimation of the gain of a phase-sensitive amplifier or equivalently
to estimate a squeezing parameter.

%% file: cap9.tex
\chapter{Classical imaging by quantum tomography}
As we showed in Chapter 2, the development of quantum tomography has
its origin in the inadequacy of classical  imaging procedures to face the
quantum problem of Wigner function reconstruction. In this chapter we 
briefly illustrate how to go back to classical imaging and profitably
use quantum tomography as a tool for image reconstruction and
compression: this is the method of  {\em Fictitious Photons
Tomography} of Ref. \cite{fict}.

\par
\begin{figure}[htb]
\begin{center}
(a)\epsfxsize=0.45\hsize\leavevmode\epsffile{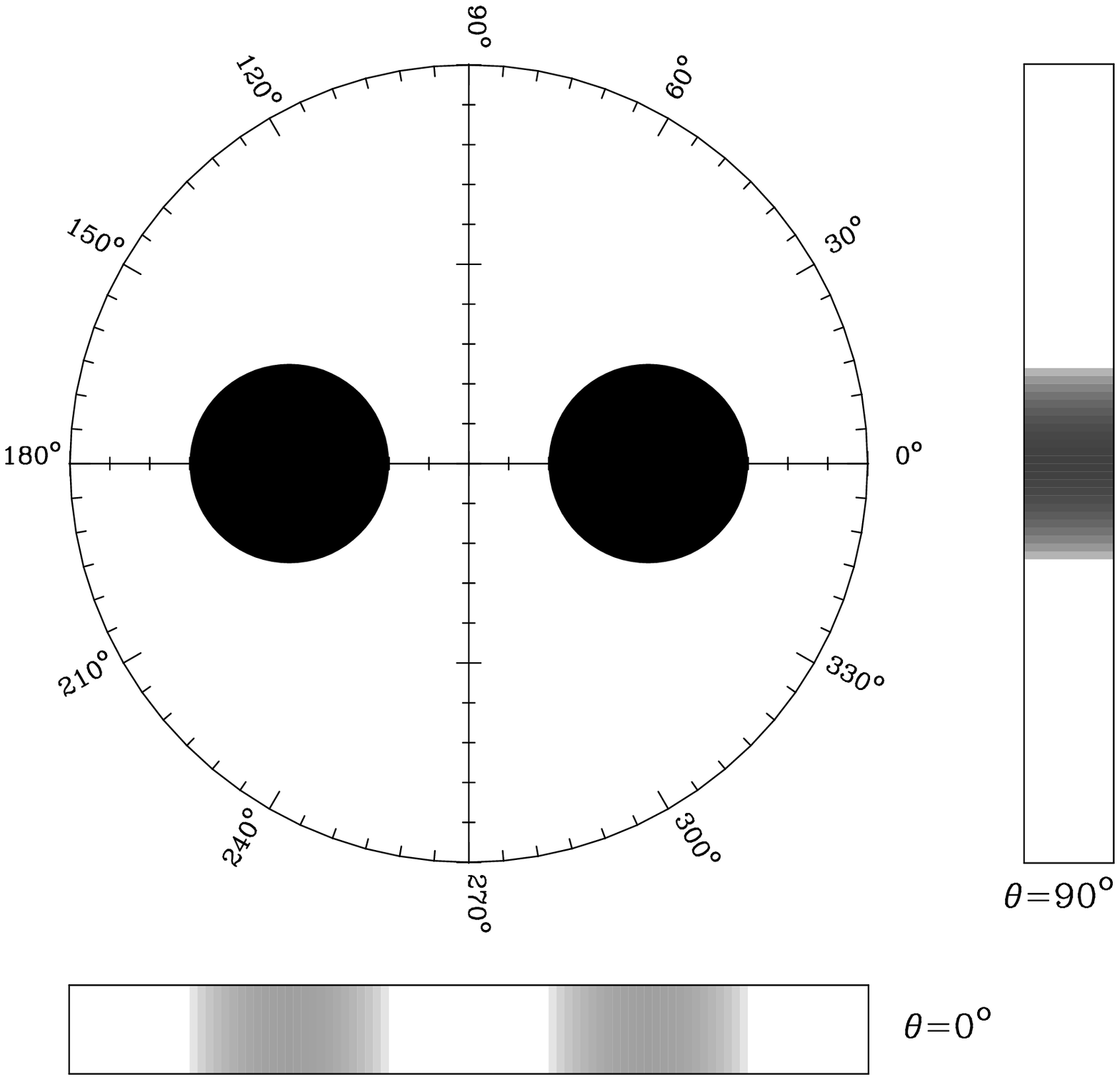}
\hfill
\epsfxsize=0.45\hsize\leavevmode\epsffile{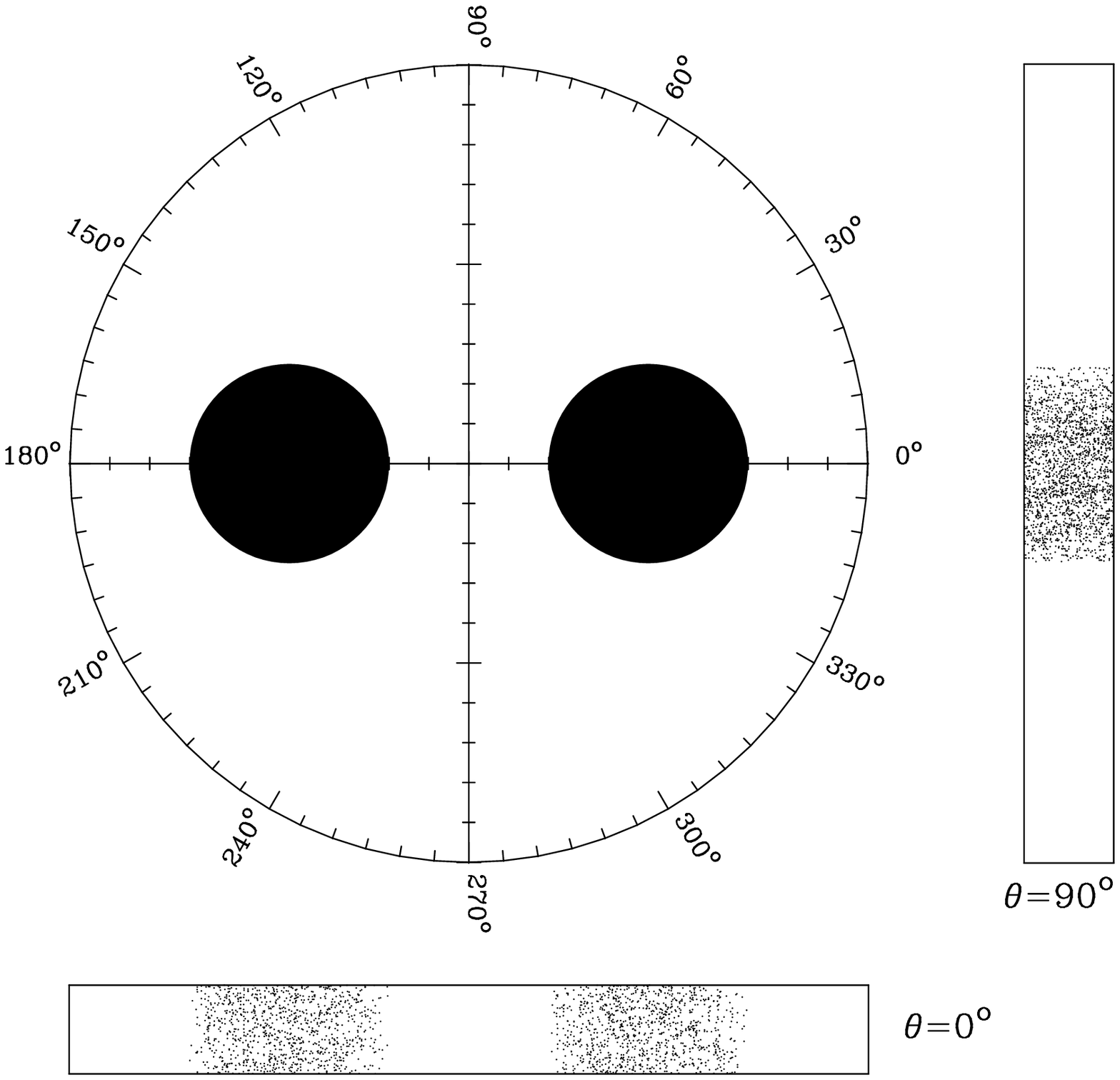}
(b) 
\end{center}
\caption{(a) Tomography of a simple object: analytical transmission
profiles are reported for $\theta =0,\pi/2$. (b) The same case of (a),
but for very weak signals: in this case the transmission profiles are
given in terms of random points on a photographic plate (here obtained
from a Monte Carlo simulation) (From Ref. \cite{fict}). }\label{f:tomo1}
\end{figure}

 The problem of tomographic imaging is to recover a mass
distribution $m(x,y)$ in a $2$-d slab from a finite collection of one
dimensional projections. The situation is schematically sketched in
Fig. \ref{f:tomo1}a where $m(x,y)$ describes two circular holes in a
uniform background. The tomographic machine, say an X-ray equipment,
collects many stripe photos of the sample from various directions
$\theta$, and then numerically performs a mathematical
transformation---the so called {\em inverse Radon transform}
\cite{natterer}---in order to reconstruct $m(x,y)$ from its radial
profiles at different $\theta$'s. The problem which is of interest for
us is when the radial profiles are not well-resolved digitalized
functions, but actually represent the density distribution of random
points, as if in our X-ray machine the beam is so weak that radial
photos are just the collection of many small spots, each from a single
X-ray photon (this situation is sketched in Fig. \ref{f:tomo1}b). It
is obvious that this case can be reduced to the previous one by
counting all points falling in a predetermined $1$-d mesh, and giving
radial profiles in form of histograms (this is what actually happens
in a real machine, using arrays of photodetectors). However, we want
to use the whole available information from each "event"---i.e. the
exact $1$-d location of each spot---in a way which is independent on
any predetermined mesh.  In practice, this situation occurs when the
signal is so weak and the machine resolution is so high (i.e. the
mesh-step is so tiny) that only zero or one photon at most can be
collected in each channel.  As we will see, this
low-signal/high-resolution case naturally brings the imaging problem
into the domain of quantum tomography. Images are identified with
Wigner functions, such to obtain a description in terms of density
matrices.  These are still trace-class matrices (corresponding to
"normalizable" images), but are no longer positive definite, because
an ``image'' generally is not a genuine Wigner function and violates
the Heisenberg relations on the complex plane (the phase space of a
single mode of radiation). Hence, such density matrices are
unphysical: they are just a mathematical tool for imaging. This is the
reason why this method has been named {\em Fictitious Photons
Tomography} \cite{fict}. As we will see in the following, the image
resolution improves by increasing the rank of the density matrix, and
in this way the present method also provides a new algorithm for image
compression, which is suited to angular image scanning.
\section{From classical to quantum imaging\label{radonsec}}
We adopt the complex notation, with $\alpha =x+iy$
representing a point in the image plane. In this way $\alpha$ and
$\alpha ^*$ are considered as independent variables, and the $2$-d
image---here denoted by the same symbol $W(\alpha,\alpha ^*)$ used
for the Wigner function---is just a generic real function of the point
in the plane. In the most general situation $W(\alpha,\alpha ^*)$ is
defined on the whole complex plane, where it is normalized to some
finite constant, and it is bounded from both below and above, with
range representing the darkness nuance. For X-ray tomography
$W(\alpha,\alpha ^*)$ roughly represents the absorption coefficient
as a function of the point $\alpha$. We consider a linear absorption
regime, i.e. the image extension is negligible with respect to the
radiation absorption length in the medium. At the same time we neglect
any diffraction effect.
 
\par As shown in Sec. \ref{convim} the customary imaging technique is
based on the inverse Radon transform. A tomography of a two
dimensional image $W(\alpha ,\alpha ^* )$ is a collection of one
dimensional projections $p(x,\theta )$ at different values of the
observation angle $\theta $. We rewrite here the definition of the
Radon transform of $W(\alpha ,\alpha ^* )$
\begin{eqnarray}
p(x,\theta ) = \int_{-\infty}^{+\infty}\!\! \frac{dy}{\pi}\;
W\left((x+iy)e^{i\theta},(x-iy)e^{-i\theta}\right) 
\;.\label{radon}
\end{eqnarray}
In Eq. (\ref{radon}) $x$ is the current coordinate along the direction
orthogonal to the projection and $y$ is the coordinate along the
projection direction.  The situation is depicted in Fig. \ref{f:tomo1}
where $W(\alpha,\alpha ^*)$ is plotted along with its $p(x,\theta )$
profiles for $\theta =0,\pi/2$ for a couple of identical circular
holes that are symmetrically disposed with respect to the origin.

\par The reconstruction of the image $W(\alpha ,\alpha ^* )$ from its
projections $p(x,\theta)$---also called ``back projection''---is given
by the inverse Radon transform, which, following the derivation in
Sec. \ref{convim}, leads to the filtering procedure
\begin{eqnarray}
W(\alpha ,\alpha ^* ) =\int_0^{\pi}\!\frac{d\theta}{2\pi} {\cal P} 
\int_{-\infty}^{+\infty}\!\! dx
{\frac {\partial p(x,\theta )/\partial x}
{x -\alpha_{\theta}}}\;,
\label{filter}
\end{eqnarray}
where $\cal P$ denotes the Cauchy principal value and $\alpha _\theta
=\hbox{Re}(\alpha e^{-i\theta})$. Eq. (\ref{filter}) is commonly used in
conventional tomographic imaging (see, for example,
Ref. \cite{mansfield}).

\par Let us now critically consider the above procedure in the case of
very weak signals, namely when $p(x,\theta)$ just represents the
probability distribution of random X-ray spots on a fine-mesh
multichannel: this situation is sketched in Fig. \ref{f:tomo1}b. From
Eq. (\ref{filter}) one can recover $W(\alpha ,\alpha ^* )$ only when
the analytical form of $p(x,\theta )$ is known.  But the experimental
outcomes of each projection actually are random data distributed
according to $p(x ,\theta)$, whereas in order to recover $W(\alpha
,\alpha ^* )$ from Eq. (\ref{filter}) one has to evaluate the first
order derivatives of $p(x ,\theta)$. The need of the analytical form for
projections $p(x ,\theta)$ requires a filtering procedure on data,
usually obtained by ``splining'' data in order to use
Eq. (\ref{filter}). 

\par The above procedure leads to approximate image reconstructions,
and the choice of any kind of smoothing parameter unavoidably affects
in a systematic way the statistics of errors. In the following we show
how quantum tomography can be used for conventional imaging in
presence of weak signals, providing both ideally controlled resolution
and reliable error statistics.  

\par The basic formula we will use is the expansion of the Wigner
function in the number representation of Eqs. (\ref{w2}) and
(\ref{w3}).  In practice, the Hilbert space has to be truncated at
some finite dimension $d_{\cal H}$, and this sets the resolution for
the reconstruction of $W(\alpha,\alpha ^*)$. However, as we will
show, this resolution can be chosen at will, independently on the
number of experimental data.

\par As previously noticed, in general an image does not correspond to
a Wigner function of a physical state, due to the fact that the
Heisenberg relations unavoidably produce only smooth Wigner functions,
whereas a conventional image can have very sharp edges. However, if
one allows the density matrix to be no longer positive definite (but
still trace class), a correspondence with images is obtained, which
holds in general. In this way every image is stored into a trace-class
matrix $\rho_{n,m}$ via quantum tomography, and a convenient
truncation of the matrix dimension $d_{\cal H}$ can be chosen.

\begin{table}[htb]
\begin{tabular}{|c|c|c|}
\hline
Symmetry & $p(x,\theta )$&$\rho$\\
\hline
Isotropy & $p(x,\theta )\equiv p(x)$            & 
$\rho_{n,m}=\rho_{n,n}\delta_{n,m}$ \\
X-axis mirror  & $p(x,\pi-\theta )= p(-x,\theta )$ & $\rho_{n,m} \in {\mathbb R}$\\  
Y-axis mirror  & $p(x,\pi-\theta )= p(x,\theta )$  & $i\rho_{n,m} \in {\mathbb R}$ \\
Inversion through the origin   & $p(x,\theta )= p(-x,\theta )$ & 
$\rho_{n,n+2d+1}=0$  \\
\hline
\end{tabular}
\caption{Geometrical symmetries of an image, analytical properties of
projections and algebraic properties of the corresponding
matrix (From Ref. \cite{fict}).
\label{t:symm}}
\end{table} 
\begin{figure}[hbt]
\begin{center}
\epsfxsize=0.65\hsize\leavevmode\epsffile{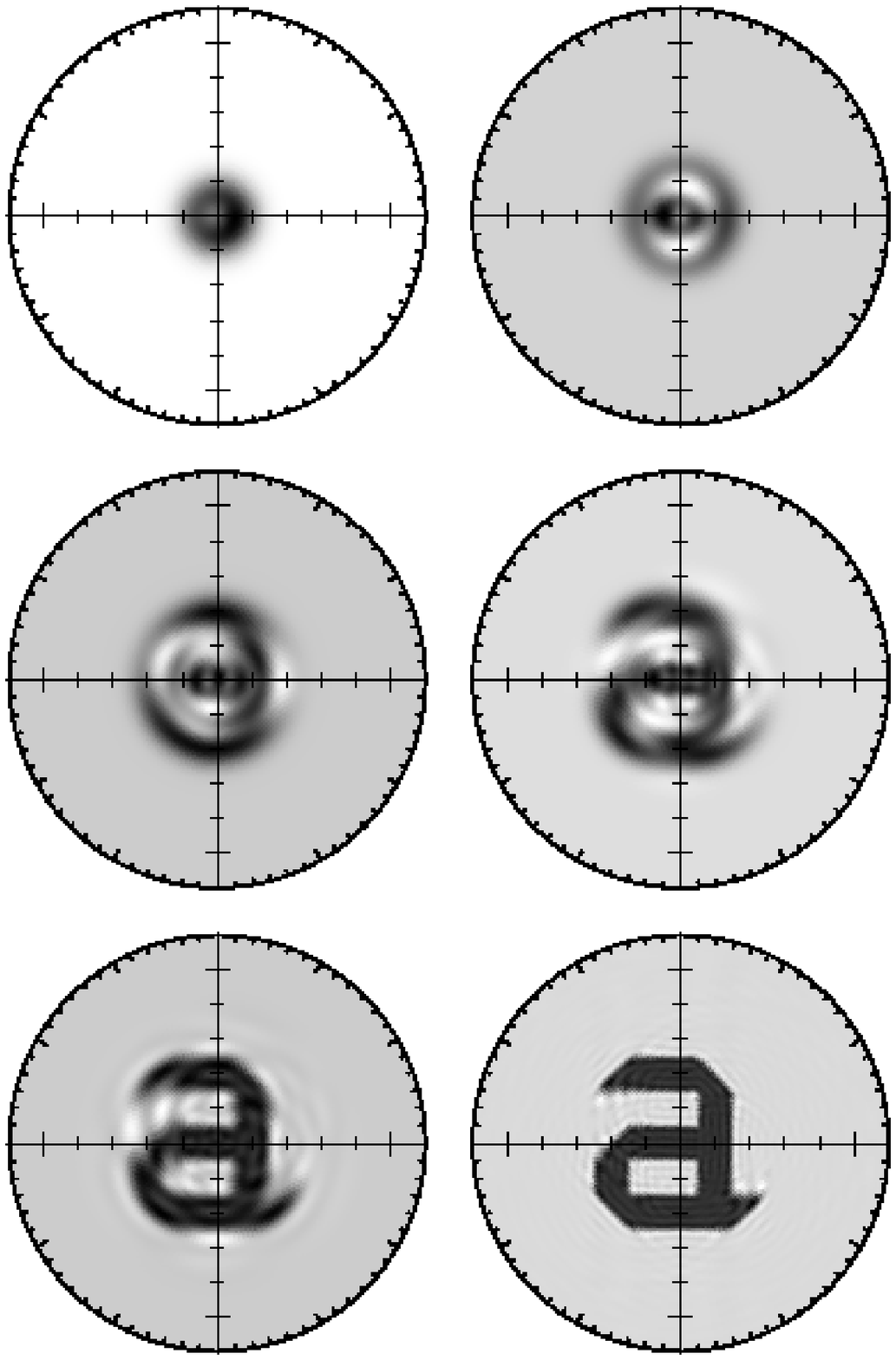}
\end{center}
\caption{Tomographic reconstruction of the font ``a'' for increasing
dimension of the truncated matrix, $d_{\cal H}=2,4,8,16,32,48$.  The
plot is obtained by averaging the kernel function $ {\cal
R}[|n+d\rangle \langle n|](x,\theta)$ of Eq. (\ref{estimat}) with
assigned analytic transmission profiles $p(x ,\theta)$, and then using
Eqs. (\ref{w2}) and (\ref{w2}) (From Ref. \cite{fict}).
}\label{f:a_analitica}
\end{figure}
\par The connection between images and matrices is the main point of
this approach: the information needed to reconstruct the image is
stored in a $d_{\cal H}\times d_{\cal H}$ matrix. For suitably chosen
dimension $d_{\cal H}$ the present method can also provide a procedure
for image compression. Notice that the correspondence between images
and trace-class matrices retains some symmetries of the image, which
manifest as algebraic properties of the matrix $\rho_{n,m}$.  For
example an isotropic image (like a uniform circle centered at the
origin) is stored in a diagonal matrix. Other symmetries are given in
Tab. \ref{t:symm}.

\par The truncated Hilbert space dimension $d_{\cal H}$ sets the
imaging resolution.  The kind of resolution can be understood by
studying the behavior of the kernels $ {\cal R}[|n+d\rangle \langle
n|](x,\theta)$ of Eq. (\ref{estimat}), which are averaged over the
experimental data in order to obtain the matrix elements
$\rho_{n,n+d}$.  Outside a region that is almost independent of $n$
and $d$, all functions $ {\cal R}[|n+d\rangle \langle n|](x,\theta)$
decrease exponentially, whereas inside this region they oscillate with
a number of oscillations linearly increasing with $2n+d$.  This
behavior produces the effects illustrated in Fig. \ref{f:a_analitica},
where we report the tomographic reconstruction of the font ``a'' for
increasing dimension $d_{\cal H}$.  The plot is obtained by
numerically integrating the kernel functions from given analytic
transmission profiles $p(x ,\theta)$.  As we see from
Fig. \ref{f:a_analitica} both the radial and the angular resolutions
improve versus $d_{\cal H}$, making the details of the image sharper
and sharper already form a relatively small truncation $d_{\cal
H}=48$.
\par
\begin{figure}[htb]
\begin{center}
\epsfxsize=0.8\hsize\leavevmode\epsffile{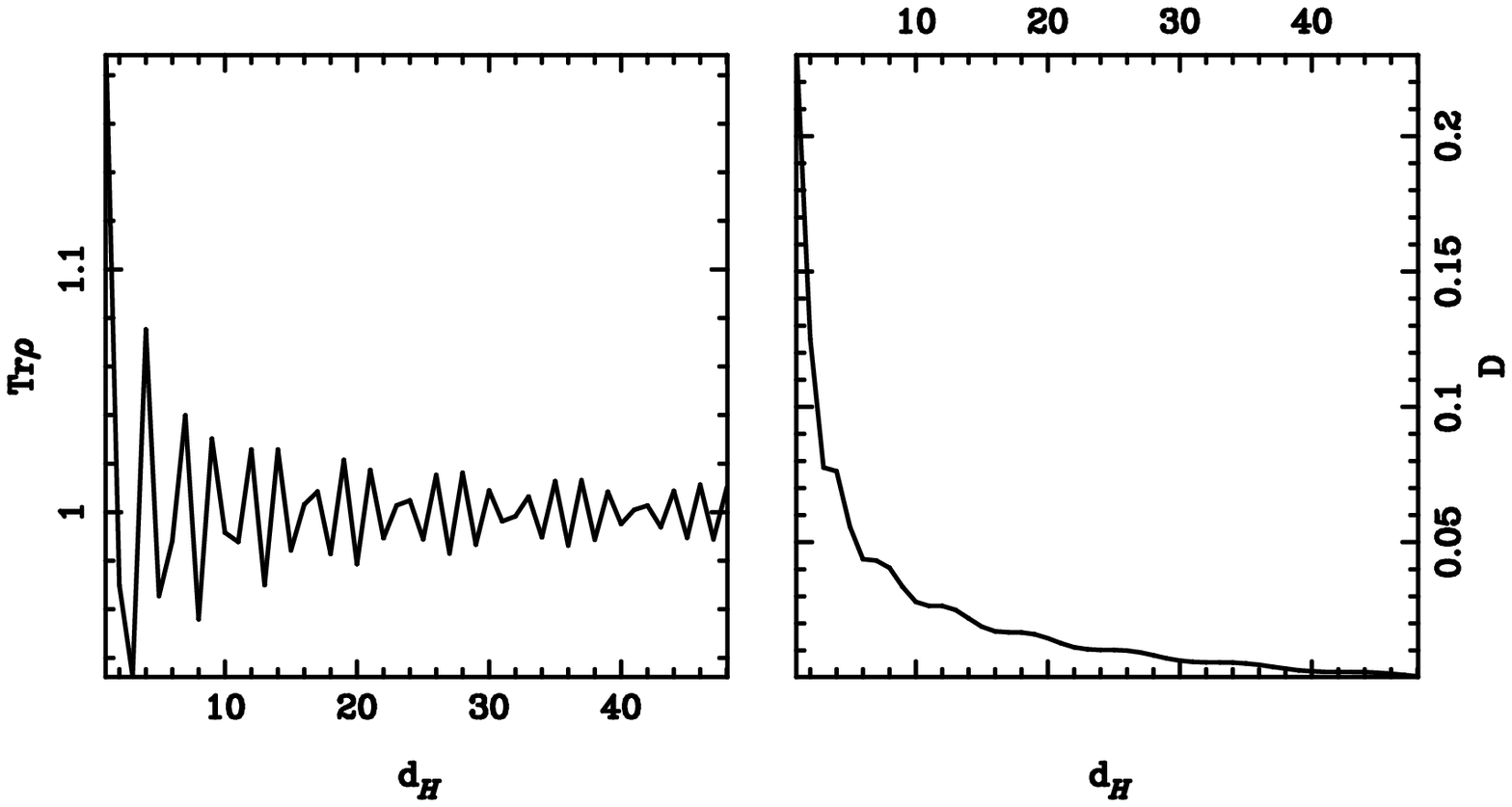}
\end{center}
\caption{Convergence of both trace and Hilbert distance $D$ in
Eq. (\ref{dist}) versus the dimensional truncation $d_{\cal H}$ of the
Hilbert space. Here the image is a uniform circle of unit radius
centered at the origin. The reconstructed matrix elements are obtained
as in Fig. \protect\ref{f:a_analitica}, whereas the exact matrix
elements are provided by Eq. (\ref{rhocircle}) (From
Ref. \cite{fict}). }\label{f:hilbert}
\end{figure}

 A quantitative measure of the precision of the tomographic
reconstruction can be given in terms of the distance $D$ between the
true and the reconstructed image, which, in turn, coincides with the
Hilbert distance $D$ between the corresponding density matrices. One
has
\begin{eqnarray}
D&=&\int d^2\alpha |\Delta W(\alpha,\alpha ^*)|^2
=\hbox{Tr}(\Delta\rho )^2 \nonumber \\&= &
\sum_{n=0}^{\infty}\Delta\rho^2_{n,n}
+2\sum_{n=0}^{\infty}\sum_{\lambda=1}^{\infty}\left|\Delta\rho^2_{n,n+
\lambda}\right|^2\;,\label{dist}
\end{eqnarray}
where $\Delta [\ldots ]=[\ldots ]_{true}-[\ldots ]_{reconstructed}$.
The convergence of $D$ versus $d_{\cal H}$ is given in
Fig. \ref{f:hilbert} for a solid circle of unit radius centered at the
origin. In this case the obtained density matrix has only diagonal
elements, according to Tab.~\ref{t:symm}. 
These are given by
\begin{eqnarray}
\rho_{n,n} = 2 \sum_{\nu=0}^{n} (-2)^{\nu}{n\choose \nu}
\Phi (1-\nu,2,2R^2)
\;,\label{rhocircle}
\end{eqnarray}
where $\Phi (\alpha ,\beta,z)$ denotes the confluent hypergeometric function
of argument $z$ and parameters $\alpha $ and $\beta$.
\begin{figure}[hbt]
\begin{center}
\epsfxsize=0.65\hsize\leavevmode\epsffile{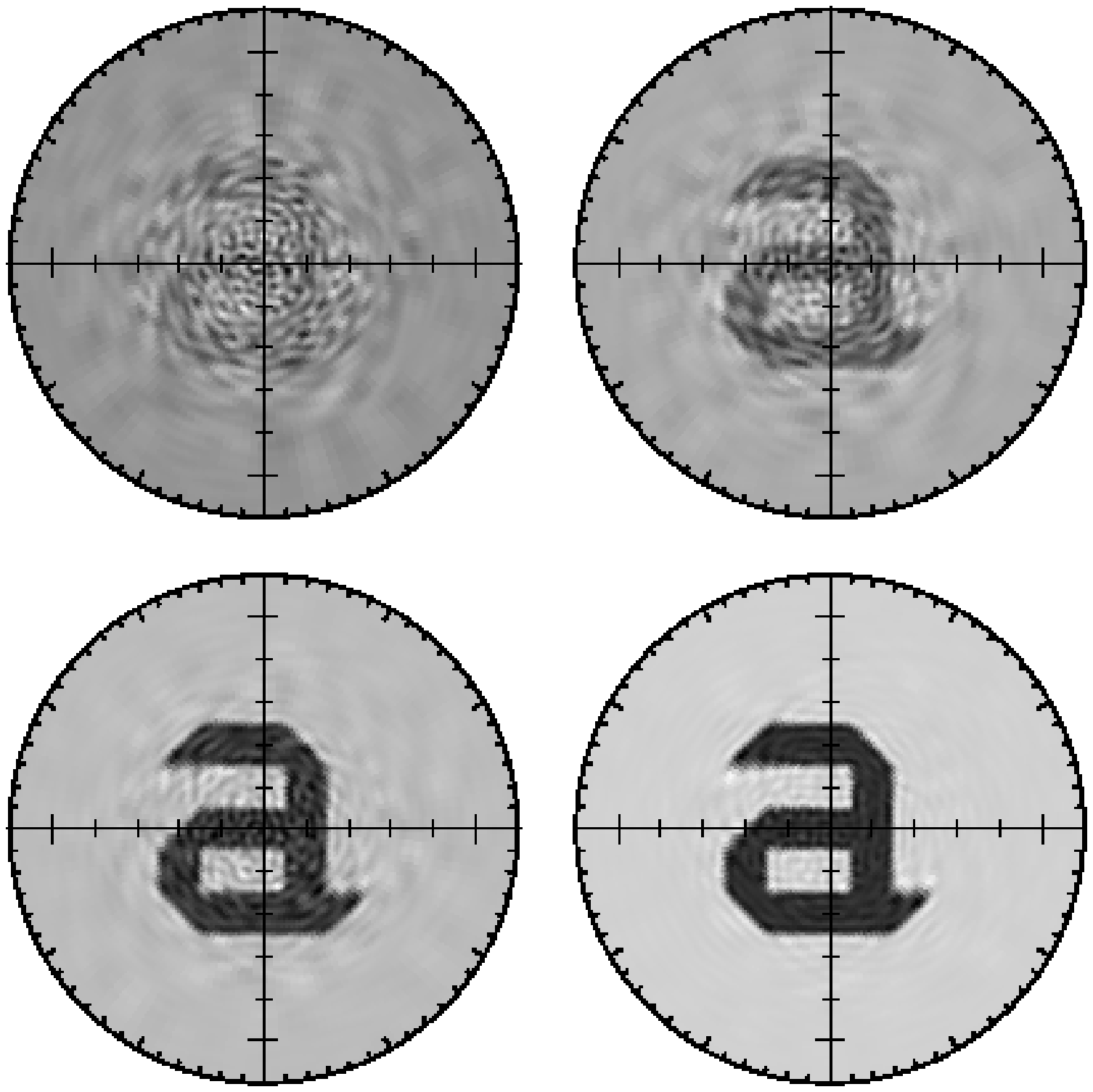}
\end{center}
\caption{Monte Carlo simulation of an experimental tomographic
reconstruction of the font ``a''. The truncation dimension is fixed at
$d_{\cal H}=48$, and the number of scanning phases is $F=100$.
The plots correspond to $10^3,10^4,10^5,10^6$ data for each phase
respectively (From Ref. \cite{fict}).}\label{f:tomo_a}
\end{figure}
\par Insofar we have analyzed the method only on the basis of given
analytic profiles $p(x,\theta)$. As already said, however, the method
is particularly advantageous in the weak-signal/high-resolution
situation, where the imaging can be achieved directly from averaging
the kernel functions on data. In this case the procedure allows to
exploit the whole available experimental resolution, whereas the image
resolution is set at will. In Fig. \ref{f:tomo_a} we report a Monte
Carlo simulation of an experimental tomographic reconstruction of the
font ``a'' for increasing number of data. All plots are obtained at
the maximum available dimension $d_{\cal H}=48$, and using $F=100$
scanning phases.  The situation occurring for small numbers of data is
given in the first plot, where the the highly resolved image still
exhibits the natural statistical fluctuations due to the limited
number of data. For larger sample the image appears sharper from the
random background, and it is clearly recognizable for a number of data
equal to $10^6$. The method is efficient also from the computational
point of view, as the time needed for image reconstruction is
quadratic in the number of elements of the density matrix, and linear
in the number of experimental data. Needless to say, imaging by
quantum homodyne tomography is at the very early stages and further
investigation is in order.